\documentclass[11pt, a4paper]{article}
\usepackage{jcappub}

\usepackage{scalerel}
\usepackage{stackengine,wasysym}

\newcommand\reallywidetilde[1]{\ThisStyle{%
  \setbox0=\hbox{$\SavedStyle#1$}%
  \stackengine{-.1\LMpt}{$\SavedStyle#1$}{%
    \stretchto{\scaleto{\SavedStyle\mkern.2mu\AC}{.5150\wd0}}{.6\ht0}%
  }{O}{c}{F}{T}{S}%
}}

\usepackage{subeqnarray}
\usepackage{lipsum}
\usepackage{accents}
\usepackage{array}
\usepackage[normalem]{ulem}

\usepackage{graphicx}
\usepackage{bm}

\usepackage{pgfplots}
\usepackage{wasysym}
\usepackage{graphicx}

\usepackage{tensor}

\newcommand{\be}{\begin{equation}}
\newcommand{\ee}{\end{equation}}
\newcommand{\bea}{\begin{eqnarray}}
\newcommand{\eea}{\end{eqnarray}}
\newcommand{\beas}{\begin{subeqnarray}}
\newcommand{\eeas}{\end{subeqnarray}}

\newcommand{\Hself}{\mathcal{J}}
\newcommand{\Anti}{\mathcal{A}}
\newcommand{\Hcal}{\mathcal{H}}
\newcommand{\Hamil}{\mathcal{V}}
\newcommand{\Hvac}{\mathcal{H}_0}
\newcommand{\vecHvac}{\vec{\mathcal{H}}_0}
\newcommand{\vecHlep}{\vec{\mathcal{H}}_{\mathrm{lep}}}
\newcommand{\Hlep}{\mathcal{H}_\mathrm{lep}}
\newcommand{\Hvacy}{\underline{\mathcal{H}}_0}
\newcommand{\Hlepy}{\underline{\mathcal{H}}_\mathrm{lep}}
\newcommand{\vecHvacy}{\vec{\underline{\mathcal{H}}}_0}
\newcommand{\vecHlepy}{\vec{\underline{\mathcal{H}}}_\mathrm{lep}}
\newcommand{\vecHeff}{\vec{\Hamil}_{\rm eff}}

\newcommand{\ATAOH}{{ATAO}-$\mathcal{V}\,$}
\newcommand{\ATAOJ}{{ATAO}-$\Hself\,$}
\newcommand{\ATAOJH}{{ATAO}-$(\Hself\pm\mathcal{V})\,$}
\newcommand{\Id}{\mathbb{I}}

\newcommand{\vrho}{\varrho}
\newcommand{\bvrho}{\bar{\varrho}}
\newcommand{\dd}{{\rm d}}

\newcommand{\com}[1]{}
\newcommand{\ii}{{\rm i}}

\newcommand{\Tcm}{T_{\rm cm}}

\newcommand{\Tr}{\mathrm{Tr}}

\newcommand{\bnu}{\bar{\nu}}

\newcommand{\x}{\mathrm{x}}
\newcommand{\y}{\mathrm{y}}
\newcommand{\z}{\mathrm{z}}
\newcommand{\me}{m_{e}}
\newcommand{\mW}{m_{\rm W}}
\newcommand{\mZ}{m_{\rm Z}}

\newcommand{\LATAO}{\langle}
\newcommand{\RATAO}{\rangle}

\newcommand{\abs}[1]{\left\lvert#1\right\rvert}
\newcommand{\norm}[1]{\left\lVert#1\right\rVert}

\definecolor{brickred}{rgb}{0.8, 0.25, 0.33}

\title{Primordial neutrino asymmetry evolution with full mean-field effects and collisions}

\author[]{Julien Froustey,}
\author[]{Cyril Pitrou}

\affiliation[]{Institut d'Astrophysique de Paris, CNRS UMR 7095, Sorbonne Universit{\'e}, 98 bis Bd Arago, 75014 Paris, France}

\emailAdd{froustey@iap.fr}
\emailAdd{pitrou@iap.fr}

\abstract{Neutrino oscillations and mean-field effects considerably enrich the phenomenology of neutrino evolution in the early Universe. Taking into account these effects, most notably the neutrino self-interaction mean-field contribution, we revisit the problem of the evolution of primordial neutrino asymmetries including for the first time the complete expression for collisions, which describe scattering and annihilations with electrons/positrons and reactions among (anti)neutrinos. We show that a generalisation of the adiabatic transfer of averaged oscillations (ATAO) scheme, a numerical method previously developed without neutrino degeneracy and based on the large separation of time scales in this problem, is sufficient to reach the same accuracy as the full quantum kinetic equation integration, but is notably faster. This approximation highlights the physics of synchronous oscillations at play in the evolution of neutrino chemical potentials, especially in the particular case with only two-neutrino mixing. In particular, it allows to understand what controls the beginning and the amplitude of oscillations, but also why there is a subsequent regime of collective oscillations with larger frequencies. We also find that it is very important to use the full collision term instead of relying on damping-like approximations, in order not to overestimate how collisions reduce these synchronous oscillations. Finally we study qualitatively how mixing parameters affect the final neutrino configuration, and in particular we show that the CP-violating Dirac phase cannot substantially affect the final $N_{\rm eff}$ nor the final electronic (anti)-neutrino spectrum, and thus should not affect cosmological observables.}

\begin{document}
\maketitle
\flushbottom


\section{Introduction}

The cosmic neutrino background, a key prediction of the standard
cosmological model, decouples around $\sim 2 \, \mathrm{MeV}$, when weak interactions become too weak
to ensure thermal equilibrium of neutrinos with the primordial plasma
of electrons, positrons and photons. Any precise determination of the
final neutrino temperatures and spectra for the various neutrino generations requires
to consider the detailed physics of this decoupling since
electrons/positrons partially annihilate into neutrinos and
antineutrinos. In the standard case where neutrinos and antineutrinos
initially follow Fermi-Dirac distributions with vanishing chemical
potentials, this has been worked out in details in the past years~\cite{Dodelson_Turner_PhRvD1992,Dolgov1992,Dolgov_NuPhB1997,Esposito_NuPhB2000,Mangano2002,Mangano2005,Grohs2015,Relic2016_revisited,Gariazzo_2019}, and the most recent predictions~\cite{Akita2020,Froustey2020,Bennett:2020zkv,FrousteyTAUP2021} recommend to consider an effective number of neutrinos $N_{\rm eff} \simeq 3.0440$ to account for the physics of this incomplete neutrino decoupling.

In thermal and chemical equilibrium, the chemical potential of a given neutrino flavour $\alpha$ and the corresponding antineutrino chemical potential are related through $\mu_\alpha =- \bar \mu_\alpha $. The initial asymmetry in a given flavour $\alpha$, defined as the difference between the neutrino and antineutrino comoving densities, is related to $\mu_\alpha$, which is not constrained {\it a priori}. One thus hopes to constrain them from their impact on cosmological observables. More specifically, to take into account the effects of momenta redshifting due to cosmological expansion, we aim at constraining the degeneracy parameters $\xi_\alpha \equiv \mu_\alpha/\Tcm$ which are conserved by expansion, where $\Tcm$ is the comoving temperature, defined as the temperature of a relativistic species that would have decoupled before electron-positron annihilations (but after muon-antimuon ones). There is first an effect of $\xi_e$ on the neutron/proton freeze-out, which affects Big-Bang Nucleosynthesis (BBN)~\cite{,Froustey2019}. Also the total energy density of a given neutrino flavour and its corresponding antineutrino is supplemented by a term $\propto \xi_\alpha^2$, leading to a modification of $N_{\rm eff}$, and this has an impact on cosmological expansion which affects both BBN and the cosmic microwave background (CMB) anisotropies. Hence, assuming a full equilibration of neutrino asymmetries with a common $\xi$, a constraint can be obtained from BBN alone~\cite{Simha:2008mt,Fields:2019pfx}, from CMB alone~\cite{Oldengott:2017tzj,Planck18}, or using a combination of both~\cite{Pitrou_2018PhysRept} to give 
\begin{equation}
    \label{eq:xi_BBN_CMB}
\xi = 0.001 \pm 0.016\,.
\end{equation}
If standard baryogenesis models involving sphalerons suggest that $\xi$ should be of the order of the baryon asymmetry $\eta = n_{\rm baryons}/n_\gamma \simeq 6.1\times 10^{-10}$~\cite{Neutrino_Cosmology,Davidson_Leptogenesis}, other proposed models like~\cite{McDonald:1999in,March-Russell:1999hpw,Gu:2010dg} manage to combine a large lepton asymmetry with the value of $\eta$. Therefore potentially “high” values of $\xi$ are not forbidden and \eqref{eq:xi_BBN_CMB} motivates why we focus in this paper on degeneracy parameters in the range $[10^{-3},10^{-1}]$. 

The total asymmetry, that is the sum over each flavour asymmetry, is preserved by the physical processes at play. However, individual asymmetries can evolve towards the average, in which case we can talk about “flavour equilibration”. The goal of this paper is to review the physics of this equilibration, that is the evolution of the degeneracy parameters, accounting for all relevant physical effects at play during neutrino decoupling. To that end, it is necessary to solve the Quantum Kinetic Equations (QKEs) that dictate the evolution of (anti)neutrino density matrices, taking into account vacuum oscillations, mean-field effects with leptons and neutrinos (the latter being referred to as self-interaction mean-field), and collision processes.

First, self-interactions have a crucial effect in delaying equilibration as they are responsible for the so-called synchronous oscillations~\cite{Pastor:2001iu,Dolgov_NuPhB2002,Abazajian2002,Wong2002}, and we find that in general there is also a second regime with {\it quasi-synchronous oscillations} having much larger frequencies. Furthermore we find that the complete form of the neutrino collision term must be used, including the full matrix structure of both reactions among neutrinos and with electrons/positrons.

Unless the chemical potential differences are very small, there is always a period when self-interactions dominate over the lepton mean-field contribution and the vacuum Hamiltonian contribution. One of the dramatic consequences is that solving the exact evolution of neutrino number densities involves very short time scales compared to the cosmological time scale, which implies that it is numerically very difficult to treat them exactly. So far, the main approach when considering non-vanishing degeneracies consisted in using a damping approximation for the collision term~\cite{Bell:1998ds,Dolgov_NuPhB2002,Pastor:2008ti,Gava:2010kz,Gava_corr,Mangano:2010ei,Johns:2016enc,Barenboim:2016shh}, either for all its components or only for its off-diagonal components. Indeed the computation of the collision term is the time-consuming step with a ${\cal O}(N^3)$ complexity, where $N$ is the number of points used to sample the neutrino spectra. We use none of these approximations, and we find from the structure of the full collision term that it cannot efficiently damp all types of synchronous oscillations, a feature that is lost when relying on damping approximations. 

In \cite{Froustey2020}, it was shown that the numerical resolution could be considerably improved, altering only subdominantly the precision of results, by using an approximate scheme which consisted in averaging over neutrino oscillations in the adiabatically evolving matter basis. In this article, we extend this method with initial degeneracies, that is taking into account the effect of the self-interaction mean-field. In section~\ref{SecTheory} we summarize the formalism used to describe the evolution of neutrino and antineutrino density matrices, and in section~\ref{SecResScheme} we detail the various numerical schemes we developed, notably the extension of the ATAO scheme when considering self-interactions. Restricting to oscillations with only two neutrinos in section~\ref{Sec2Neutrinos}, we derive analytic expressions for synchronous and quasi-synchronous oscillations. Two physically motivated cases with two-neutrino flavours are then investigated in details in section~\ref{SecRelevant2Neutrinos}. They allow to understand the evolution of neutrino asymmetry in the general case with three neutrinos, which is presented in section~\ref{Sec:3neutrinos}, along with an assessment of the dependence on the main mixing parameters (mass ordering, mixing angles, Dirac phase). Finally we discuss the main differences with existing results in the literature in section~\ref{SecDiscussion}.

\section{Neutrino evolution in the primordial Universe}\label{SecTheory}

In order to determine neutrino evolution in the early Universe, one must solve a set of quantum kinetic equations in the expanding Universe, involving both neutrino oscillations and collisions. We present in this section all the variables relevant to this problem, with a particular emphasis on the physical quantities related to the presence of a neutrino/antineutrino asymmetry.

\subsection{Reduced variables}

To account for cosmological expansion and the associated redshifting of momenta, it is standard practice to consider the comoving variables
\begin{equation}
x=\me/\Tcm\, , \qquad y=p/\Tcm\, ,\qquad z=T_\gamma/\Tcm\, ,
\end{equation}
where $\Tcm \propto a^{-1}$.  We define density matrices (in flavour
space) $\vrho_{\alpha\beta}$ for neutrinos and $\bvrho_{\alpha\beta}$
for antineutrinos with $\alpha,\beta$ running on flavour indices
$e,\mu,\tau$, or simply $\vrho$ and $\bvrho$ in indexless notation.\footnote{We adopt the same convention as
  \cite{Froustey2020} to define these density matrices.} In homogeneous and isotropic cosmology they depend only on time and energy, hence in reduced variables they depend exclusively on $x$ and $y$. We then define the comoving number density as
\begin{equation}
\mathbb{N}_\nu \equiv \int \vrho \,  {\cal D}y\,,\qquad \text{where }\qquad  {\cal D}y
\equiv \frac{y^2 \dd y }{2\pi^2} \, , 
\end{equation}
such that the true number density matrix is $(\me/x)^3 \mathbb{N}_\nu$, and
similarly for antineutrinos we define $\mathbb{N}_{\bnu}$ from $\bvrho$.\footnote{Note that, contrary to reference~\cite{Froustey2020}, we omit the bar on comoving thermodynamical quantities since we use exclusively these dimensionless variables.}

\subsection{Dynamical equations}

The dynamical evolution of the density matrices is given by the Quantum Kinetic Equations (QKEs), whose
structure is~\cite{SiglRaffelt,McKellarThomson,BlaschkeCirigliano,Froustey2020} 
\begin{equation}
  \label{eq:QKE_compact}
\begin{aligned}  
\frac{\partial \vrho}{\partial x} &= - \ii [\Hamil +
\Hself,\vrho] + \mathcal{K} \, ,\\
\frac{\partial \bvrho}{\partial x} &= + \ii [\Hamil -\Hself,\bvrho] + \bar{\mathcal{K}} \, .
\end{aligned}
\end{equation}
We distinguish three terms on the right-hand side of these equations.
\begin{itemize}
	\item The collision terms $\mathcal{K}$ and $\bar{\mathcal{K}}$ account for momentum-exchanging scattering processes. They depend on $x$, $\vrho(x,y)$, $\bvrho(x,y)$ and also on $z$ for interactions with electrons/positrons. Their full expressions can be found in~\cite{SiglRaffelt,BlaschkeCirigliano,Froustey2020}.
	\item The effective Hamiltonian, which contributes only via its traceless part, is split into its vacuum and lepton mean-field contributions $\Hamil = \Hvac + \Hlep$, with
\begin{equation}
\label{eq:Hamil_general}
  \begin{aligned}
     \Hvac &\equiv \frac{1}{xH}\left(\frac{x}{\me}\right) U
 \frac{\mathbb{M}^2}{2y}U^\dagger\,,\\
     \Hlep &\equiv - \frac{1}{x H}  \left(\frac{\me}{x}\right)^5 \, 2 \sqrt{2} G_F y \, \frac{{\mathbb{E}}_{\rm lep} + {\mathbb{P}}_{\rm lep}}{\mW^2} \,.
  \end{aligned}
\end{equation}
Here, $H \equiv \dd (\ln x) / \dd t$ is the Hubble rate, $U$ is the Pontercorvo-Maki-Nakagawa-Sakata (PMNS) \cite{GiuntiKim} mixing
matrix, $\mathbb{M}^2$ is the diagonal matrix of mass differences with
$\mathbb{M}_{ii}^2 = \Delta m_{i1}^2 \equiv m_i^2- m_1^2$. ${\mathbb{E}}_{\rm lep} = {\rm
  diag}(\rho_{e^\pm}, \rho_{\mu^\pm},0)$ is
the diagonal matrix of lepton comoving energy densities in which we can safely
ignore the contributions of $\tau^\pm$ leptons, too heavy to have a significant density in the temperature range of our study. Likewise, ${\mathbb{P}}_{\rm lep}$ is the diagonal matrix of lepton pressures.

The Hubble rate $H$ is given by the Friedmann equation, that we recall here to highlight its dependence on $x$,
\begin{equation}
    \label{eq:scaling_Hubble_x}
    H = \frac{\me}{M_{\rm Pl}} \times \frac{\me}{x^2} \times \sqrt{\frac{\rho}{3}} \qquad \text{where} \quad \rho =\rho_\gamma + \rho_{\nu, \bar{\nu}} + \rho_{e^\pm} + \rho_{\mu^\pm}  \, ,
\end{equation}
where we stress again that the energy densities are the comoving ones, differing by a factor $(\me/x)^4$ from the physical ones. $M_{\rm Pl} \simeq 2.435 \times 10^{18} \, \mathrm{GeV}$ is the reduced Planck mass.

It proves convenient to separate the $y-$dependence in the Hamiltonian---see section \ref{Sec2Neutrinos}. Hence we define 
\begin{equation}
    \label{eq:Hamil_y}
    \mathcal{H}_0 \equiv \Hvacy / y \, , \qquad  \mathcal{H}_{\rm lep} \equiv \Hlepy y \, .
\end{equation}

	\item The self-interaction Hamiltonian $\Hself$ must be included when considering neutrino asymmetries. We introduce the notation
\begin{equation}
\label{DefJ}
\Hself = \frac{1}{xH} \left(\frac{\me}{x}\right)^3  \sqrt{2} G_F \Anti \ ,  \ \  \text{where} \quad \Anti \equiv (\mathbb{N}_\nu - \mathbb{N}_{\bnu}) = \int (\vrho-\bvrho)\mathcal{D}y\,.
\end{equation}
Note that $\Anti$, referred to as the “(integrated) neutrino asymmetry", is simply proportional to the lepton number matrix $\eta_\nu \equiv \Anti/n_\gamma = \pi^2/[2 \zeta(3) z^3] \times \Anti$.
\end{itemize}

Compared to the more general form of the equation (cf. for instance equation~(2.20) in \cite{Froustey2020}), we neglected two terms. First, an asymmetric term identical to $\Hself$ but for charged leptons, that is with $\Anti \to {\mathbb{N}}_{\rm lep}$ where ${\mathbb{N}}_{\rm lep} = {\rm diag}(n_{e^-}-n_{e^+},0,0)$ is the diagonal matrix of lepton number density asymmetries after $\mu^\pm$ annihilations. Due to global charge neutrality we have $(n_{e^-} - n_{e^+}) = n_{\rm baryons}$, and given that $n_{\rm baryons}/n_\gamma \simeq 6.1\times 10^{-10}$ after electron/positron annihilations~\cite{Pitrou_2018PhysRept}, it is considerably smaller than neutrino number densities, hence completely negligible. Finally, we also neglected a symmetric term proportional to (anti)neutrino energy densities, which has the same form as $\Hlep$ but with the replacements $\mW \to \mZ$ and ${\mathbb{E}}_{\rm lep} \to {\mathbb{E}}_{\nu,\bar\nu}$, and is thus always smaller in magnitude than $\Hlep$.\footnote{For initial Fermi-Dirac distributions at the same temperature and without degeneracies, it is purely proportional to the identity matrix, hence it does not contribute to the dynamics of density matrices. If we consider instead initial degeneracies, we have $\rho_{\nu_\alpha}+\rho_{\bar{\nu}_\alpha} \propto \xi_\alpha^2$, therefore this contribution is typically smaller than $\Hlep$ (for relativistic leptons) by a factor which is of the order of the $\xi_\alpha^2$ differences.}

\subsection{Mixing parameters}
\label{subsec:Mixing_param}

We adopt the same parameterization of the PMNS matrix as in reference~\cite{Froustey2020}, that is
\begin{equation}
\label{eq:PMNS}
U = R_{23}(\theta_{23}) S R_{13}(\theta_{13}) S^\dagger R_{12}(\theta_{12}) \,,
\end{equation}
with the rotation matrices $R_{12}(\theta_{12}) \equiv R_\z(-\theta_{12})$, $R_{13}(\theta_{13}) \equiv R_\y(\theta_{13})$, $R_{23}(\theta_{23}) \equiv R_\x(-\theta_{23})$ (see appendix \ref{AppSU2SO3} for definitions). The matrix $S = \mathrm{diag}(1,1,e^{i \delta})$ is responsible for potential CP-violating effects when a non-vanishing Dirac phase $\delta$ is considered. In the standard case and except in the section~\ref{SecDiracPhase} dedicated to the effect of the Dirac phase, we set the $\delta=0$.

In all cases, and unless otherwise specified, we shall use the most recent values of the mixing parameters from the Particle Data Group \cite{PDG}, which read for the normal ordering of masses: 
\begin{align}\label{ValuesStandard}
\left(\frac{\Delta m_{21}^2}{\rm 10^{-5} \, eV^2},\frac{\Delta m_{31}^2}{\rm 10^{-3} \, eV^2},s_{12}^2,s_{23}^2,s_{13}^2\right)_{\rm NO} &= \left(7.53, 2.53, 0.307, 0.545, 0.0218 \right) \, .
\end{align}
that is we use $\theta_{12} = 0.587$, $\theta_{13}=0.148$ and
$\theta_{23}=0.831$. When considering inverted mass ordering, we keep these parameters except for the different preferred values $\theta_{23} = 0.824$ and $\Delta m_{31}^2 \simeq - 2.46 \times 10^{-3} \, \mathrm{eV}^2$.

\subsection{Neutrino asymmetry matrix $\Anti$}
\label{subsec:Anti}

Long before neutrino decoupling, that is for temperatures much larger than $2\,{\rm MeV}$, neutrinos and antineutrinos are maintained at kinetic and chemical equilibrium, thus generally following Fermi-Dirac (FD) distributions with a chemical potential $g(T_\nu, \mu, p) \equiv \left[e^{(p - \mu)/T_\nu}+1\right]^{-1}$.

Introducing the reduced variables $z_\nu = T_\nu/\Tcm$ and $\xi = \mu/\Tcm$, we rewrite this FD distribution $g(z_\nu, \xi, y) = \left[e^{(y - \xi)/z_\nu}+1\right]^{-1}$. In most of the temperature range of interest, since electrons and positrons have not annihilated and all species are coupled, we can consider\footnote{We only take $z_\nu = 1$ for the analytical discussion in order to simplify the presentation. In the numerical resolution, the spectra evolve following the QKEs and $e^\pm$ annihilations increase the neutrino temperatures.} $z_\nu = 1$. We thus define $g(\xi, y) \equiv g(1, \xi, y)$, and the initial conditions read
\begin{equation}\label{rhoinit}
\vrho_{\rm init} = {\rm  diag}\left[g(\xi_\alpha,y)\right] \, , \quad\bvrho_{\rm init} = {\rm
  diag}\left[g(-\xi_\alpha,y)\right]\,.
\end{equation}
From equation~\eqref{Intg1}, the asymmetry matrix is initially
\begin{equation}
\label{eq:init_Anti}
\Anti_{\rm init} = \frac{1}{6}{\rm diag}\left[\xi_\alpha+\frac{\xi_\alpha^3}{\pi^2}\right] \, .
\end{equation}

For reasons detailed in section~\ref{SecResScheme}, we also introduce the evolution equation for $\Anti$. It is obtained by combining the QKE \eqref{eq:QKE_compact} with the definition~\eqref{DefJ},
\begin{equation}\label{dJdx}
\frac{\dd \Anti}{\dd x} = -\ii \int \left[\Hamil,
\vrho + \bvrho\right] {\cal D}y+ \int \left(\mathcal{K} -
  \bar{\mathcal{K}}\right) {\cal D}y\,.
\end{equation}
In principle there is no need to solve this equation for $\Anti$ because it is a simple consequence of the definition~\eqref{DefJ} with equations~\eqref{eq:QKE_compact}. However, some approximate resolution schemes promote $\Anti$ to an independent variable, thus requiring this additional equation to ensure the overall consistency.

\subsection{MSW transitions}

\begin{figure}[!ht]
  \centering
  \includegraphics[]{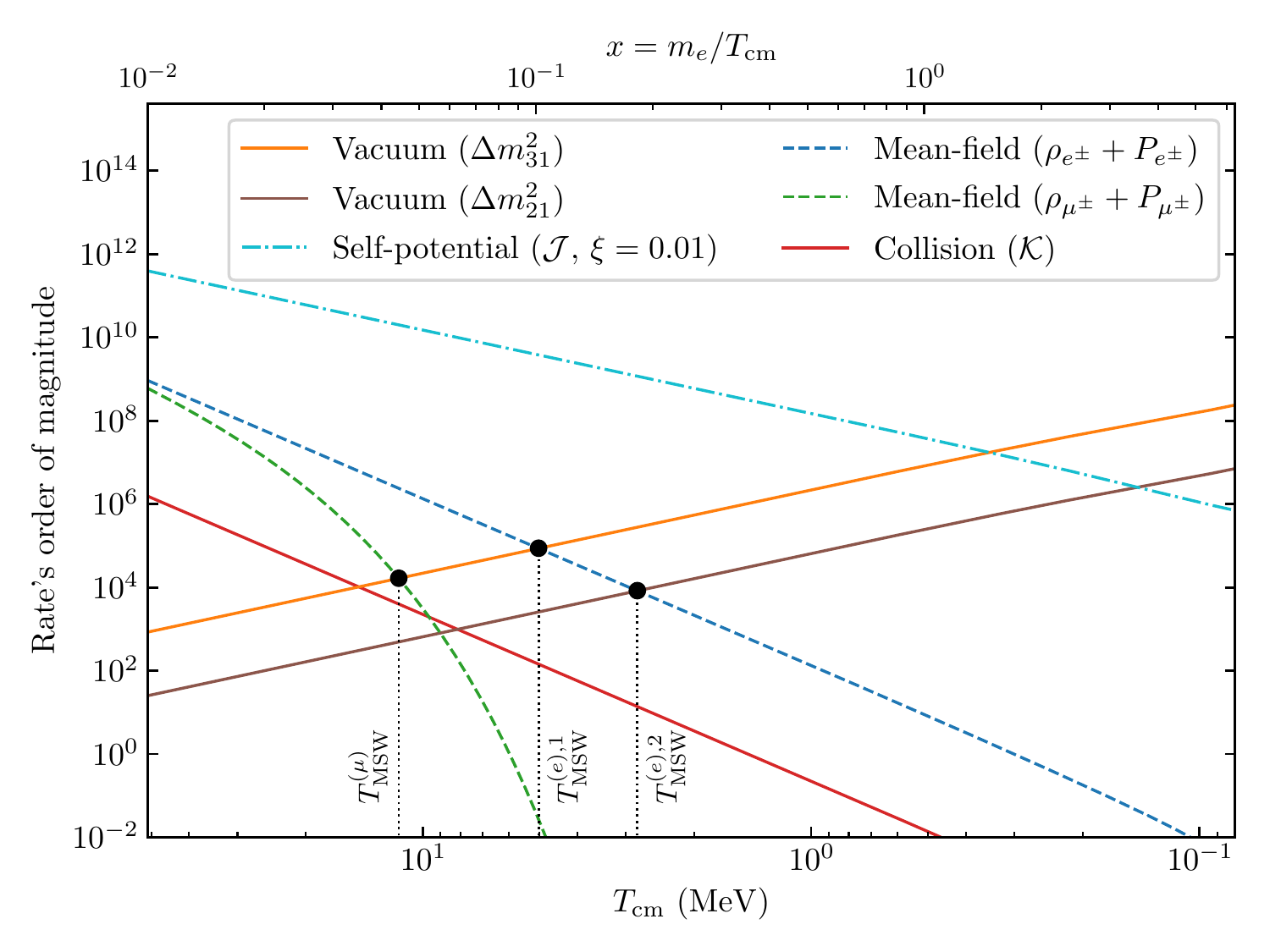}
	\caption{\label{fig:ODG_QKE} Orders of magnitude of the different rates involved in the QKE, for $y=y_{\rm eff}=3.15$ (this averaged value will be justified in section~\ref{FreqSyncOsc}). $\Hself$ is plotted with $\Anti$ given by \eqref{eq:init_Anti} and $\xi = 0.01$. The oscillation frequencies, set by the Hamiltonian eigenvalues, are very large compared to the collision rate and its variation (see reference~\cite{Froustey2020} for this discussion). As the temperature decreases, the dominant contribution in the Hamiltonian changes from $\Hself$ to $\Hamil$, $\Hamil$ itself being dominated first by $\Hlep$ and then by $\Hvac$. We estimate the magnitude of the collision rate as in figure 1 of~\cite{Mirizzi2012}.}
\end{figure}

Schematically, the lepton mean-field term scales as $\Tcm^5$, whereas the vacuum oscillation Hamiltonian scales as $1/\Tcm$ (discarding the common $1/xH$ scaling). Hence there is always a Mikheev-Smirnov-Wolfenstein (MSW) transition~\cite{MSW_W,MSW_MS} from lepton mean-field domination to vacuum domination, which can be resonant or not depending on the mixing angles and the mass ordering. There are two differences with the MSW transition in stars. First, the lepton mean-field term in stellar environments is $\sqrt{2}G_F n_{e^-}$, but it is cancelled here by the positron contribution $-\sqrt{2}G_F n_{e^+}$ since the electron/positron asymmetry is negligible. Hence, in the cosmological case the dominant lepton mean-field contribution is given by~\eqref{eq:Hamil_general}. Second, the role of the electron density profile crossed by emitted neutrinos in a star is now played by the thermal evolution of the Universe. In the cosmological context, there are three transitions which are illustrated in figure~\ref{fig:ODG_QKE}.

\begin{enumerate}
\item Since $m_\mu/\me \simeq 207$, the first MSW transition, that we call the muon-driven MSW transition, occurs when the $\mu^\pm$ mean-field effects become of the same order as the vacuum Hamiltonian associated with the large mass gap $\Delta m_{31}^2$ (or equivalently $\Delta m_{32}^2$), and this occurs around $T_{\rm MSW}^{(\mu)} \simeq 12\,{\rm MeV}$ (see section~\ref{SecDescriptionMuonTransition}), when muons are not relativistic.
\item When the $e^\pm$ mean-field effects also become of the same order as the vacuum Hamiltonian associated with the large mass gap $\Delta m_{31}^2$, we encounter the first electron-driven MSW transition around $T_{\rm MSW}^{(e), 1} \simeq 5\,{\rm MeV}$ (see section~\ref{SecElectronMSW}).
\item Finally, when the same mean-field term becomes of the same order as the vacuum Hamiltonian associated with the small mass gap $\Delta m_{21}^2$, we reach the second electron-driven MSW transition around $T_{\rm MSW}^{(e), 2} \simeq 2.8\,{\rm MeV}$ (see section~\ref{SecElectronMSW}).
\end{enumerate}
The presence of a neutrino asymmetry modifies this picture because self-interaction mean-field effects (abbreviated as {\it self-interactions} when it is clear that we do not refer to collisions between (anti)neutrinos) scale as $\Tcm^3$ and the traceless part of the neutrino asymmetry is proportional to $|\xi_\alpha-\xi_\beta|$. Unless the degeneracy differences are very small, there is always a period when self-interactions dominate over the lepton mean-field contribution until they become smaller than the vacuum contribution (see figure~\ref{fig:ODG_QKE}). At the beginning of this period of self-interaction mean-field domination we can encounter a Matter Neutrino Resonance (MNR)~\cite{Malkus:2014iqa,Johns:2016enc}, when lepton mean-field effects become smaller than self-interaction effects. However in that early phase all matrix densities and all mean-field contributions (save the negligible vacuum one), are diagonal in flavour space, therefore no conversion can occur. Conversely, describing the end of the self-interaction domination, when the vacuum Hamiltonian takes over the self-interaction effects, is rather complicated owing to the physics of synchronous oscillations which takes place, and which depends on the lepton-driven MSW transitions. One of the goals of this article is precisely to revisit the physics of these oscillations and their consequences for the equilibration of asymmetries.

Finally, note that cases where degeneracies are so small that self-interactions are at most of the order of the vacuum or lepton mean-field contributions around the MSW transition, lead to rather different physical effects since this condition is largely dependent on the magnitude of neutrino momenta. These low degeneracy regimes have been investigated in~\cite{Johns:2016enc}, but we will not explore such small values, motivated by the fact that BBN constraints are of the order of $10^{-2}$ on $\xi_e$, see equation~\eqref{eq:xi_BBN_CMB}.

\section{Resolution schemes}\label{SecResScheme}

The QKEs~\eqref{eq:QKE_compact} are challenging to solve for various reasons, the main one being the coexistence of multiple time scales: the different terms in the Hamiltonian correspond to different oscillation frequencies, that need to be compared to the collision rate---the latter being in addition particularly computationally expensive. The orders of magnitude of the different terms involved in the QKE~\eqref{eq:QKE_compact} are shown on figure~\ref{fig:ODG_QKE}.

\paragraph{ATAO approximation}

The separation of these time scales allows for the use of effective resolution schemes. 
In general, for a given Hamiltonian $\mathcal{H}$ governing the evolution of a density matrix $\vrho$, i.e., if $\partial_x \vrho = - \ii [\mathcal{H},\vrho]$, the eigenvalues of $\mathcal{H}$ give the oscillation frequencies of $\vrho$. More precisely, noting $U_\mathcal{H}$ the unitary matrix which diagonalizes $\mathcal{H}$ (that is $\mathcal{H} = U_\Hcal D_\Hcal U_\Hcal^\dagger$ with $D_\Hcal$ diagonal), the density matrix in the “$\Hcal$-basis” is $U_\Hcal^\dagger \vrho U_\Hcal$. The off-diagonal components of this matrix have oscillatory phases equal to the differences of the diagonal components of $D_\Hcal$.

If $U_\Hcal$ evolves slowly enough\footnote{$U_\Hcal$ must evolve slowly compared to the inverse oscillation frequency, that is schematically $\abs{(U_\Hcal^\dagger \partial_x U_\Hcal)^i_j} \ll \abs{(D_\Hcal)^i_i - (D_\Hcal)^j_j}$, which corresponds roughly to comparing the Hubble rate to the oscillation frequencies, see~\cite{Froustey2020} for details.}, the oscillation frequencies are so large that the off-diagonal components of $U_\Hcal^\dagger \vrho U_\Hcal$ are averaged out. Therefore, transforming back to the flavour basis, we define the averaged matrix $\LATAO \vrho \RATAO_\Hcal $ by
\begin{equation}
\label{DefAverage}
\LATAO\vrho \RATAO_\Hcal \equiv U_\Hcal \reallywidetilde{\left( U^\dagger_\Hcal \vrho U_\Hcal \right)} U_\Hcal^\dagger  \, .
\end{equation}
The wide overtilde notation means that we keep only the diagonal part---thus neglecting the fast off-diagonal oscillatory evolution which averages to zero. This procedure requires that the diagonalizing basis changes slowly relative to the oscillations, which is a standard case of adiabatic approximation. Since oscillations are averaged throughout the adiabatic evolution of the Hamiltonian, the \emph{adiabatic transfer of averaged oscillations} (ATAO) consists in the approximation
\begin{equation}
\label{eq:def_ATAO}
 \vrho \simeq \LATAO \vrho \RATAO_\Hcal  \quad \text{i.e.} \quad \vrho = U_\Hcal \widetilde{\vrho}_\Hcal U_\Hcal^\dagger \, ,
 \end{equation}
with $\widetilde{\vrho}_\Hcal$ diagonal. When including collisions, we account for their effects on time scales much larger than the one set by $\mathcal{H}$, which leads to the evolution equations
\begin{equation}\label{ATAOrhodot}
 \partial_x \widetilde{\vrho}_\Hcal = U^\dagger_\mathcal{H} \LATAO\mathcal{K}\RATAO_\mathcal{H}  U_\mathcal{H} = \widetilde{\mathcal{K}}_\Hcal\,. 
\end{equation}
Note that the collision term depends on $\vrho$, which is evaluated with the approximation~\eqref{eq:def_ATAO}. 

Such a situation is encountered by neutrinos in the early universe: the results of figure~\ref{fig:ODG_QKE} show that the Hamiltonian governing the evolution of $\vrho$ is progressively dominated, as the temperature decreases, by the self-potential (and the lepton mean-field), then by the vacuum contribution, and we now detail the associated approximation schemes.

\subsection{\ATAOH}

If the self-potential can be ignored (for instance if we consider a case without neutrino asymmetries), the fast scale is set by the Hamiltonian $\Hamil$ and we will call this situation the \ATAOH approximation, which was used in \cite{Froustey2020}. As previously explained, we thus approximate\footnote{Concerning $\bvrho$, it is equivalent to average it around $\pm \Hamil$, hence our choice to use $\Hamil$ for both $\vrho$ and $\bvrho$.} $\vrho \simeq \LATAO \vrho \RATAO_{\Hamil}$ and $\bvrho \simeq \LATAO \bvrho \RATAO_{\Hamil}$, such that
\begin{equation}\label{rtildetorhoATAOH}
\vrho = U_{\Hamil} \, \widetilde{\vrho}_{\Hamil} \,U^\dagger_{ \Hamil}\,,\qquad \bvrho = U_{\Hamil}\, \widetilde{\bvrho}_{\Hamil} \,U^\dagger_{\Hamil}\,,
\end{equation}
with $\widetilde{\vrho}_{\Hamil}$ and $\widetilde{\bvrho}_{\Hamil}$ being diagonal. Therefore, it is convenient to solve for the $N_\nu$ diagonal components of these variables instead of the $N_\nu^2$ variables of the density matrices in flavour basis (which, in this approximation, are not independent). The evolution equation~\eqref{ATAOrhodot} leads to
\begin{equation}\label{BasicATAOH}
\partial_x \widetilde{\vrho}_\Hamil = \widetilde{\mathcal{K}}_\Hamil[\vrho,\bvrho]\,,\qquad \partial_x \widetilde{\bvrho}_\Hamil = \widetilde{\bar{\mathcal{K}}}_\Hamil[\vrho,\bvrho]\,.
\end{equation}
Since the collision term depends on $\vrho,\bvrho$, this means that the evolved variables $\widetilde{\vrho}_\Hamil$ and $\widetilde{\bvrho}_\Hamil$ are transformed to the flavour basis with \eqref{rtildetorhoATAOH}, so as to evaluate the collision term whose values  in flavour space are eventually transformed back into the matter basis. We then keep only their diagonal components through 
\begin{equation}
\widetilde{\mathcal{K}}_\Hamil \equiv \reallywidetilde{\left( U^\dagger_\Hamil \mathcal{K} U_\Hamil \right)}\,,\qquad \widetilde{\bar{\mathcal{K}}}_\Hamil \equiv \reallywidetilde{\left( U^\dagger_\Hamil \bar{\mathcal{K}} U_\Hamil \right)}\,.
\end{equation}

Actually, $\Hamil$ depends on both $x$ and $y$, and so does $U_\Hamil$. Hence this averaging scheme is momentum-dependent, which is a central feature to understand the evolution of density matrices. When lepton mean-field effects can be ignored, then the $y$ dependence is the same for all momenta (a $1/y$ prefactor in $\Hamil \simeq \Hvac$) and the unitary matrices $U_{\Hamil}$ do not depend on $y$ anymore since they all reduce to the PMNS matrix. 

\subsection{\ATAOJH}

When neutrino asymmetries cannot be ignored, we see on figure~\ref{fig:ODG_QKE} that there is a range of temperatures for which $\Hself$ must necessarily be included in the Hamiltonian. As can be seen in the QKEs~\eqref{eq:QKE_compact}, the Hamiltonian for $\vrho$ is then $\Hself + \Hamil$ while it is $\Hself - \Hamil$ for $\bvrho$. Therefore, the \ATAOJH approximation reads $\vrho \simeq \LATAO\vrho \RATAO_{\Hself+\Hamil}$ and $\bvrho \simeq \LATAO\bvrho \RATAO_{\Hself-\Hamil}$, such that 
\begin{equation}
\vrho = U_{\Hself + \Hamil} \, \widetilde{\vrho}_{\Hself + \Hamil} \,U^\dagger_{\Hself + \Hamil}\,,\qquad \bvrho = U_{\Hself - \Hamil}\, \widetilde{\bvrho}_{\Hself - \Hamil} \,U^\dagger_{\Hself - \Hamil}\, ,
\end{equation}
where $\widetilde{\vrho}_{\Hself + \Hamil}$ and $\widetilde{\bvrho}_{\Hself - \Hamil}$ are diagonal. We solve the evolution of $\vrho, \bvrho$ on timescales much larger than the one set by $\Hself \pm \Hamil$, on which oscillations are averaged, hence the evolution equation is given by~\eqref{ATAOrhodot}
\begin{equation}\label{BasicATAOJH}
\partial_x \widetilde{\vrho}_{\Hself +\Hamil} = \widetilde{\mathcal{K}}_{\Hself +\Hamil}[\vrho,\bvrho]\,,\qquad \partial_x \widetilde{\bvrho}_{\Hself - \Hamil} = \widetilde{\bar{\mathcal{K}}}_{\Hself - \Hamil}[\vrho,\bvrho]\,.
\end{equation}
The method is similar to the \ATAOH case, but we need to handle the fact that the Hamiltonian itself depends on $\vrho$, through the self-potential $\Hself$. In order to compute it at each time step, we would need to keep track of the $N_\nu^2$ entries of each density matrix in the flavour basis. A better possibility, which we choose, consists in promoting $\Hself$ (actually, $\Anti$) to be an independent variable with its own evolution equation~\eqref{dJdx}. Equation~\eqref{DefJ} is then only used to set the initial value of $\Hself$ from the initial conditions on $\vrho$, $\bvrho$. In doing so, we go from $2 \times N\times N_\nu^2$ to $2 \times N \times N_\nu + N_\nu^2$ variables, with $N$ the number of momentum nodes (cf.~section~\ref{subsec:numerics}). We stress that the evolution of $\Anti$ depends on the full collision terms in flavour space, and not just on the diagonal components in the matter basis $\widetilde{\mathcal{K}}_{\Hself \pm\Hamil}$, as is the case for $\widetilde{\vrho}_{\Hself +\Hamil}$ and $\widetilde{\bvrho}_{\Hself -\Hamil}$.

\paragraph{High temperatures: \ATAOJ} It is clear from figure~\ref{fig:ODG_QKE} that at large temperatures, $\Hself$ largely dominates $\Hamil \simeq \Hlep$ (except for very small $\xi$ that lie outside the range of values we span here). That is why one could consider an even simpler \ATAOJ approximation, where $\Hself \pm \Hamil$ is replaced by $\Hself$. In that case, the changes of basis for $\vrho$ and $\bvrho$ are achieved with the same matrix $U_\Hself$. In section~\ref{FreqSyncOsc}, we show that this “leading order” Hamiltonian leads to theoretical estimates of synchronous oscillations frequencies in agreement with the existing literature, while using the full \ATAOJH allows to get an important correction which is responsible for quasi-synchronous oscillations.
The weight of $\Hamil$ in the \ATAOJH scheme becomes more important when the temperature decreases (i.e., $x$ increases), since $\Hself \propto x^{-2}$ and $\Hvac \propto x^2$. 

\subsection{QKE}

The QKE method is not an approximation scheme, but consists instead in solving exactly the neutrino and antineutrino evolutions, that is equations~\eqref{eq:QKE_compact}. However
these equations are very stiff at early times given that all terms except the vacuum one increase for
large temperatures. Therefore, integration times are typically much longer, in addition to the fact that we need to keep track of the $N_\nu^2$ entries of each density matrix in the flavour basis, contrary to the $N_\nu$ diagonal ones in the matter basis when using an ATAO framework.

\subsection{Numerical methods}
\label{subsec:numerics}

The general method used to solve for the time evolution of density matrices is described in section~4.1 of~\cite{Froustey2020}. The neutrino spectra are sampled on a grid and we have several possible choices for the spacing of the reduced momenta $y$ in this grid. We found that in the context of asymmetry equilibration, a linear spacing is much more adequate than the Gauss-Laguerre quadrature. All numerical results presented in this article are performed with an extension of the code {\tt NEVO}, using a linear grid with $N=40$ points, the minimum and maximum momenta being chosen as described in \cite{Froustey2020}. We start the numerical resolutions at $\Tcm = 20\,{\rm MeV}$, the final temperature depending on the particular configuration investigated. For initial conditions, we set $z_{\rm init}$ using that photons, $e^\pm$ and neutrinos are fully thermalized with a common temperature, as detailed in appendix~\ref{SectionPlasma}. In the case of vanishing degeneracies, this determines $z_{\rm init}-1 \simeq 7.42 \times 10^{-6}$. The subsequent evolution of $z$ is determined with~\eqref{D1Froustey2020}.

In the general QKE method, the only difference in the code is the contribution of commutators of the type $[\Anti, \vrho]$ and  $[\Anti, \bvrho]$ in \eqref{eq:QKE_compact}. However, when using the \ATAOJH method, one needs to add $N_\nu^2$ variables corresponding to the degrees of freedom of $\Anti$ whose evolution is determined by \eqref{dJdx}. 

When equations are stiff, we must rely on implicit methods that
require the computation of the Jacobian of the system of differential
equations. The default method consists in using a finite difference
estimation. The complexity of the calculation of the collision term is ${\cal O}(N^3)$ since for each momentum one must compute on a two-dimensional integral \cite{Dolgov_NuPhB1997}. Hence with finite differences the complexity for the Jacobian is ${\cal O}(N^4)$. However we can provide its explicit form to the solver and it reduces its evaluation to ${\cal O}(N^3)$. This method was used in \cite{Froustey2020} in both the QKE and the \ATAOH schemes. 

This powerful numerical technique can be extended to the \ATAOJH scheme, and the essential steps are described in appendix~\ref{AppJacobJ}.
Since we only add $N_\nu^2$ variables, the complexity remains ${\cal O}(N^3)$. All in all, we found that the code was at least ten times faster with the \ATAOJH scheme, and even more at low temperatures where the fast oscillations (see  next section) slow even more the QKE algorithm.

\section{Synchronous oscillations with two neutrinos}\label{Sec2Neutrinos}

The presence (and domination) of the self-interaction mean-field in the QKEs radically changes the phenomenology of neutrino evolution. This non-linear term notably leads to oscillations of all momentum-modes at a common frequency, a phenomenon named \emph{synchronous oscillations}, studied both numerically \cite{Pastor:2001iu,Dolgov_NuPhB2002,Mangano:2010ei} and analytically \cite{Abazajian2002,Wong2002}. In this section, we extend this theoretical work in the framework of the ATAO approximations we developed: this allows to explicitly calculate the next-to-leading order contribution to the oscillation frequency that was not considered in previous works, and that we check numerically in the next section.

We restrict to a two-flavour case, which allows to easily perform the following calculations thanks to the vector representation of $2 \times 2$ Hermitian matrices. We do not specify yet the values of the mixing parameters, as they will be set for different physical setups in section~\ref{SecRelevant2Neutrinos}.

Let us thus consider in this section the vacuum Hamiltonian of the form
\begin{equation}
\label{eq:Hvac_2nu}
    \Hvac = \frac{1}{xH} \left( \frac{x}{\me} \right) U \begin{pmatrix} 0 & 0 \\ 0 & \Delta m^2/2y
    \end{pmatrix} U^\dagger \quad \text{with} \quad U = \begin{pmatrix} \cos{\theta} & \sin{\theta} \\ 
    - \sin{\theta} & \cos{\theta} \end{pmatrix} \, ,
\end{equation}
along with the lepton mean-field contribution of the type
\begin{equation}
\label{eq:Hlep_2nu}
    \Hlep = - \frac{1}{xH} \left(\frac{\me}{x}\right)^5 \frac{2 \sqrt{2} G_F y}{\mW^2}
    \begin{pmatrix}
    \rho_{l^\pm} + P_{l^\pm} & 0 \\
    0 & 0
    \end{pmatrix} \, .
\end{equation}
In order to maintain a similar expansion history as in the case of three neutrinos, we add one fully decoupled thermalised neutrino flavour to the energy content of the Universe when studying the case of only two neutrino oscillations.

\subsection{Transformation to vectors}

It is customary to rephrase the density matrix evolution as an evolution for vectors using the relation between a Hermitian $2\times2$ matrix $P$, and a vector of $\mathbb{R}^3$ $\vec{P}$
\begin{equation}\label{MatrixToVector}
P = \frac{1}{2}P^0 \Id  +\frac{1}{2} \vec{P} \cdot \vec{\sigma}  \, ,
\end{equation}
where $\vec{\sigma} = \left(\sigma_\x, \sigma_\y, \sigma_z \right)$ is the “vector” of Pauli matrices. Commutators of matrices are then handled using $[\sigma_i,\sigma_j] = 2 \ii
\epsilon_{ijk} \sigma_k$ as we obtain
\begin{equation}
-\ii [P, Q] =  \frac{1}{2} \left( \vec{P} \wedge \vec{Q} \right) \cdot \vec{\sigma} \,.
\end{equation}
The evolution of the neutrino and antineutrino density matrices in vector notations\footnote{For consistency, we write the “vector part” of the two-neutrino density matrix $\vec{\vrho}$, while it is common in the literature to call this the \emph{polarization vector} $\vec{P}$ \cite{SiglRaffelt,Dolgov_NuPhB2002,Johns:2016enc}.} are immediately obtained to be
\begin{equation}
\label{eq:QKE_2nu}
\partial_x \vec{\vrho} =  \left(\vec{\Hamil}  + \vec{\Hself}\right) \wedge \vec{\vrho} +
\vec{\mathcal{K}}\,,\qquad 
\partial_x \vec{\bvrho} = \left( -\vec{\Hamil}  +
\vec{\Hself}\right) \wedge \vec{\bvrho} + \vec{\bar{\mathcal{K}}}\,,
\end{equation}
which we must supplement by
\begin{equation}
\partial_x \vrho^0 = \mathcal{K}^0\,,\qquad \partial_x \bvrho^0 = \bar{\mathcal{K}}^0\,,
\end{equation}
to account for the evolution of the trace part of density matrices. 

In the QKE~\eqref{eq:QKE_2nu}, the vector form of the Hamiltonian $\vec{\Hamil}= \vec{\mathcal{H}}_0 + \vec{\cal H}_{\rm lep}$ is the sum of the vacuum contribution obtained from \eqref{eq:Hvac_2nu}
\begin{equation}
\label{eq:Hvec}
\vec{\mathcal{H}}_0 = \frac{1}{xH} \left(\frac{x}{\me}\right) \frac{\Delta m^2}{2 y} \begin{pmatrix} \sin(2 \theta) \\ 0 \\ - \cos(2 \theta) \end{pmatrix} \, ,
\end{equation}
and the lepton mean-field one, derived from \eqref{eq:Hlep_2nu},
\begin{equation}
\label{eq:Hlepvec}
\vec{\mathcal{H}}_{\rm lep} = - \frac{1}{xH} \left(\frac{\me}{x}\right)^5 \frac{2 \sqrt{2} G_F y}{\mW^2} \begin{pmatrix} 0 \\ 0 \\ \rho_{l^\pm} + P_{l^\pm} \end{pmatrix} \, .
\end{equation}
Finally, the asymmetry vector evolves as
\begin{equation}\label{ddxvecJ}
\frac{\dd \vec{\Anti}}{\dd x} = \int{\left(\vec{\Hamil} \wedge [\vec{\vrho } + \vec{\bvrho}] \right) \mathcal{D}y} + \int{\left(\vec{\mathcal{K}} - \vec{\bar{\mathcal{K}}}\right) \mathcal{D} y}\,.
\end{equation}

This vector formalism allows for a more visual representation of the ATAO schemes. Averaging $\vrho$ with respect to an Hamiltonian $\Hcal$ corresponds to \emph{projecting $\vec{\vrho}$ onto $\vec{\Hcal}$}. To see this, we first note that the restriction to the diagonal part of an Hermitian two-by-two matrix corresponds to a projection along $\vec{e}_\z$ in vector notation. Hence when applying the averaging definition~\eqref{DefAverage}, the first step is the rotation which aligns $\vec{\Hcal}$ with $\vec{e}_\z$, then the diagonal part restriction selects only the $\z$-component of this rotation $\vec{\vrho}_\Hcal$, and finally it is rotated back into the initial frame. As a result one has, in the case of two neutrinos,
\begin{equation}
\overrightarrow{\LATAO \vrho \RATAO}_{\Hcal} = (\vec{\vrho} \cdot
\hat{\Hcal}) \hat{\Hcal}
\end{equation}
where $\hat{\mathcal{H}}$ is the unit vector in the direction of $\vec{\mathcal{H}}$. Since the equations of motion~\eqref{eq:QKE_2nu} correspond to instantaneous precessions set by $\vec{\mathcal{H}}$ (up to the collision term), the averaging procedure corresponds to projecting along that precession vector, i.e. removing the fast rotating part that is orthogonal to it.

\subsection{Frequency of synchronous oscillations}
\label{FreqSyncOsc}

\begin{figure}[!ht]
  \centering
  \includegraphics[]{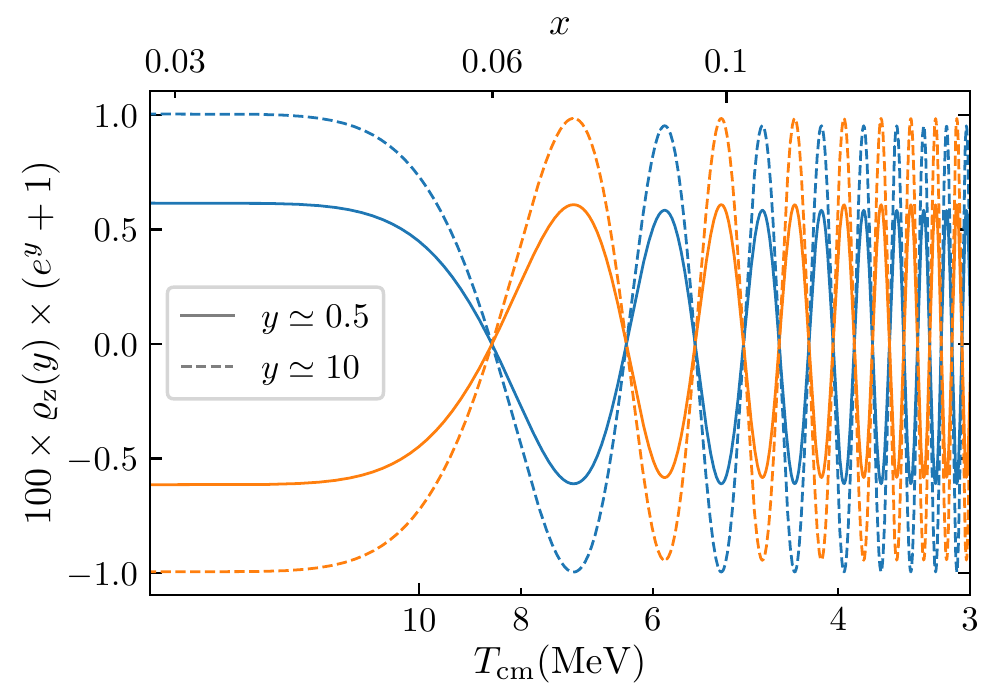}
	\caption{\label{fig:SyncOsc} Synchronous oscillations in a two-neutrino $\nu_\mu - \nu_\tau$ case with $\Delta m^2 = 2.45\times10^{-3} \, \mathrm{eV}^2$, $\theta=0.831$, without collisions. In blue $\vrho_\z = \vrho_{\mu \mu} - \vrho_{\tau \tau}$ and in orange $\bvrho_\z$. The initial degeneracy parameters are $\xi_\mu = 0.01$ and $\xi_\tau = 0$.}
\end{figure}

In some setups where the non-linear self-potential term in the QKEs dominates, such as dense neutrino gases or the early universe (for not too small asymmetries), it has been shown in references~\cite{Samuel:1993uw,Kostelecky:1993yt,Kostelecky:1993dm,Kostelecky:1993ys,Pastor:2001iu,Dolgov_NuPhB2002} that neutrinos develop so-called momentum-independent \emph{synchronous oscillations}, with all $y-$modes being “locked” on the asymmetry vector $\vec{\Anti}$. This is shown on figure~\ref{fig:SyncOsc}, where the physical parameters are the same as in the upcoming section~\ref{SecMuonMSW}.

To understand this phenomenon and make quantitative predictions regarding the behaviour of the system of neutrinos and antineutrinos in different setups, we will first ignore the effect of collisions. The initial density matrices are given by equations~\eqref{rhoinit}.
Hence the initial vector components are $\vrho_\z(y) = g(\xi_1,y) - g(\xi_2,y)$ and $\bvrho_\z(y) = g(-\xi_1,y) - g(-\xi_2,y)$, and $\vrho_{\x, \y}(y) = \bvrho_{\x, \y}(y) = 0$. Since we neglect collisions in this section, it is clear from~\eqref{eq:QKE_2nu} that the norms of $\vec{\vrho}$ and $\vec{\bvrho}$ are conserved. The adiabatic evolution of these vectors thus consists in a rotation so as to follow the direction of their Hamiltonian. Therefore, in the \ATAOJH approximation, we can write the density matrix vectors
\begin{equation}
\label{eq:vrho_ATAOJV}
\begin{aligned}
\vec{{\vrho}} &= \abs{g(\xi_1,y)- g(\xi_2,y)} \widehat{\Hself+\Hamil} \, , \\
\vec{{\bvrho}}  &= - \abs{g(-\xi_1,y)- g(-\xi_2,y)} \widehat{\Hself-\Hamil} \, ,
\end{aligned}
\end{equation}
where $\widehat{\Hself+\Hamil}$ is the unit vector in the direction of $\vec{\Hself}+\vec{\Hamil}$. One must remember that at the initial temperatures we consider ($\Tcm \sim 20 \, \mathrm{MeV}$), the Hamiltonian is largely dominated by $\Hself$. The \ATAOJ approximation then corresponds to discarding $\Hamil$ in the above expressions, and this will give the leading order behaviour of the asymmetry.

\paragraph{Leading order} Let us then focus on this high temperature region first, when the misalignment between $\vec{\vrho}$ and $\vec{\bvrho}$ is negligible, i.e.,
\begin{equation}
    \vec{\vrho} =  \abs{g(\xi_1,y)- g(\xi_2,y)} \widehat{\Hself} \, , \qquad
\vec{{\bvrho}}  = - \abs{g(-\xi_1,y)- g(-\xi_2,y)} \widehat{\Hself} \, .
\end{equation}
Hence, the asymmetry vector is obtained from~\eqref{DefJ} and~\eqref{IntegralsFD} and reads
\begin{equation}
\label{JtohatJ}
\vec{\Anti} = \frac{1}{6}\left\lvert \xi_1-\xi_2 \right\rvert \left(1+\frac{\xi_1^2+\xi_2^2+\xi_1\xi_2}{\pi^2}\right)\widehat{\Anti} \, ,
\end{equation}
with the unit vector definition $\widehat{\Anti} = \widehat{\Hself}$, which is equal initially to ${\rm sgn}(\xi_1 - \xi_2) \vec{e}_\z$. We can use the expressions of $\vec{\vrho}, \, \vec{\bvrho}$ to explicitly compute the $y-$integral appearing in~\eqref{ddxvecJ}. It is then particularly convenient to use the quantities~\eqref{eq:Hamil_y} which isolate the momentum dependence of the Hamiltonian. Therefore, using the integrals given in \eqref{IntegralsFD}, we can rewrite~\eqref{ddxvecJ} as
\begin{equation}
\label{eq:derivA_leading}
\frac{\dd \vec{\Anti} }{\dd x}= F(\xi_1,\xi_2)
\vec{\underline{\mathcal{V}}}_{\rm eff}\wedge \vec{\Anti}\qquad \text{where} \qquad
\vec{\underline{\mathcal{V}}}_{\rm eff} \equiv \left(\vecHvacy + y^2_{\rm eff}
\vecHlepy \right) \, ,
\end{equation}
where we defined the slowness factor
\begin{equation}
\label{eq:slowness}
F(\xi_1,\xi_2) \equiv \frac{3}{2} \frac{\xi_1+\xi_2}{\pi^2 +\xi_1^2 +
  \xi_2^2+\xi_1 \xi_2}\,,
\end{equation}
in agreement with \cite{Abazajian2002,Wong2002}. The typical “average” momentum is
\begin{equation} 
y_{\rm eff} \equiv \pi\sqrt{1+\frac{\xi_1^2+\xi^2_2 }{2\pi^2}} \simeq \pi \, ,
\end{equation}
in agreement with equation~(33) in \cite{Wong2002} or equation~(2.19) in \cite{Abazajian2002} (derived in the particular case $\xi_1 = 0$).

This standard result allows to recover the key features of synchronous oscillations. The evolution of all $y-$modes is locked on the evolution of $\vec{\Anti}$, which precesses around the effective Hamiltonian computed for $y = y_{\rm eff}$. However, the oscillation frequency is greatly reduced compared to standard oscillations set by the Hamiltonian $\Hamil(y_{\rm eff})$, since for small degeneracies $F \propto (\xi_1 + \xi_2) \ll 1$.

Initially, all unit vectors are aligned $\widehat{\Anti} \parallel \hat{\Hamil} \simeq \hat{\cal H}_{\rm lep} \parallel \vec{e}_\z$. Then, as the temperature decreases, $\Hself$ dominates less compared to $\Hamil$ and the vectors $\vec{\vrho}$ and $\vec{\bvrho}$ become aligned with different directions (namely, $\widehat{\Hself + \Hamil}$ and $\widehat{\Hself - \Hamil}$), leading to~\eqref{eq:vrho_ATAOJV}.

\paragraph{Next-to-leading order} Let us therefore now account for the effect of $\Hamil$, in that $\vrho$ and $\bvrho$ do not get projected on the exact same directions. We will assume that $\lvert \vec{\Hamil} \rvert \ll \lvert \vec{\Hself} \rvert$, such that we can perform an expansion of the unit vector
\begin{equation}
\label{ExpandSH}
\widehat{\Hself+\Hamil} \simeq \widehat{\Hself} + \frac{\vec{\Hamil} }{ |\vec{\Hself}| } - \left(\frac{\vec{\Hamil}\cdot \vec{\Hself}}{|\vec{\Hself}|^2}\right) \widehat{\Hself}+\cdots
\end{equation}  
For $\widehat{\Hself-\Hamil} $ the expression is identical up to $\vec{\Hamil} \to -\vec{\Hamil}$. This expansion gives, in the \ATAOJH approximation, the next-to-leading order (NLO) terms that were not explicited in previous works (cf., for instance, equation~(25) in~\cite{Wong2002}). Including this expansion in equation~\eqref{ddxvecJ}, we can once again recast the evolution of the asymmetry as a precession equation
\begin{equation}
\label{eq:precession}
    \frac{\dd \vec{\Anti}}{\dd x} = \vec{\Omega} \wedge \vec{\Anti} \, ,
\end{equation}
where the oscillation frequency (in $x$ variable) reads, retaining only the vacuum contribution $\Hamil = \Hvac$ for simplicity,\footnote{This is not an oversimplification. Indeed, as long as $\Hlep$ dominates over $\Hvac$, all vectors are aligned along $\vec{e}_\z$ and no precession takes place.}
\begin{equation}
\label{eq:Omega_JH}
\vec\Omega = F(\xi_1, \xi_2 ) \vecHvacy \left[1
  -\frac{12}{\sqrt{2} G_F}\left(\frac{x}{\me}\right)^3 \frac{x H \vecHvacy \cdot
    \widehat{\Anti}}{|\xi_1-\xi_2|(\xi_1 + \xi_2)} \right]\,.
\end{equation}
The second term between brackets is the NLO term, which accounts for the difference between purely synchronous and {\it quasi}-synchronous oscillations, given that its origin is rooted in the orientation differences between $\vec{\vrho}$ and $\vec{\bvrho}$. Note that we also took the lowest order contribution $\lvert \vec{\Anti} \rvert \simeq \abs{\xi_1 - \xi_2}/6$, valid for small degeneracy parameters (we can use the leading order expression~\eqref{JtohatJ} since the evolution of $\vec{\Anti}$ is a precession, hence its norm is unchanged). The expression~\eqref{eq:Omega_JH} leaves a priori the possibility of divide-by-zero if $\xi_1 = \pm \, \xi_2$. We discuss these special cases at the end of this section.

To estimate the precession frequency, we consider as before that initially $\widehat{\Anti} = {\rm sgn}(\xi_1 - \xi_2) \vec{e}_z$ and assume the transition between $\Hlep$ and $\Hvac$ to be abrupt enough such that we can estimate, using \eqref{eq:Hvac_2nu}, $x H \vecHvacy \cdot \widehat{\Anti} = - {\rm sgn}(\xi_1 - \xi_2) (x/\me) (\Delta m^2 / 2) \cos(2\theta)$. Therefore we get
\begin{equation}
\label{eq:Omega_norm}
    \lvert \vec\Omega \rvert = \frac{\abs{F(\xi_1, \xi_2 ) \Delta m^2}}{2 \me H} \times \left\lvert 1
  + \left(\frac{x}{x_{\rm tr}}\right)^4 \frac{{\rm sgn}(\Delta m^2 \cos(2 \theta))}{\xi_1^2 - \xi_2^2} \right \rvert \,,
\end{equation}
where we defined
\begin{equation}
x_{\rm tr} \equiv \me \left(\frac{\sqrt{2} G_F}{6 |\Delta m^2
    \cos(2\theta)|}\right)^{1/4}\simeq 3.7 \left(\frac{10^{-3}\,{\rm eV}^2}{|\Delta m^2
    \cos(2\theta)|}\right)^{1/4}\,.
\end{equation}
Given the scaling $H \propto x^{-2}$ recalled in~\eqref{eq:scaling_Hubble_x}, the frequency of synchronous oscillations keeps increasing as the Universe expands, first as $\Omega \propto x^2$ (leading order) and then as $\Omega \propto x^6$ (NLO domination).

This second behaviour is a novel result of this paper. Although one might expect that the effect of $\Hamil$ would be completely subdominant compared to $\Hself$, the particular form of the equation of motion changes this picture. Keeping the dominant term in the expansion~\eqref{ExpandSH} (i.e., $\widehat{\Hself}$) leads in the evolution equation~\eqref{ddxvecJ} to an integral symmetric under $\xi_1 \to - \xi_1$ and similarly for $\xi_2$. Therefore, the associated contribution is proportional to $\xi_1^2 - \xi_2^2$, which after dividing by the norm of $\vec{\Anti}$ accounts for the precession frequency $\propto \xi_1 + \xi_2$ obtained in \eqref{eq:slowness}. However, the first order correction in~\eqref{ExpandSH} is odd with respect to $\vec{\Hamil}$, hence an antisymmetric integral with respect to $\xi_{1,2} \to - \xi_{1,2}$. The corresponding result is $\propto \xi_1 - \xi_2$, which is enhanced compared to the leading order term. 

The transition from leading to next-to-leading order is then found to be around
\begin{equation}
\label{eq:xNLO}
x_{\rm NLO} \equiv x_{\rm tr }|\xi_1^2-\xi_2^2|^{1/4}\,.
\end{equation}
Note that depending on the sign of $\Delta m^2 \cos(2\theta)/(\xi_1^2-\xi_2^2)$ the frequency can go through zero, which means that $\vec{\Anti}$ can precess in one direction, slow down, and then precess in the opposite direction with a frequency increasing as $\propto x^6$.

\paragraph{Summary: evolution of $\vec{\Anti}$} Initially, at high temperatures (typically $\Tcm \sim 20 \, \mathrm{MeV}$), the lepton term dominates over the vacuum one and $\vec{\Anti} \propto \vecHlep \parallel \vec{e}_z$. All vectors are aligned, and this situation does not change until the MSW transition between $\Hlep$-domination to $\Hvac$-domination. If this transition is slow enough compared to the precession frequency, then $\vec{\Anti}$ keeps following $\vec{\underline{\cal V}}_{\rm eff}$  and ends up aligned with $\vecHvacy$. This corresponds to an \emph{adiabatic} evolution of the asymmetry vector itself. Conversely, if the transition is too abrupt (that is much shorter than the precession timescale), $\vec{\Anti}$ gets brutally misaligned with $\vecHvac$ and oscillations develop. Let us stress that the evolution of $\vrho$ and $\bvrho$ is in general adiabatic, but it is the evolution of the vector that they track, namely $\vec{\Hself}$, which can be non-adiabatic depending on the value of the slowness factor~\eqref{eq:slowness}.

One of the key parameters to estimate the (non-)adiabaticity of this transition is therefore the precession frequency for $\vec{\Anti}$ at the time of the MSW transition. Given the values in figure~\ref{fig:ODG_QKE}, we see that we can estimate this frequency deep into the \ATAOJ regime, that is using~\eqref{eq:derivA_leading}. After muons and antimuons have annihilated (their remaining asymmetry is completely negligible here), and before electrons and positrons did so, that is in the range $200 \,{\rm MeV}\ge \Tcm \ge 0.5 \,{\rm MeV}$, the Hubble parameter~\eqref{eq:scaling_Hubble_x} reads
\begin{equation}\label{Htox2}
H \simeq \frac{\me}{M_{\rm Pl}} \times \frac{\me}{x^2} \sqrt{\frac{\pi^2}{45} \times \left[1 +
    (N_\nu+2)\frac{7}{8}\right]}\simeq \frac{\me}{x^2}\times 2.278 \cdot 10^{-22}\,,
\end{equation}
where in the last step we have taken $N_\nu=3$. When entering the correct numbers and approximating the slowness
factor by its lowest order in the $\xi_\alpha$, that is $F(\xi_1,\xi_2)\simeq 3(\xi_1+\xi_2)/(2\pi^2)$, we estimate the precession frequency to be
\begin{equation}\label{EvaluationOmega}
\Omega(x) \simeq 1.28\times 10^6 \,x^2\, \abs{\xi_1+\xi_2} \, \frac{\Delta m^2}{10^{-3}\,{\rm eV}^2} \, .
\end{equation}
If oscillations do develop, initially at the frequency~\eqref{EvaluationOmega}, the increasing influence of $\Hamil$ compared to $\Hself$ leads to a new behaviour: beyond $x_{\rm NLO}$ given by~\eqref{eq:xNLO}, the frequency increases faster and $\Omega \propto x^6$. These features are illustrated in section~\ref{SecRelevant2Neutrinos}. Note that the calculation of the NLO assumes $\lvert \vec{\Hamil} \rvert \ll \lvert \vec{\Anti} \rvert$, but at some point the vacuum term becomes dominant over the self-potential one (cf.~figure~\ref{fig:ODG_QKE}) and the \ATAOJH regime breaks down. This is discussed in more detail in section~\ref{SecTransitionToATAOV}.

\paragraph{Particular cases} While the previous calculation seemed fairly general, there are two specific cases that deserve to be discussed.
\begin{itemize}
    \item \emph{Equal asymmetries ---} if $\xi_1 = \xi_2$ the asymmetry vector $\vec{\Anti}$ is strictly zero and the previous formalism is inadequate (namely, \eqref{eq:derivA_leading} cannot be obtained anymore since initially $\vec{\vrho} = \vec{\bvrho} = \vec{0}$). 
    \item \emph{Equal but opposite asymmetries ---} if $\xi_1 = - \xi_2$, the leading order term in~\eqref{eq:Omega_JH} vanishes (since $F(\xi_1,\xi_2) \propto \xi_1 + \xi_2$), but not the next-to-leading order contribution, a special case investigated at the end of section~\ref{SecMuonMSW}.
\end{itemize}

\section{Relevant two-neutrino cases for the primordial Universe}\label{SecRelevant2Neutrinos}

The previous results derived with only two neutrinos can shed some light on the physics at play in the standard case with three neutrinos and a general PMNS matrix. After the muon-driven MSW transition and before the electron-driven one, the oscillations only take place in the $\nu_\mu-\nu_\tau$ subspace since the unitary matrix $U_{\rm eff}$ that diagonalizes $\Hamil$ is approximately
\begin{equation}\label{UtoU}
U_{\rm eff} =  R_{23}(\theta_{23}^{\rm eff}) = \begin{pmatrix}
1 & 0 & 0\\
0 & \cos \theta_{23}^{\rm eff} &\sin \theta_{23}^{\rm eff} \\
0 & -\sin \theta_{23}^{\rm eff} & \cos \theta_{23}^{\rm eff}
\end{pmatrix} \,,
\end{equation}
this form being rigorously valid in the limit $m_\mu/m_e \to \infty$. Expanding in the ratio $\epsilon = \Delta m_{21}^2  / \Delta m_{32}^2$, we find 
\begin{equation}\label{th23th23}
\tan(2 \theta_{23}^{\rm eff}) = \tan(2 \theta_{23}) - \epsilon
\frac{ \sin (\theta_{13}) \sin(2 \theta_{12})}{\cos^2 (\theta_{13}) \cos^2(2 \theta_{23})} +{\cal O}(\epsilon^2)\,.
\end{equation}
Given the values~\eqref{ValuesStandard} we find $\theta_{23}^{\rm eff}  \simeq \theta_{23}$ with a difference of order $0.25\,\%$. Hence, we investigate the case $\Delta m^2 = \Delta m_{32}^2$ and $\theta=\theta_{23}$ in section~\ref{SecMuonMSW} to study the evolution of density matrices after the muon-driven transition.
 
Unfortunately, the system is not so easily reduced to a two-neutrino system when it comes to the description of the subsequent electron-driven MSW transitions. For simplicity, we choose to consider a fictitious configuration where $\theta_{13}=\theta_{23}=0$ such that oscillations take only place in the $\nu_e-\nu_\mu$ subspace, with the electrons/positrons being the relevant contribution to the lepton mean-field effects \eqref{eq:Hlep_2nu}. This configuration is detailed in section~\ref{SecElectronMSW}. For numerical applications we consider $\Delta m^2 = \Delta m_{21}^2$ and $\theta=\theta_{12}$. Although it is an ideal setup, it will provide important insight for the full three-neutrino case in section~\ref{Sec:3neutrinos}.

\subsection{Muon-driven MSW transition}\label{SecMuonMSW}

Let us consider a muon-driven MSW transition with $\theta = \theta_{23} \simeq 0.831$ and $\Delta m^2 = \Delta m_{32}^2 \simeq 2.453 \times 10^{-3} \, \mathrm{eV}^2$ for numerics \cite{PDG}. We restrict to normal ordering for simplicity, and do not include collisions yet. This means that electrons and positrons are absent from this description, except for their contribution to the energy density and thus the Hubble parameter~\eqref{eq:scaling_Hubble_x}.

\subsubsection{Description of the transition}\label{SecDescriptionMuonTransition}

As outlined before, synchronous oscillations of the neutrino ensemble can develop when the MSW transition occurs, provided this transition is abrupt enough for $\vec{\Anti}$ to get suddenly misaligned with $\vecHeff$ and precess around it. Let us first estimate the location of this transition. The energy density of muons/antimuons drops very rapidly once they become non-relativistic. In this limit, we get
\begin{equation}
\rho_{\mu^\pm}+P_{\mu^\pm} = 4 \sigma^{5/2}
\left(\frac{x}{2\pi}\right)^{3/2} {\rm e}^{-\sigma x} \times \frac{\me^4}{x^3}
\end{equation}
with $\sigma = m_\mu/\me\simeq 206.77$. Approximating $y_{\rm  eff} \simeq \pi$, we find that the vacuum term is equal in magnitude to the lepton term --- which is our definition of the transition (see appendix~\ref{AppAdiabatic}), --- for $x_{\rm MSW}\simeq 0.043$, that is for $\Tcm \simeq 12\,{\rm MeV}$. Given the exponential drop $\exp(-\sigma x)$ of muons energy density, one can note that $x_{\rm MSW}$ is very mildly sensitive to the value of $\Delta m^2$.

The adiabaticity parameter given by \eqref{Defgammatr2} is then
\begin{equation}
\label{eq:gammaMSW_muon}
\gamma_{\rm MSW} \simeq 100 \times \abs{\xi_1 + \xi_2} \,.
\end{equation}
For $\xi_1+\xi_2$ of a few percent or smaller, we find $\gamma_{\rm MSW}<1$ and the transition is abrupt, that is the evolution of $\vec\Anti$ during the transition is very non-adiabatic. Hence we expect that as the direction of the effective Hamiltonian moves away from the vertical axis, $\vec{\Anti}$ will develop oscillations at the frequency $\Omega$. For much larger $\xi_1+\xi_2$ such that $\gamma_{\rm MSW} > 1$, and considering the Landau-Zener estimation for the degree of adiabaticity~\eqref{LZ}, $\vec{\Anti}$ should tend to follow adiabatically the transition to the vacuum Hamiltonian.

\begin{figure}[!ht]
  \centering
  \includegraphics[]{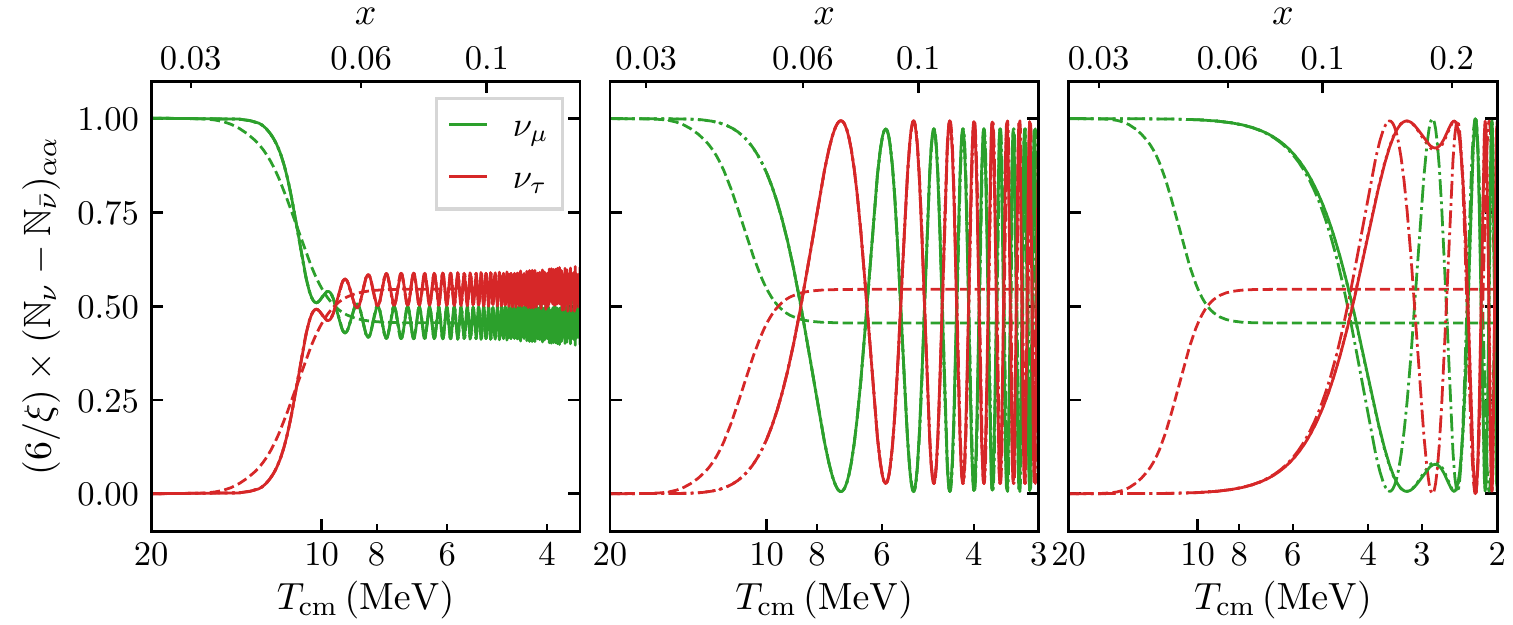} \\
  \includegraphics[]{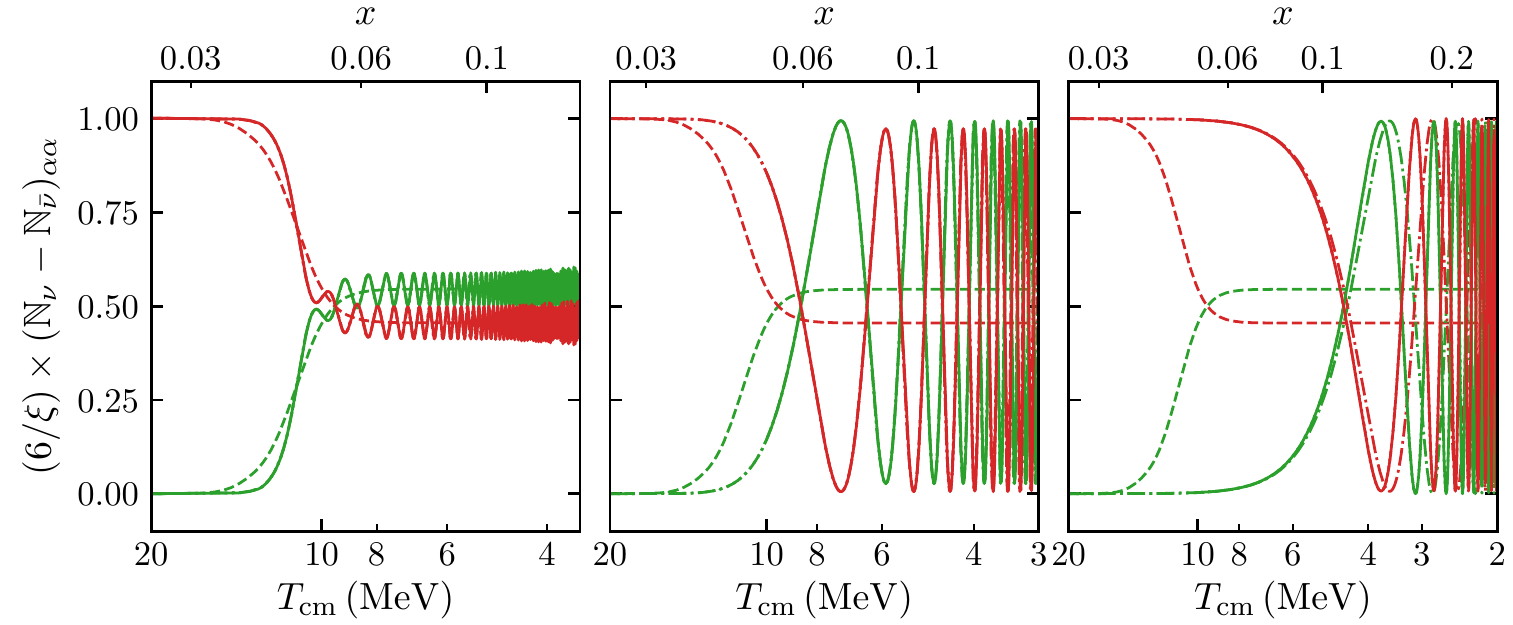}
	\caption{\label{fig:2nu_mu-tau} Evolution of the flavour asymmetries for a two-neutrino $\nu_\mu$ (green) - $\nu_\tau$ (red) system without collisions, with $\Delta m^2 = 2.45 \times 10^{-3} \, \mathrm{eV}^2$ and $\theta = 0.831$. We compare different numerical schemes: in solid line QKE, in dots \ATAOJH (hidden behind QKE), in dot-dashes \ATAOJ, and in dashes \ATAOH. The initial degeneracy parameters are on the first row $\xi_2 =0$ and $\xi_1=0.1,0.01,0.001$ from left to right ; on the second row $\xi_1 =0$ and $\xi_2=0.1,0.01,0.001$. On the $\y$-axis label, $\xi$ stands for the non-zero initial $\xi_i$.}
\end{figure}

The evolution of asymmetry is illustrated in figure~\ref{fig:2nu_mu-tau}. It is clear that the evolution with the self-interaction mean-field is completely different from the evolution where this has been ignored and which corresponds to the \ATAOH line: no synchronous oscillations take place in this scheme. These oscillations, in agreement with our adiabaticity estimate, do develop significantly for initial degeneracies smaller than one percent. On the contrary, for $\xi_1 + \xi_2 = 0.1$, the transition is quasi-adiabatic and $\vec{\Anti}$ follows the direction set by $\vecHeff$ with oscillations of much smaller amplitude compared to the smaller $\xi$ cases (right plots). Furthermore, it appears that at small degeneracies, the \ATAOJ results differ from the more accurate \ATAOJH scheme, the latter matching perfectly the QKE method. The difference between \ATAOJ and \ATAOJH can be understood by considering the NLO contribution to the precession frequency: this extra contribution explains the “wrong” frequency in the \ATAOJ case (see for instance the bottom right plot on figure~\ref{fig:2nu_mu-tau}), or even the wrong qualitative behaviour of the asymmetry (top right plot).

\paragraph{Synchronous oscillation frequency} 

\begin{figure}[!ht]
	\centering
	\includegraphics[]{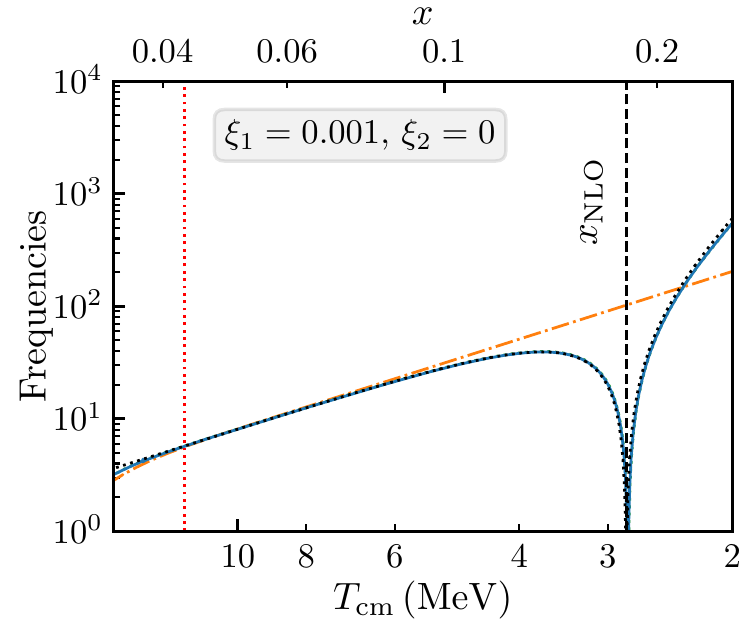}
        \includegraphics[]{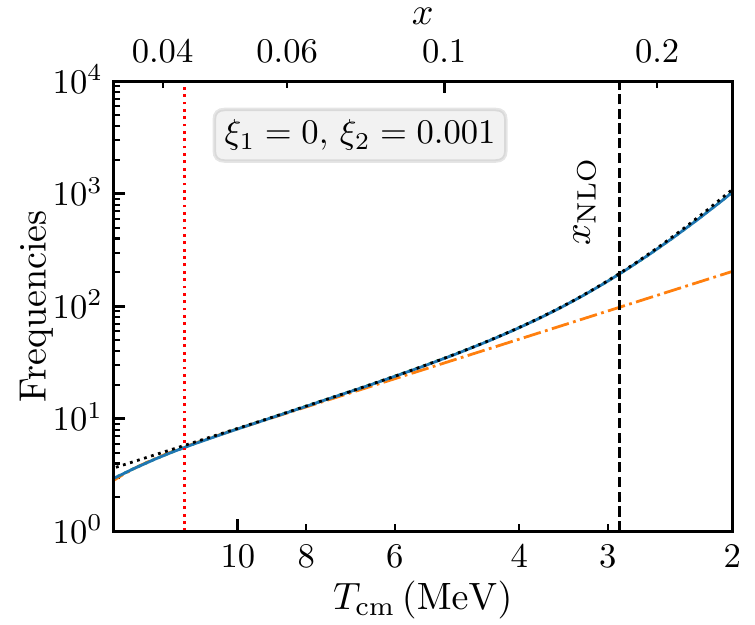}
	\caption{\label{fig:Omega_2nu_mu-tau} Frequency of synchronous oscillations in the case $\xi_1=0.001$, $\xi_2=0$
          (left) and the case $\xi_1=0$, $\xi_2=0.001$ (right). We consider a $\nu_\mu - \nu_\tau$ system with $\Delta m^2 = 2.45 \times 10^{-3}\,{\rm eV}^2$ and $\theta = 0.831$. The vertical red line is the location of the MSW
          transition. The dashed black line is $|\Omega \sin(2\theta)|$, that is the analytic approximation. The coloured lines correspond to $\lvert (\partial_x \vec{\Anti} \wedge
          \vec{\Anti}) \rvert /|\vec{\Anti}|^2$, in the QKE method (blue), the
          \ATAOJ scheme in orange and the \ATAOJH in green (hidden behind QKE).}
\end{figure}

To estimate the frequency from our runs we compute $(\partial_x \vec{\Anti} \wedge \vec{\Anti})/|\vec{\Anti}|^2$ which gives, given the precession equation~\eqref{eq:precession}, the projection of the rotation vector $\vec{\Omega}$ orthogonally to $\vec{\Anti}$. If the MSW transition is abrupt, the precession takes place around $\vecHvac$ with an angle $2 \theta$, hence the former quantity should be equal to $\lvert \sin(2 \theta) \vec{\Omega} \rvert$. Both frequencies are shown on figure~\ref{fig:Omega_2nu_mu-tau}. We clearly see the transition from the regime $\Omega \propto x^2$ to
$\Omega \propto x^6$, that is the transition to the NLO regime, and in particular how the \ATAOJH scheme fits the QKE results, while (as expected by construction) the \ATAOJ scheme completely misses this change of regime. In the region $x_{\rm MSW} < x \ll x_{\rm NLO}$, $\Hself$ largely dominates over $\Hamil$ and all three schemes coincide.

Also, since $\theta=0.831 > \pi/4$, $\cos(2\theta)<0$ and according to~\eqref{eq:Omega_norm} the frequency can go through zero for normal ordering ($\Delta m^2>0$) with $|\xi_1|>|\xi_2|$ or inverted ordering ($\Delta m^2<0$) with $|\xi_2|>|\xi_1|$. That is the case in the top right plot of figure~\ref{fig:2nu_mu-tau} ($\xi_1 = 0.001$, $\xi_2 = 0$ and normal ordering): we observe a back and forth motion of $\Anti$ for\footnote{The value predicted using \eqref{eq:xNLO} is slightly different from the one obtained numerically, because \eqref{eq:xNLO} assumes zero adiabaticity, such that the angle between $\vec{\Anti}$ and $\vecHvac$ is exactly $2 \theta$, whereas in reality $\vec{\Anti}$ partially follows the direction of $\vec{\Omega}$ during the MSW transition (see the bottom right plot on figure~\ref{fig:Coll_e-mu} for a similar behaviour in an electron-driven transition).} $T_{\rm NLO} \simeq 2.8 \, \mathrm{MeV}$, which corresponds to $\Omega = 0$ as visible on figure~\ref{fig:Omega_2nu_mu-tau}, left plot. This transition between two frequency regimes with a change of rotation direction is a feature also seen in figures 9 and 10 of \cite{Johns:2016enc}.

If we consider smaller degeneracies, such that
\begin{equation}
|\xi_1^2-\xi_2^2| < \frac{x^4_{\rm MSW}}{x^4_{\rm tr}} \simeq
1.8\times 10^{-8}\left(\frac{|\Delta m^2
    \cos(2\theta)|}{10^{-3}\,{\rm eV}^2}\right)
\end{equation}
then the NLO contribution to the precession frequency will dominate already when the MSW transition occurs. The adiabaticity parameter $\gamma_{\rm MSW}$ must be rescaled by multiplying it by the factor in square brackets in equation~\eqref{eq:Omega_JH}, and we get approximately 
\begin{equation}
\gamma_{\rm MSW} =\frac{4.1 \times 10^{-7}}{\abs{\xi_1 - \xi_2}} \, ,
\end{equation}
assuming the NLO contribution does dominate in~\eqref{eq:Omega_JH}, which amounts to multiplying~\eqref{eq:gammaMSW_muon} by $(x_{\rm MSW}/x_{\rm tr})^4 /\abs{\xi_1^2 - \xi_2^2}$. Therefore, if we satisfy the condition $|\xi_1-\xi_2| \gg 4.1 \times10^{-7}$, the transition is still abrupt and oscillations do develop. For even smaller degeneracies, there is no clear region where $|\vec{\Hself}| \gg |\vec{\Hamil}|$, and the subsequent phenomenology can only be captured by a full QKE resolution as in~\cite{Johns:2016enc}. 

\paragraph{Beginning of oscillations} Provided the MSW transition is non-adiabatic, oscillations of $\vec{\Anti}$ appear, driving each individual mode. However, one can see on figure~\ref{fig:2nu_mu-tau} that depending on the value of $(\xi_1, \xi_2)$, the apparent “start” of these oscillations looks shifted while $x_{\rm MSW}$ is the same. We can estimate how oscillations develop, which provides an additional check of our analytical developments.

The asymmetry evolves with a frequency $\Omega(x)$, therefore the phase of the oscillations is at any time given by
\begin{equation}
    \frac{\dd \Phi}{\dd x} = \Omega(x)  \qquad \text{hence} \qquad \Phi(x) = \frac{1}{3} \Omega(x) x \, ,
\end{equation}
where we used the fact that $\Omega \propto x^2$, keeping only the leading order contribution~\eqref{EvaluationOmega}. Half a period of oscillation is reached when $\Phi(x_\pi) = \pi$, which happens for
\begin{equation} 
\label{eq:x_startosc}
x_\pi = \left( \frac{1}{1.28 \times 10^6} \times \frac{10^{-3} \, \mathrm{eV}^2}{\abs{\Delta m^2}} \times \frac{1}{\abs{\xi_1 + \xi_2}} \times 3 \pi \right)^{1/3} \simeq 0.067 \, ,
\end{equation}
for $\xi_1 = 0.01$ and $\xi_2=0$, which agrees with figure~\ref{fig:2nu_mu-tau}, top middle plot. 

\subsubsection{Particular cases}\label{SecParticular}

In this subsection, we use the $\nu_\mu - \nu_\tau$ framework to discuss the particular cases of equal and equal but opposite asymmetries, for which the calculations of section~\ref{FreqSyncOsc} are no longer valid. 

If $\xi_1 = \xi_2$, the vector parts of $\vrho(y)$, $\bvrho(y)$ and $\Anti$ are all equal to zero, and will therefore remain so. The self-potential term cancels in the QKE and the \ATAOH scheme describes accurately the neutrino evolution.

The case $\xi_1 = - \xi_2$ would correspond to a vanishing total lepton number density, while each flavour could display large asymmetries. This would result in a possibly significant contribution to the total energy density, hence the interest for this particular case. It was shown in \cite{Pastor:2001iu,Dolgov_NuPhB2002} that in this scenario synchronous oscillations are hampered as long as $\Hself$ dominates. This is in perfect agreement with our theoretical analysis: at leading order, as $F(\xi, -\xi) = 0$ the first term in \eqref{eq:Omega_JH} vanishes. However, our calculation of the NLO contribution shows that oscillations can still take place, but directly with a frequency $\propto x^6$. 

\begin{figure}[!ht]
  \centering
  \includegraphics[]{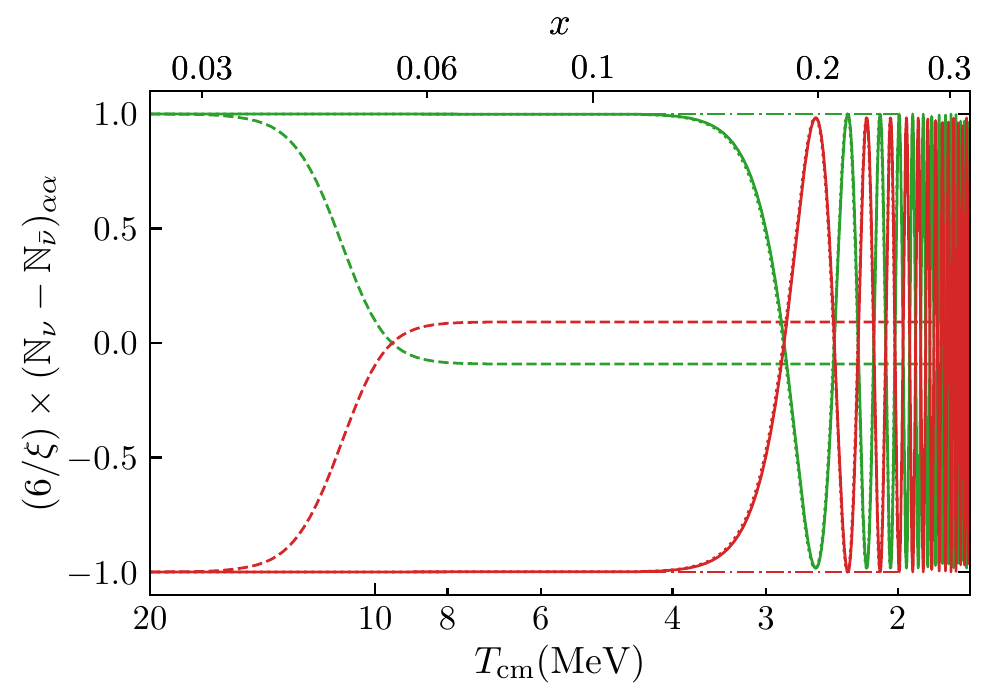}
	\caption{\label{fig:xi1mxi2} Equal but opposite asymmetries $\xi_1 = -\xi_2 = \xi = 0.001$. We clearly see that, if we discarded the next-to-leading order contribution in~\eqref{eq:vrho_ATAOJV} (\ATAOJ curve, in dashed-dots), oscillations would be switched off. However, the lowest order term in the frequency expression is now $\propto x^6$ and gives rise to quasi-synchronous oscillations beyond $x\simeq 0.2$, see~equation~\eqref{eq:x_xi1mxi2}.}
\end{figure}

To check this prediction, we plot the evolution of asymmetries for $\xi_1 = -\xi_2 = \xi = 0.001$ on figure~\ref{fig:xi1mxi2}. In the \ATAOJ scheme (which ignores the NLO contribution), oscillations never appear, contrary to the actual QKE evolution, correctly captured by the \ATAOJH scheme. The onset of synchronous oscillations is delayed compared for instance to the right plots of figure~\ref{fig:2nu_mu-tau}, and we can estimate the location of this starting point exactly as in the previous section, the only difference being that we use the NLO part of~\eqref{eq:Omega_norm} $\Omega \propto x^6$. We find that the location of the first half-oscillation is
\begin{equation} 
\label{eq:x_xi1mxi2}
x_\pi = \left( \frac{8 \pi ^2}{3} \times 2.278 \times 10^{-22} \times \frac{\me^2}{\abs{\Delta m^2}} \times x_{\rm tr}^4 \times \xi \times 7 \pi \right)^{1/7} \simeq 0.20 \, ,
\end{equation}
for $\xi = 0.001$, in excellent agreement with figure~\ref{fig:xi1mxi2}. 

\newpage

\subsection{Electron-driven MSW transition}\label{SecElectronMSW}

We now consider an electron/positron driven transition in the (fictitious) $\nu_e - \nu_\mu$ subspace, with the mixing angle $\theta = \theta_{12} \simeq 0.587$ and the small mass gap $\Delta m^2 = \Delta m_{21}^2 \simeq 7.53 \times 10^{-5} \, \mathrm{eV}^2$ for numerics. The difference with the previous case comes from the fact that the MSW transition now takes place when electrons are still relativistic. Moreover, we show in the following section how the collision term is very different from the one in the $\nu_\mu - \nu_\tau$ subspace. 

\begin{figure}[!ht]
  \centering
  \includegraphics[]{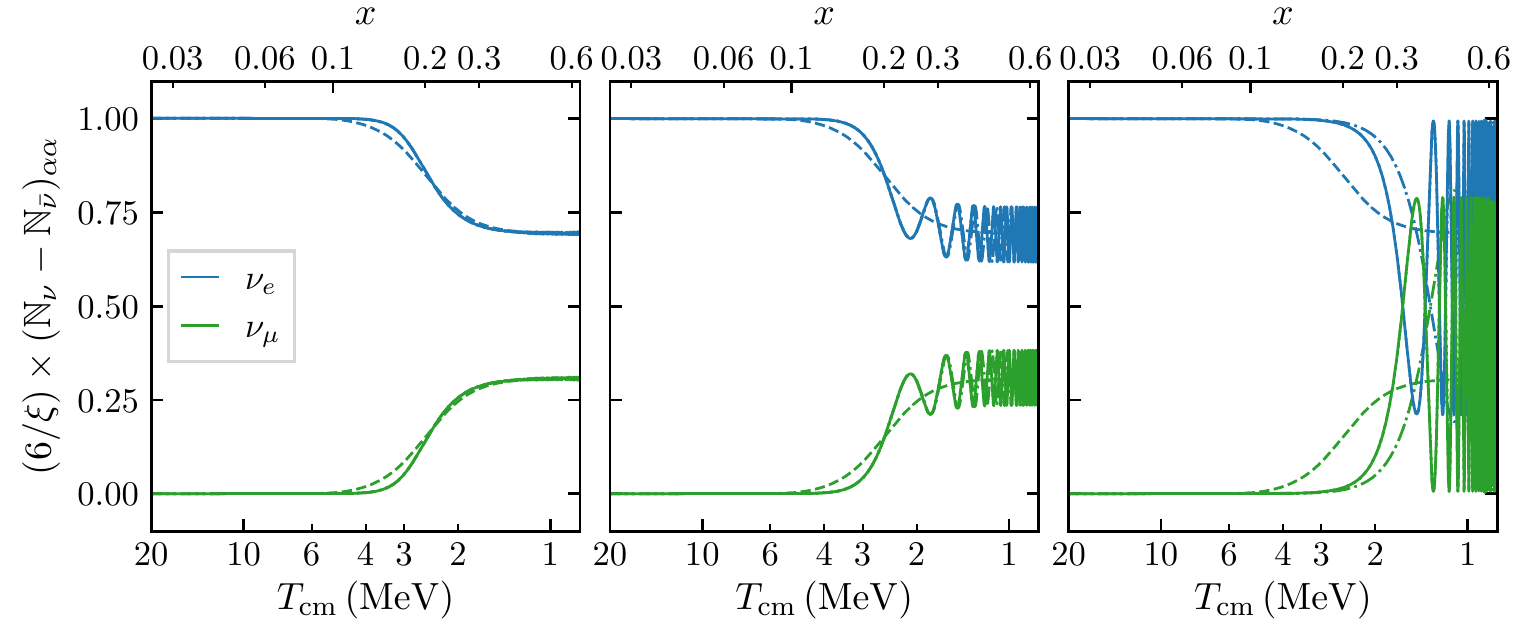} \\
  \includegraphics[]{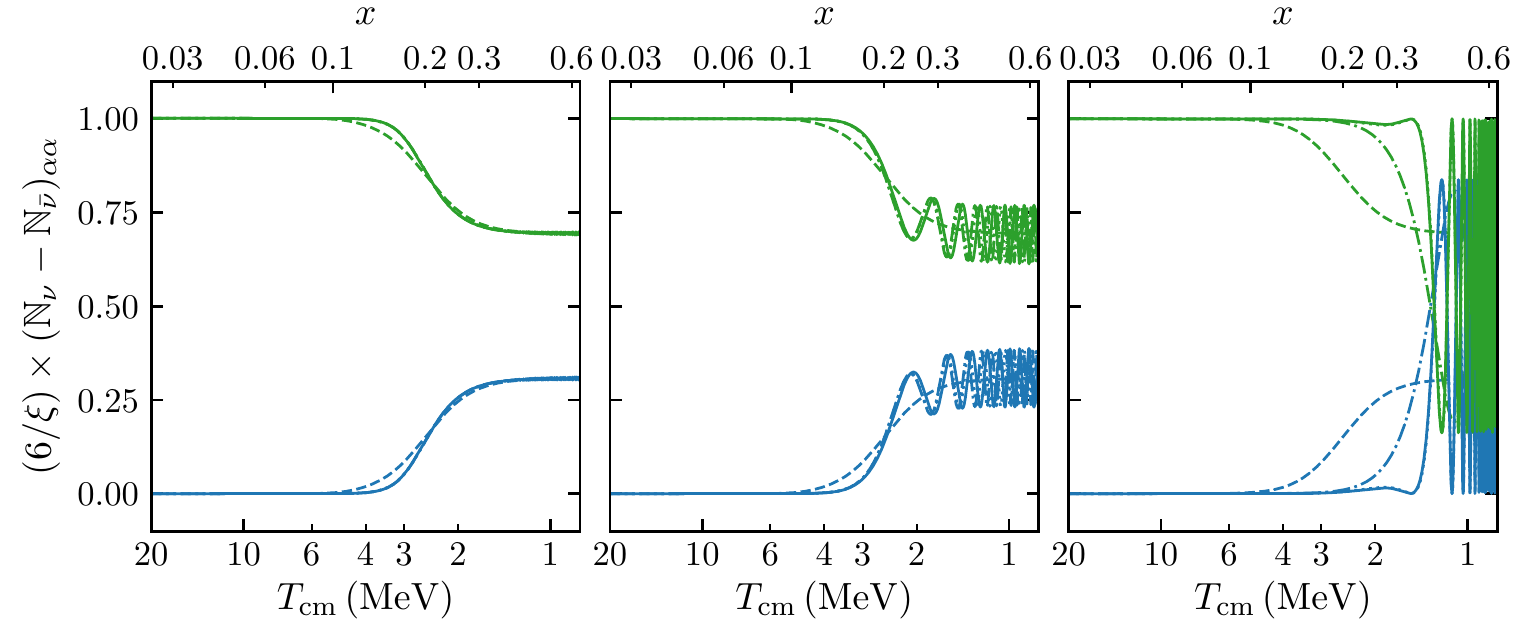}
	\caption{\label{fig:2nu_e-mu}Evolution of the flavour asymmetries for a two-neutrino $\nu_e$ (blue) - $\nu_\mu$ (green) system without collisions, with $\Delta m^2 = 7.53 \times 10^{-5} \, \mathrm{eV}^2$ and $\theta = 0.587$. We compare different numerical schemes: in solid line QKE, in dots \ATAOJH (hidden behind QKE), in dot-dashes \ATAOJ, and in dashes \ATAOH. The initial degeneracy parameters are on the first row $\xi_2 =0$ and $\xi_1=0.1,0.01,0.001$ from left to right ; on the second row $\xi_1 =0$ and $\xi_2=0.1,0.01,0.001$.}
\end{figure}

The relativistic limit is sufficient to estimate the location of the MSW transition, therefore we use (we take the comoving plasma temperature $z=1$ for simplicity, which is justified since $e^\pm$ annihilations are just beginning at this stage):
\begin{equation}
\rho_{e^\pm}+P_{e^\pm}  = \frac{7 \pi^2}{45} (\me/x)^4\, ,
\end{equation}
to deduce that the transition takes place for
\begin{equation}
x_{\rm MSW} = \left(\frac{\me^6 G_F}{\mW^2 \abs{\Delta
    m^2}}\frac{28\sqrt{2}\pi^2 y^2_{\rm eff}}{45}\right)^{1/6}=0.118\,\left(\frac{10^{-3} \, {\rm eV}^2}{\abs{\Delta m^2}}\right)^{1/6}\,.
\end{equation}
For the numerical values chosen, we find $x_{\rm MSW} \simeq 0.18$, that is $\Tcm \simeq 2.8\,{\rm MeV}$\footnote{Note also that the first electron-driven transition associated with the large mass gap should be around $\Tcm = 5\,{\rm MeV}$ by application of this estimate with $\Delta m^2 = \Delta m_{31}^2$.}. We estimate the adiabaticity of this transition with~\eqref{Defgammatr2}, and find
\begin{equation}
\gamma_{\rm MSW} \simeq 485 \times \abs{\xi_1 + \xi_2} \, .
\end{equation}
The larger prefactor compared to the estimate~\eqref{eq:gammaMSW_muon} in the muon-driven case makes the transition adiabatic up to smaller degeneracies. This is in agreement with the results of figure~\ref{fig:2nu_e-mu}: for instance, the transition is much more adiabatic (small amplitude of synchronous oscillations) for $\xi_1 + \xi_2 = 0.01$ compared to figure~\ref{fig:2nu_mu-tau}.

\begin{figure}[!ht]
	\centering
	\includegraphics[width=0.49 \textwidth]{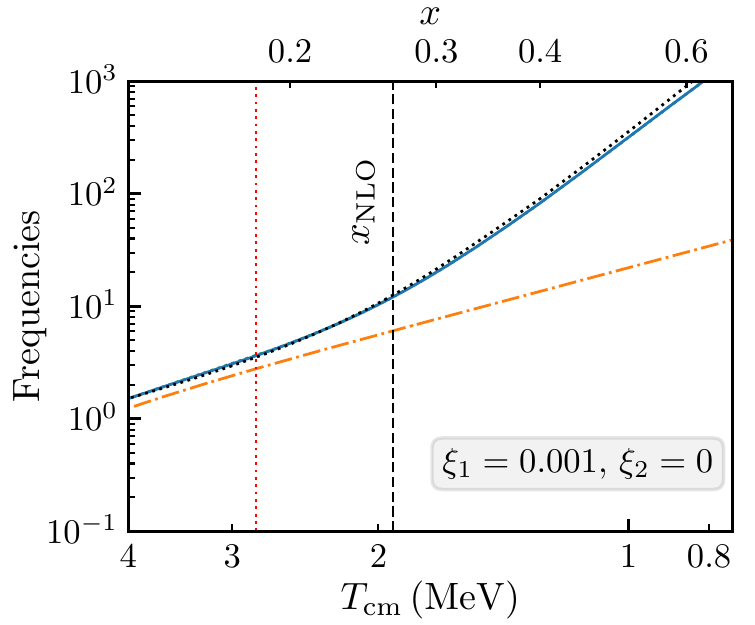}
        \includegraphics[width=0.49 \textwidth]{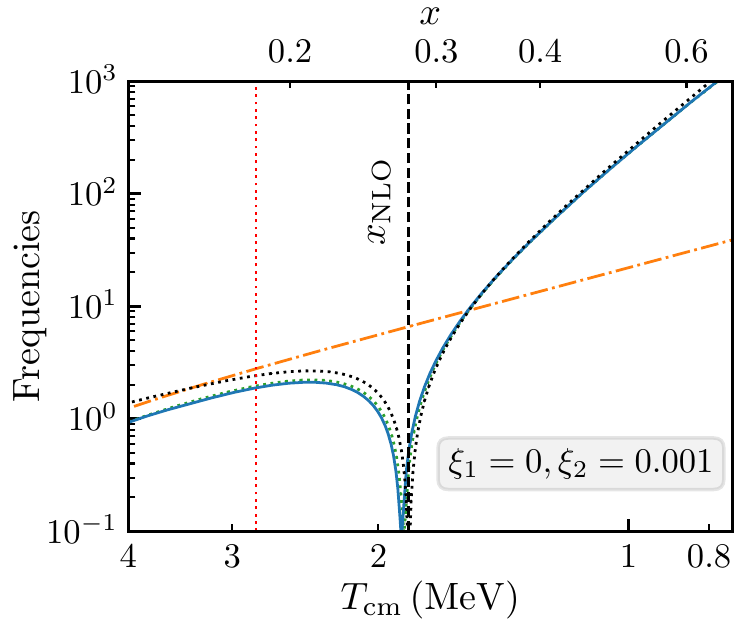}
	\caption{\label{fig:Omega_2nu_e-mu} Same plot as~\ref{fig:Omega_2nu_mu-tau} for the $\nu_e - \nu_\mu$ system with mixing parameters $\theta=0.587$ and $\Delta m^2 = 7.53\times10^{-5}\,{\rm eV}^2$.}
\end{figure}

The frequency regimes outlined in section~\ref{FreqSyncOsc} are once again observed on figure~\ref{fig:Omega_2nu_e-mu}, where we plot the quantities $|\Omega \sin(2\theta)|$ and $\lvert (\partial_x \vec{\Anti} \wedge \vec{\Anti}) \rvert /|\vec{\Anti}|^2$. We see the transition from $\Omega \propto x^2$ to $\Omega \propto x^6$ and the possible cancellation of the frequency at this transition. Contrary to the case studied in section~\ref{SecMuonMSW}, it happens now in normal ordering for $\abs{\xi_2}>\abs{\xi_1}$ because $\cos(2 \theta) > 0$.

\subsection{Effect of collisions}\label{SecCollisions}

In the previous sections, we systematically discarded the collision term in the QKEs in order to focus on the synchronous oscillation phenomenon and how approximate numerical schemes (namely, the \ATAOJH procedure) accurately capture the physics at play.

Taking into account the scattering and annihilation processes is nevertheless crucial for a precision calculation, not only since these processes will determine neutrino decoupling and partial reheating \cite{Dolgov_NuPhB1997,Esposito_NuPhB2000,Mangano2002,Mangano2005,Relic2016_revisited,Grohs2015,Grohs:2016cuu,Froustey2019,Froustey2020,Bennett:2020zkv}, but also because they can reduce flavour asymmetry differences. This second effect was notably shown in references~\cite{Dolgov_NuPhB2002,Johns:2016enc}, but these works used approximate expressions for the collision term (so-called damping approximation). We aim at showing the effect of the exact collision term, whose expression was derived for instance in~\cite{SiglRaffelt,BlaschkeCirigliano,Froustey2020}.

For the following discussion, we only recall that the collision term $\mathcal{K}[\vrho,\bvrho]$ is an integral whose matrix structure is determined by the statistical factors associated to two-body reactions $(1) + (2) \to (3) + (4)$. They read typically (we write $\vrho_i = \vrho(y_i)$ for particle $i$):
\begin{equation}
\label{eq:stat_fact1}
    \left[\vrho_4 (1- \vrho_2) + \mathrm{Tr}\left(\vrho_4 (1-\vrho_2) \right) \right] \vrho_3 (1- \vrho_1) - \{ \text{loss} \} + \mathrm{h.c.} \, ,
\end{equation}
for neutrino elastic scattering (this term corresponds to the process $\nu^{(1)} + \nu^{(2)} \to \nu^{(3)} + \nu^{(4)}$) and
\begin{equation}
\label{eq:stat_fact2}
    f_4^{(e)} (1- f_2^{(e)})G^{L/R} \vrho_3 G^{L/R} (1- \vrho_1) - \{ \text{loss} \} + \mathrm{h.c.} \, ,
\end{equation}
for reactions with electrons and positrons (this particular terms stands for a scattering $\nu^{(1)} + e^{(2)} \to \nu^{(3)} + e^{(4)}$). In the above expressions, the loss part corresponds to the exchange $\left\{ \vrho_i \leftrightarrow (1 - \vrho_i) \right\}$ for all distributions, and $\mathrm{h.c.}$ stands for “hermitian conjugate”. The coupling matrices $G^L$ and $G^R$ are diagonal in flavour space, and read in the full three-neutrino framework
\begin{equation}
    \label{eq:coupling_matrices}
    G^L = \begin{pmatrix} g_L +1 & 0 & 0 \\ 0 & g_L & 0 \\ 0 & 0 & g_L \end{pmatrix} \, , \quad
    G^R = \begin{pmatrix} g_R & 0 & 0 \\ 0 & g_R & 0 \\ 0 & 0 & g_R \end{pmatrix} \, ,
\end{equation}
with $g_L = -\frac12 + \sin^2{\theta_W}$, $g_R = \sin^2{\theta_W}$ where $\theta_W$ is Weinberg's angle. In the $ee$ entry of $G^L$, the extra factor of $1$ accounts for the charged currents between $e^\pm$ and $\nu_e$. Since we do not consider collisions with other charged leptons (due to their negligible density in the range of temperatures of interest), this is the only additional factor in $G^L$. Therefore the collision terms satisfy the general property
\begin{equation}\label{MagicKFundamental}
\mathcal{K}[U_{\rm s} \vrho U^\dagger_{\rm s}, U_{\rm s} \bvrho U^\dagger_{\rm s}] = U_{\rm s}\mathcal{K}[\vrho,\bvrho]  U^\dagger_{\rm s}\,,\qquad \bar{\mathcal{K}}[U_{\rm s} \vrho U^\dagger_{\rm s}, U_{\rm s} \bvrho U^\dagger_{\rm s}] = U_{\rm s}\bar{\mathcal{K}}[\vrho,\bvrho]  U^\dagger_{\rm s}
\end{equation}
for constant unitary matrices of the type
\begin{equation}\label{UtoU2}
U_{\rm s}  = \begin{pmatrix}
1 & 0 \\
0 & {\cal U}
\end{pmatrix} \,,\qquad {\cal U} \in {\rm U}(2) \,.
\end{equation}

In general, the collision term $\mathcal{K}[\vrho, \bvrho]$, being made of statistical factors like \eqref{eq:stat_fact1} and \eqref{eq:stat_fact2}, tends to make the density matrices in flavour basis diagonal, with entries being Fermi-Dirac distributions --- or $\vrho$ and $\bvrho$ must be obtained from conjugation of such matrices with a unitary matrix of the type \eqref{UtoU2}. The degeneracies are not constrained by processes like $\nu_\alpha + \nu_\beta \to \nu_\alpha + \nu_\beta$ or $\nu_\alpha + \bar{\nu}_\alpha \to \nu_\beta + \bar{\nu}_\beta$. The only constraint is due to the processes $\nu_\alpha + \bar{\nu}_\alpha \to e^- + e^+$, which impose $\xi_\alpha = - \bar{\xi}_\alpha$ at equilibrium. Therefore, if collisions are strong enough, the density matrices are pushed towards
\begin{equation}
\label{eq:vrho_coll0}
\vrho \sim {\rm diag}[ g(\xi_\alpha,y) ]\,, \qquad  \bvrho \sim {\rm
  diag}[ g(-\xi_\alpha,y) ]\, ,
\end{equation}
where $\sim$ stands for the possible conjugation by a matrix of the form~\eqref{UtoU2}.

\paragraph{Muon-driven transition} In the framework of section~\ref{SecMuonMSW}, we considered a two-neutrino case with only $\nu_\mu$ and $\nu_\tau$. When focusing on the $\nu_\mu-\nu_\tau$ subspace of \eqref{eq:coupling_matrices}, the $G^L$ and $G^R$ matrices are proportional to the identity matrix.

Initially, the collision term vanishes since $\vrho$ and $\bvrho$ are in the form~\eqref{eq:vrho_coll0}. What may come as a surprise is the fact that it keeps vanishing even though $\vrho$ and $\bvrho$ evolve. Indeed, at high temperature the \ATAOJ scheme is valid and both $\vrho$ and $\bvrho$ are diagonalized by the same matrix $U_{\Hself}$, that is furthermore identical for all momenta $y$. This means that we have\footnote{For clarity, we omit the subscript $\Hself$ for the matter density matrix $\widetilde{\vrho}_{\Hself}$. More generally in this section, $\widetilde{\vrho}$ will be the diagonal density matrix, whether the Hamiltonian is $\Hself$, $\Hself + \Hamil$, \dots} $\vrho = U_{\Hself} \widetilde{\vrho} U_{\Hself}^\dagger$ (and similarly for $\bvrho$), so from the restriction of the general property~\eqref{MagicKFundamental} we deduce the relations 
\begin{equation}\label{MagicKUJ}
\mathcal{K}[\vrho,\bvrho] = U_{\Hself}
\mathcal{K}\left[\widetilde\vrho,\widetilde\bvrho\right]  U^\dagger_{\Hself} \ , \qquad 
\bar{\mathcal{K}}(\vrho,\bvrho) = U_{\Hself} \bar {\mathcal{K}}\left[\widetilde\vrho,\widetilde\bvrho\right]  U^\dagger_{\Hself}  \, .
\end{equation}
Thanks to this peculiar “factorization”, the collision term keeps vanishing as long as $\widetilde{\vrho}$ and $\widetilde{\bvrho}$ are diagonal matrices (which they are by definition) of Fermi-Dirac distributions. This is much less restrictive, and remains satisfied as long as $\Hself \gg \Hamil$ since $\widetilde{\mathcal{K}} = 0$ leads to $\partial_x \widetilde{\vrho} = 0$, hence the collision term keeps vanishing, and so on. With or without collisions, the evolution of $\vrho, \bvrho$ is purely due to the change of direction of $\vec{\Anti}$, which oscillates more or less around $\vec{\Hamil}$ depending on the adiabaticity of the MSW transition.

Note that the previous argument is only exact for the part of $\mathcal{K}$ corresponding to neutrino self-interactions. It extends to the scattering with electrons/positrons as long as all particles share the same temperature. But even beyond this, when $e^\pm$ annihilations populate the neutrinos, they do so in creating pairs of neutrinos/antineutrinos, so the collision term acts to maintain thermal distributions, but not to equilibrate asymmetries.

\begin{figure}[!ht]
	\centering
	\includegraphics[]{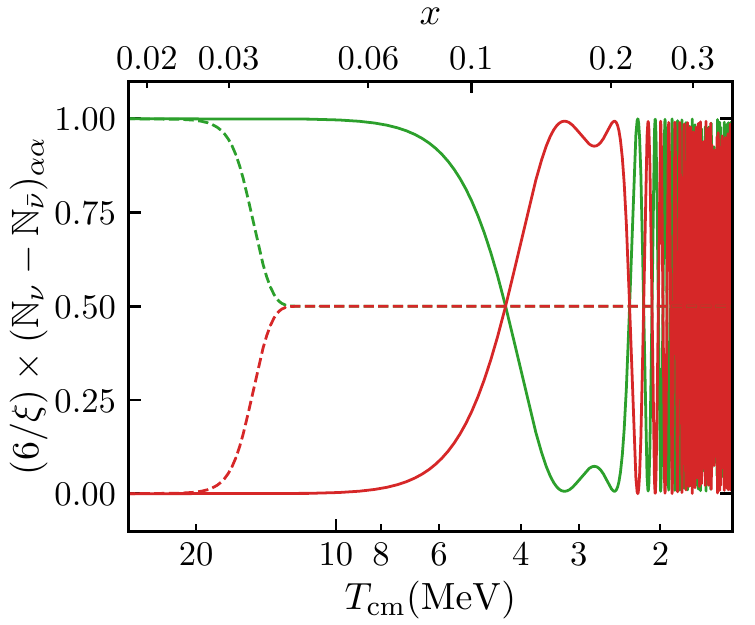}
    \includegraphics[]{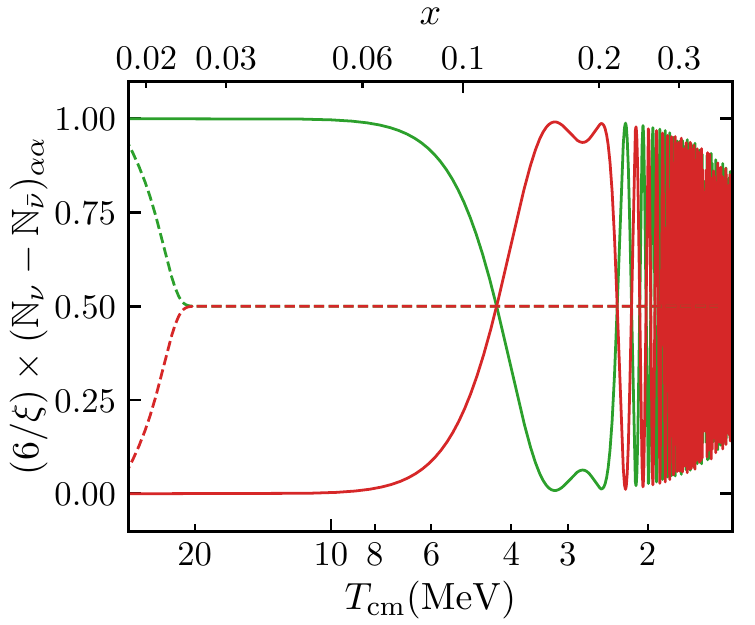}
	\caption{\label{fig:Collisions_mu} Effect of collisions on
          the muon-driven transition for $\xi_1=0.001$, $\xi_2=0$, with the collision term set to its actual value (left) and artificially multiplied by $1000$ (right). We only plot the result of two numerical schemes: \ATAOH (dashes) and \ATAOJH (solid, equivalent to QKE).}
\end{figure}

All in all, the asymmetries are not affected at all by the collision term as long as the \ATAOJ scheme is a good description of neutrino evolution. However, we have shown that below $\sim 10 \, \mathrm{MeV}$ the refined \ATAOJH scheme is necessary to capture the physics. The very fact that $\vrho$ and $\bvrho$ are not diagonalized with the same unitary matrix (either $U_{\Hself+\Hamil}$ or $U_{\Hself-\Hamil}$), and furthermore the $y$-dependence of these matrices, means that we lose the property \eqref{MagicKUJ}, that is
\begin{equation}\label{MagicKUJLost}
\mathcal{K}[\vrho,\bvrho] \neq U_{\Hself+\Hamil}(y)
\mathcal{K}[\widetilde\vrho,\widetilde\bvrho]  U^\dagger_{\Hself+\Hamil}(y) \ , \qquad
\bar{\mathcal{K}}[\vrho,\bvrho] \neq U_{\Hself-\Hamil}(y) \bar {\mathcal{K}}[\widetilde\vrho,\widetilde\bvrho]  U^\dagger_{\Hself-\Hamil}(y) \,. 
\end{equation}
When this non-equality is not meaningless --- it is necessarily suppressed by a factor $|
\vec{\Hamil}|/|\vec{\mathcal{\Hself}}| \propto x^4 /|\xi_1-\xi_2|$ --- the collision term starts to have a mild effect, which is even smaller for large $\xi_\alpha$ differences. Therefore, only for rather small $\xi_\alpha$ differences can a slight equilibration effect due to collisions be expected. However, since the frequency $\Omega$ of synchronous oscillations is then smaller, the actual start of oscillations is delayed until a moment when collisions are inefficient. This is why we expect collisions to have a negligible effect throughout the evolution in this $\nu_\mu - \nu_\tau$ system. We check this on figure~\ref{fig:Collisions_mu}, left plot, where the evolution is indistinguishable from the one without collisions (figure~\ref{fig:2nu_mu-tau}, top right plot). On the right plot, we artificially multiplied the collision term by $1000$, and we do observe in that case the damping of quasi-synchronous oscillations when $\abs{\Hamil} \sim \abs{\Hself}$, which corresponds to a reduction of asymmetry differences between the two flavours.

In principle, taking into account scattering and annihilations with muons and antimuons invalidates this picture since the charged currents with the $\nu_\mu$ and $\bar \nu_\mu$ make $G^L$ non-proportional to the identity, and the general property \eqref{MagicKUJ} would be lost. However their density is so suppressed in this range of temperature that we were able to check that the previous results are not affected.

Note also that the behaviour is very different if we ignore the self-interaction mean-field and rely on the pure \ATAOH scheme. In that case, the system is made of pure mass states of $\Hvac$ after the MSW transition (since each $\vrho$ is diagonal in the mass basis). However, given the $y$-dependence of $\Hamil$, this transition does not happen at the same time for all momenta. Thus there cannot be a property of the type \eqref{MagicKUJ} with $U_{\Hself} \to U_{\Hamil}$ because there is no unique $\Hamil$, and furthermore $U_{\Hamil}(y)$ depends on $y$ which prevents its factorization out of the collision integral. Therefore the collision term will tend to restore diagonality in flavour space (that is reduce flavour coherence), and this can only be compatible with pure mass states (a requirement of the \ATAOH approximation) when all flavours have reached the same distributions, that is when the asymmetry matrix $\Anti$ is proportional to the identity and thus $\vec{\Anti}=\vec{0}$. In other words, the collision term is strongly incompatible with the evolution of asymmetries dictated by the \ATAOH scheme, and thus damps them. We observe this behaviour on figure~\ref{fig:Collisions_mu}: $\Anti_\z \to 0$ in the presence of collisions, which was not the case on figure~\ref{fig:2nu_mu-tau}.

To conclude, if we ignore the self-interaction mean-field, the collision term efficiently damps the asymmetry differences, because the unitary adiabatic evolution is not the same for density matrices at various momenta. When including the additional self-interaction potential, as long as it dominates the vacuum and lepton mean-fields, the density matrices at various momenta evolve adiabatically with the common
unitary matrix $U_{\Hself}$ and this preserves the initial absence of
effect of the collision term. It is only when \ATAOJ is insufficient and one has to rely on \ATAOJH that one starts to see the effect of the unitary evolution differing between momenta, but also between neutrinos and antineutrinos. This allows the collision term to damp the asymmetry vector. But this comes with a very large delay and the collision term is only able to barely damp $\Anti_\z$.

\paragraph{Electron-driven transition} In the framework of section~\ref{SecElectronMSW}, the difference with the $\nu_\mu - \nu_\tau$ case is the fact that $G^L$ is no longer proportional to the identity: $G^L = {\rm diag}(g_L +1, g_L)$. Once oscillations develop and $U_\Hself \neq \mathbb{I}$, there is no property like~\eqref{MagicKUJ}. In other words, the matrix $G^L$ sets the direction $\vec{e}_\z$ towards which the collision term now unavoidably attracts $\vec{\vrho}$ and $\vec{\bvrho}$ (whereas before, $\mathcal{K}$ was blind to any global rotation of axes). Therefore, the collision term tends to erase flavour coherence much more efficiently: it damps oscillations but does not necessarily allow to fully reach a state where the two neutrino flavours have identical distributions, because the collision term becomes too weak at temperatures below the MSW transition.

\begin{figure}[!ht]
	\centering
	    \includegraphics[]{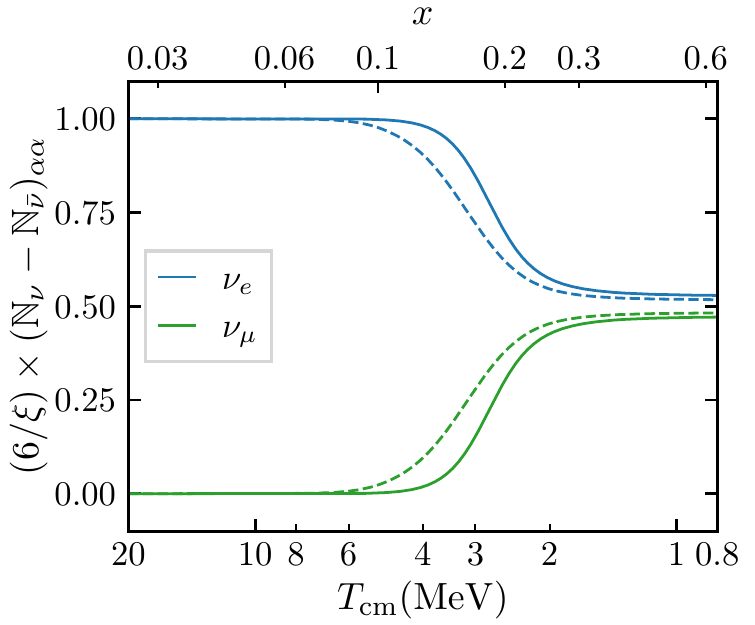}
	    \includegraphics[]{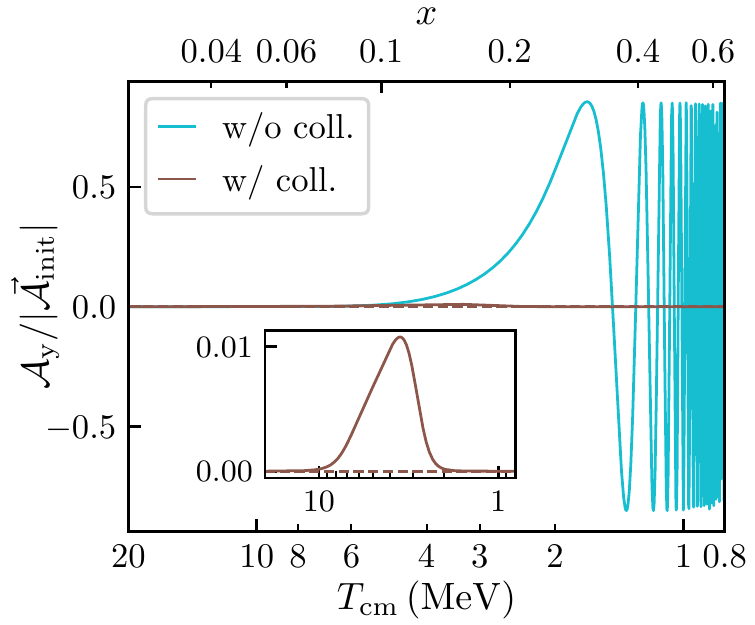}
	    \includegraphics[]{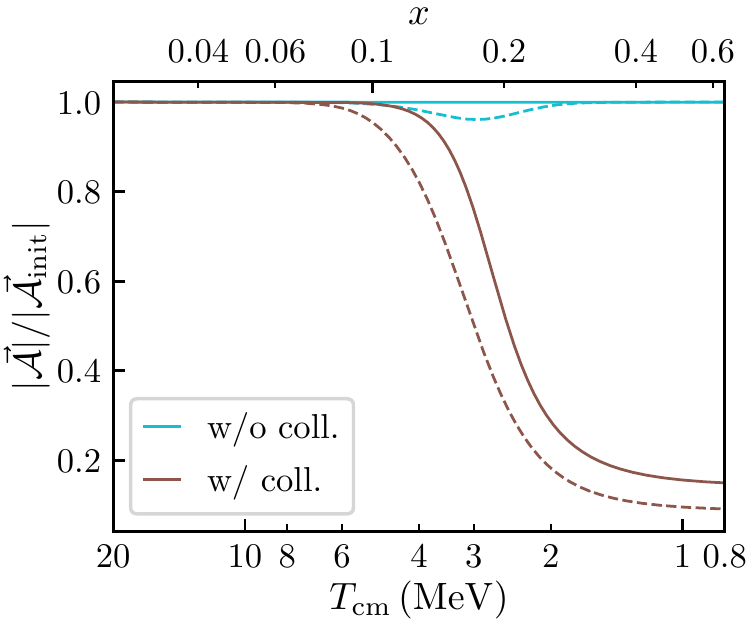}
        \includegraphics[]{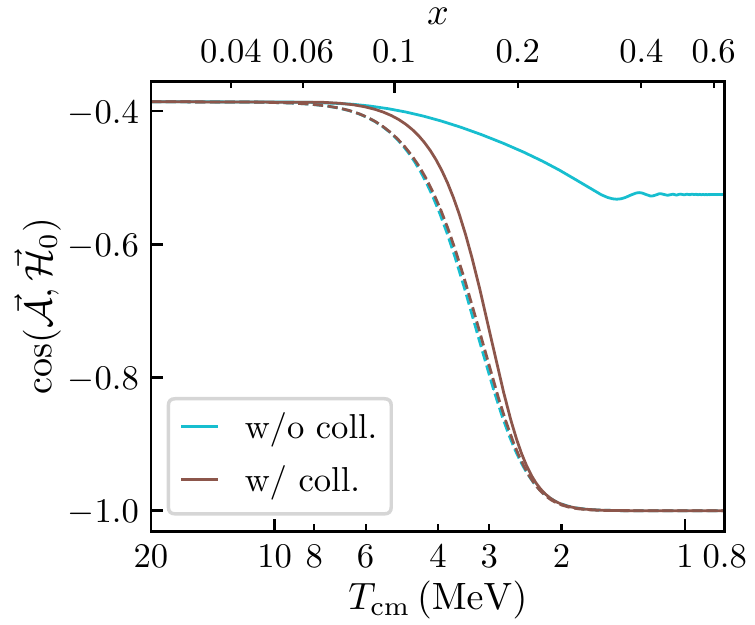}
	\caption{\label{fig:Coll_e-mu} Effect of collisions on the evolution of the $\nu_e - \nu_\mu$ system, with $(\xi_1 = 0.001, \, \xi_2 = 0)$. The dashed lines correspond to the \ATAOH scheme (no self-interactions in the mean-field), and the solid lines to the \ATAOJH scheme (equivalent to the full QKE resolution). \emph{Top left plot:} evolution of electron and muon flavour asymmetries. \emph{Top right plot:} $\y-$component of the asymmetry vector $\vec{\Anti}$. \emph{Bottom left plot:} norm of $\vec{\Anti}$. \emph{Bottom right plot:} angle between $\vec{\Anti}$ and the final precession direction $\vecHvac$.}
\end{figure}

The top left plot of figure~\ref{fig:Coll_e-mu} is equivalent to the top right plot of figure~\ref{fig:2nu_e-mu} ($\xi_1 = 0.001$, $\xi_2 = 0$), but including collisions. As expected, the asymmetry is damped by $\mathcal{K}$ in both the \ATAOH and \ATAOJH schemes, and the evolution looks quite similar, suggesting that one could be satisfied with the simpler \ATAOH resolution scheme. However, this misses some important physical features, as the other plots on figure~\ref{fig:Coll_e-mu} show. First, if one neglects the self-potential there is no precession of $\vec{\Anti}$ around $\vec{\Hamil}$, but simply the alignment of all $\vec{\vrho}, \, \vec{\bvrho}$ with $\vec{\Hamil}$ which evolves from $\Hlep$ domination to $\Hvac$ domination. This is clearly seen on the top right plot of figure~\ref{fig:Coll_e-mu} (dashed lines): the $\y$-component of $\vec{\Anti}$ is constantly equal to zero, which is expected as $\widehat{\Anti}$ evolves from $\vec{e}_\z$ to $\hat{\Hcal}_0$ that lies in the $(\x-\z)$ plane. In the correct \ATAOJH scheme however, synchronous oscillations do develop when collisions are discarded. When one takes them into account, $\Anti_\y$ does not take sizeable values but starts an oscillation (that is strongly damped by $\mathcal{K}$), see the insert plot.

The bottom plots of figure~\ref{fig:Coll_e-mu} show respectively the norm of $\vec{\Anti}$ and its alignment with $\vecHvac$. In the no-collision case (blue curves), the final angle between $\vec{\Anti}$ and $\vecHvac$ is non-zero (bottom right plot), but different from its initial value due to the very small adiabaticity of the MSW transition: $\vec{\Anti}$ slightly rotates towards $\vec{\Hamil}_{\rm eff}$, and precesses with an angle slightly different from $2 \theta$. Concerning its norm, $\lvert \vec{\Anti} \rvert$ is conserved without collisions.\footnote{The small “trough” in the \ATAOH case (dashed blue line) around $3 \, \mathrm{MeV}$ is not a numerical artefact. Since in that case each individual $\vec{\vrho}(y)$ changes its direction from $\vecHlep(y)$ to $\vecHvac(y)$ at different times (instead of being all locked on $\vec{\Anti}$), the norm of $\vec{\Anti}$ can only be compared at early and late times.} In contrast, we observe that including collisions (brown curves) $\vec{\Anti}$ gets aligned with $\vecHvac$ (but in the opposite direction due to the value of $\theta$), while the asymmetry differences are damped---a result of the competition between precession (which sets the preferred direction $\hat{\Hcal}_0$) and collisions (with the preferred direction $\vec{e}_\z$).

\section{Evolution with three flavours of neutrinos}\label{Sec:3neutrinos}

Having presented in the previous sections the salient features of two-neutrino evolution in the presence of flavour asymmetries, we can now turn to the full three-neutrino framework. Our goal is not to provide a thorough exploration of the parameter space, but instead to highlight the main physical characteristics of neutrino evolution with non-zero asymmetries.

\subsection{Method}

We have shown the accuracy of the \ATAOJH scheme that we can confidently use instead of a full QKE resolution, more computationally expensive. Therefore, the results are here obtained with this method and are compared with the \ATAOH scheme where we recall that the self-interactions are ignored in the mean-field, so as to highlight how the self-potential changes the dynamics.

However, it is impossible to integrate correctly the evolution at low temperatures. First, oscillations become too fast as their frequency grows as $x^6$ when the NLO dominates. Then, we reach the point where the \ATAOJH scheme starts to fail and oscillations must become gradually not synchronized. Eventually the system must converge to a state where fast oscillations disappear, that is an \ATAOH scheme. We chose to switch to an \ATAOH scheme at low temperature to effectively capture this transition from \ATAOJH to \ATAOH. In principle one should use the full QKE scheme to integrate numerically this phase, but for the same reasons it is numerically daunting. We chose to switch to the \ATAOH scheme around $2\,{\rm MeV}$, since the final MSW transition is over and collisions become rather inefficient (except for electron/positron annihilations). This method of instantaneous switching to the \ATAOH scheme necessarily misses some physics since it hides the complexity of the transition, but as we discuss in section~\ref{SecTransitionToATAOV} we expect that this does not significantly affect the results obtained for the evolution of asymmetry.

\subsection{Results with standard mixing parameters}

Given the numerous possibilities for the values of the initial degeneracy parameters, we chose to restrict to two types of initial conditions. First, we consider the case where electronic flavour neutrinos have the largest degeneracy ($\xi_e=\xi_{\rm max}$), with $\xi_\tau=\xi_e/10$. Second we consider the case where muonic neutrinos have the largest non vanishing initial degeneracy ($\xi_\mu=\xi_{\rm max}$), with $\xi_\tau=\xi_\mu/10$. We do not report results where $\xi_\tau$ is the largest, because it is qualitatively very similar to the case where $\xi_\mu$ is the largest potential, since oscillations develop in exactly the same way. In any case, the third initial degeneracy parameter is set to zero.

\begin{figure}[!ht]
	\centering
	\includegraphics[]{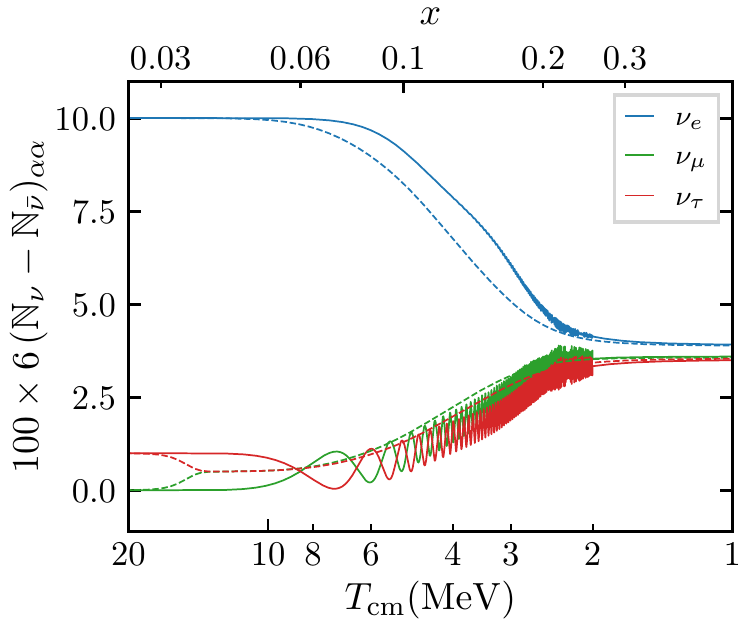}
	\includegraphics[]{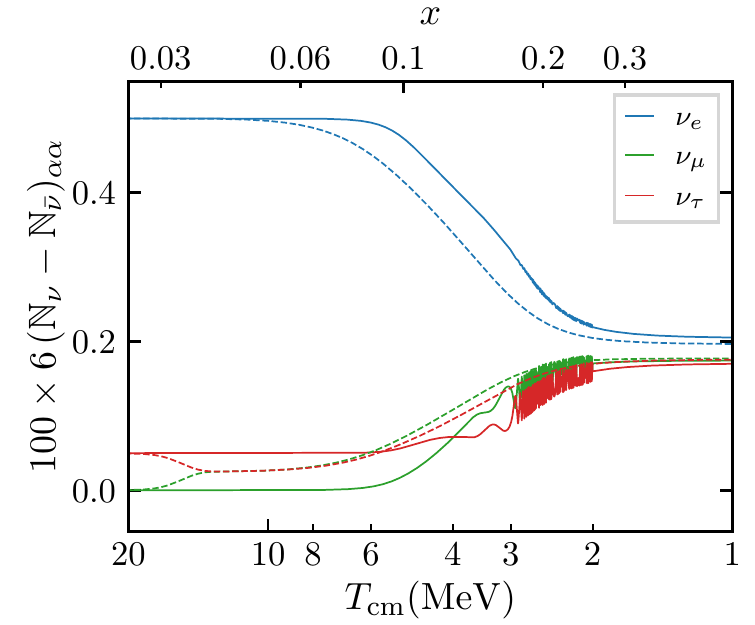}\\
	\includegraphics[]{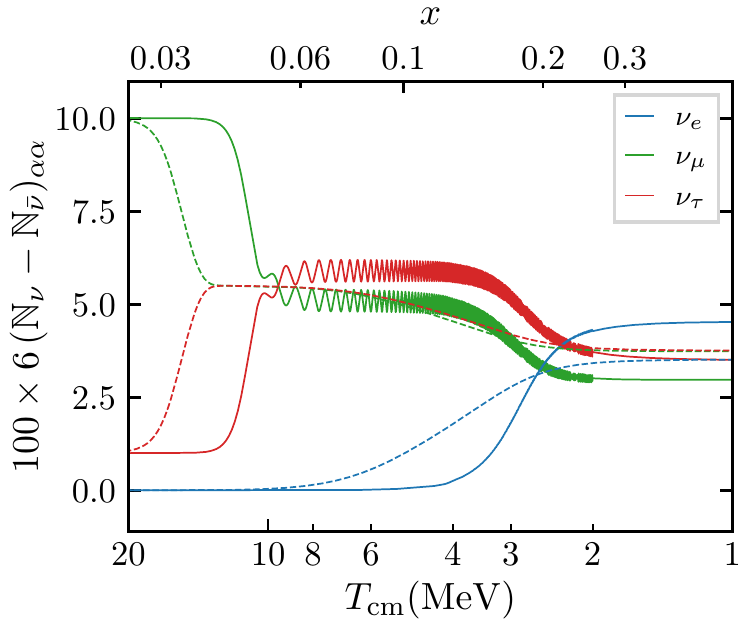}
	\includegraphics[]{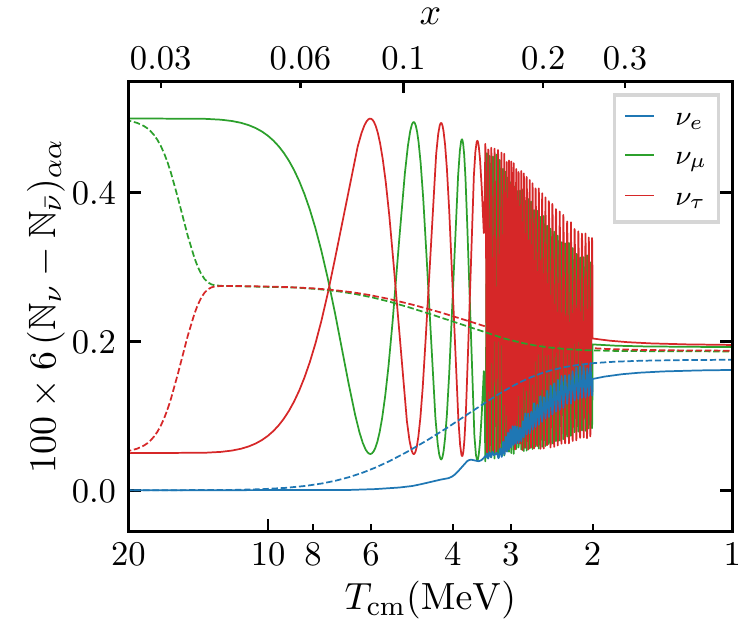}
	\caption{\label{figStandard} Initial conditions are $\xi_\alpha=(0.1,0,0.01)$ (top left), $\xi_\alpha=(0.005,0,0.0005)$ (top right), $\xi_\alpha=(0,0.1,0.01)$ (bottom left) and $\xi_\alpha=(0,0.005,0.0005)$ (bottom right). 
	The solid lines are the \ATAOJH schemes (extended into a simple \ATAOH below $2\,{\rm MeV}$), and the dashed lines are \ATAOH schemes throughout.}
\end{figure}

Results are depicted in figure~\ref{figStandard} for both typically large ($\xi_{\rm max}=0.1$) and typically small ($\xi_{\rm max}=0.005$) potentials. We can observe how synchronous oscillations develop in the $\nu_\mu-\nu_\tau$ space after the muon-driven transition. Their amplitude is reduced for large initial potentials since the transition is then more adiabatic in that case, as detailed in section~\ref{SecMuonMSW}. In the case of small initial potentials, the transition from leading order to NLO oscillations is also clearly visible (right plots of figure~\ref{figStandard}). Note that even though the general behaviour is that asymmetries tend to converge, this trend stops before equilibration is complete, and is in general less complete than in the case where the self-interaction mean-field is ignored. Furthermore in some cases, the ordering of final asymmetry is not the same as the ordering of initial ones.

If we consider cases with much larger initial asymmetries in the muonic and tauic neutrinos, typically such that $\xi_\mu+\xi_\tau \gg 0.1$, the muon-driven transition is very adiabatic, and oscillations do not develop at that transition as the asymmetry vector closely follows the evolution of $\Hamil(y_{\rm eff})$. This is illustrated in the left plot of figure~\ref{figComparison}.

\begin{figure}[!ht]
	\centering
    \includegraphics[]{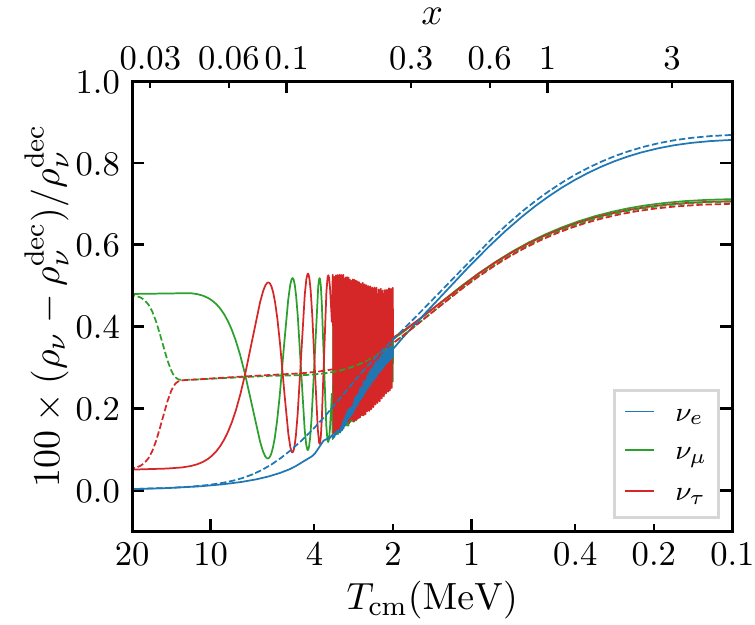}
    \includegraphics[]{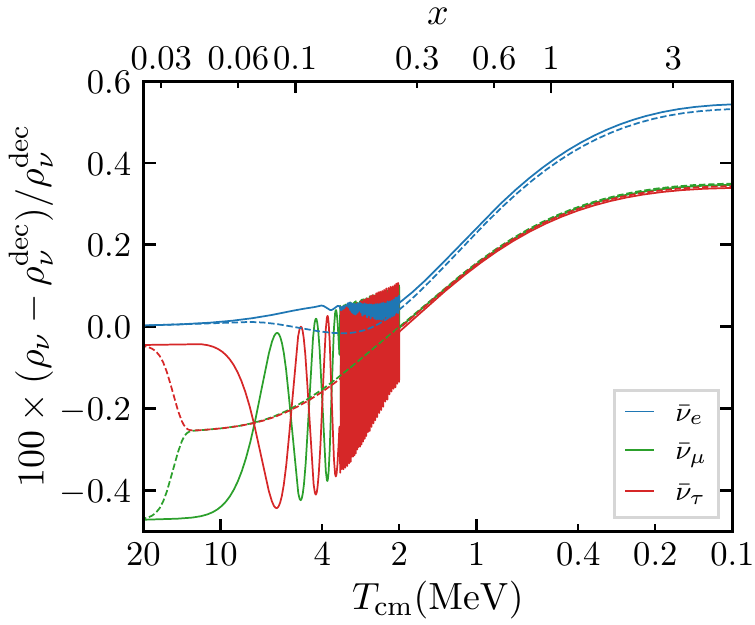}
	\caption{\label{figrho} Fractional difference of the energy density with respect to a decoupled neutrino without chemical potential. The solid lines are the \ATAOJH schemes (extended into a simple \ATAOH below $2\,{\rm MeV}$), and the dashed lines are \ATAOH schemes throughout. The left plot is for neutrinos and the right plot for antineutrinos. Initial conditions are $\xi_\alpha=(0,0.005,0.0005)$.}
\end{figure}

We stress that in this section we always consider the full collision term, and even though we show the results for the evolution of the asymmetry (since it is the relevant quantity to discuss synchronized oscillations), neutrinos are also partially reheated by electron/positron annihilations, which preserve the asymmetry. In figure~\ref{figrho} we show the energy density fractional difference with respect to the one of a completely decoupled neutrino species with vanishing chemical potential $\rho_\nu^{\rm dec} = 7 \pi^2/240$. In general the final effective temperature and distortions of electronic (anti-)neutrinos, which have a direct effect on neutron/proton freeze-out and thus BBN, depend on initial degeneracy parameters. Hence the impact on BBN predictions is not straightforward and one should perform a full BBN analysis for each set of initial conditions~\cite{Grohs:2016cuu}, as was done for the standard case of vanishing potentials in \cite{Froustey2019}.

It is beyond the scope of this paper to perform a full exploration of parameters with all initial degeneracy parameters and all mixing parameters. However we aim here at highlighting how results are qualitatively modified when considering different mixing parameters.

\subsection{Dependence on neutrino mass ordering}

\begin{figure}[!ht]
	\centering
	\includegraphics[]{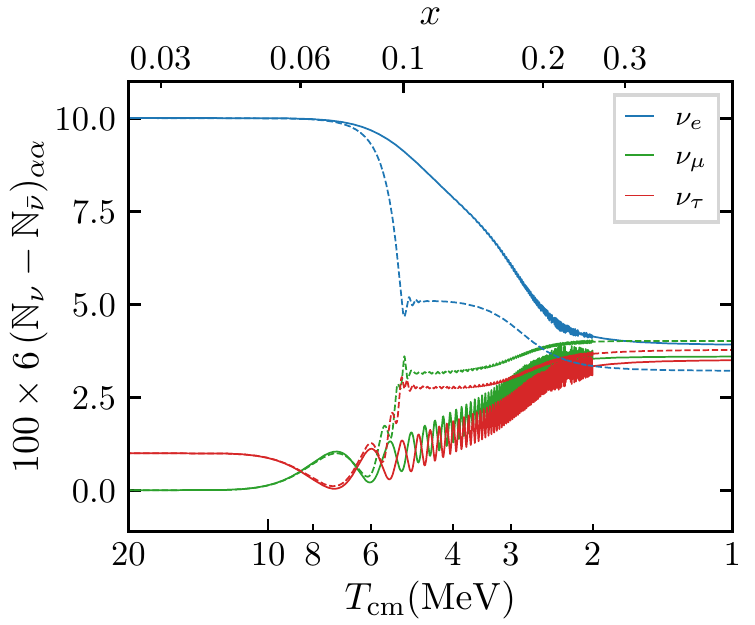}
        \includegraphics[]{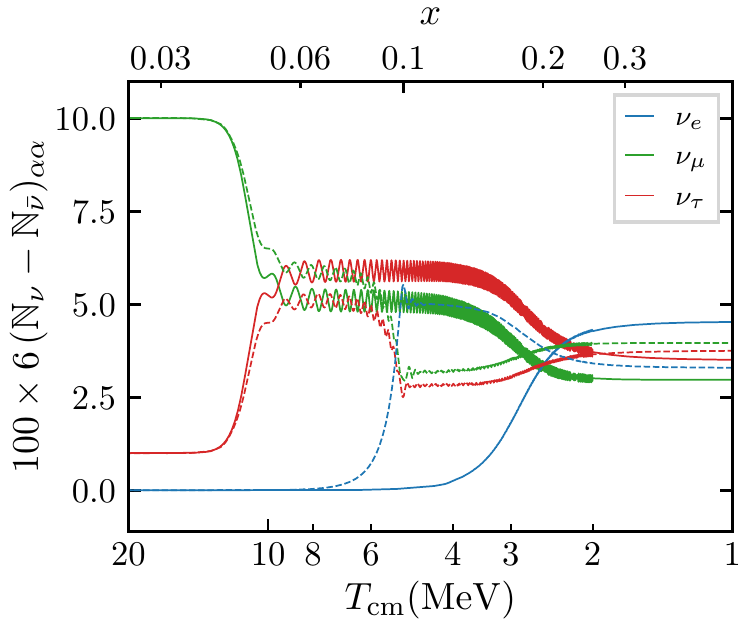}
	\caption{\label{figcomp4} The normal ordering case in solid lines, is compared to the inverted ordering case in dashed lines. Initial conditions are $\xi_\alpha=(0.1,0,0.01)$ on the left, and $\xi_\alpha=(0,0.1,0.01)$ on the right.}
\end{figure}

In figure~\ref{figcomp4} we show the dependence on the neutrino mass ordering on two examples. The main difference is the resonant nature of the first electron-driven MSW transition at $T_{\rm MSW}^{(e),1} \simeq 5\,{\rm MeV}$ in IO. It leads to a much faster evolution, a feature also observed in figure 1 of~\cite{Mangano:2011ip}. On the examples of figure~\ref{figcomp4}, we see two consequences of this resonant transition: the ordering of degeneracy parameters can be modified, and collisions are much more efficient in damping the synchronous oscillations right after the transition.

\subsection{Dependence on mixing angles}\label{sec:mixing_angles}

In the two-neutrino case we have shown that $\vecHvac$ sets the precession direction of $\vec{\Anti}$; in other words, the values of the mixing angles are key parameters to determine the final asymmetry differences. Even though they are now better and better constrained~\cite{PDG}, let us explore qualitatively in this section their influence on the equilibration process.

\begin{figure}[!ht]
	\centering
	\includegraphics[]{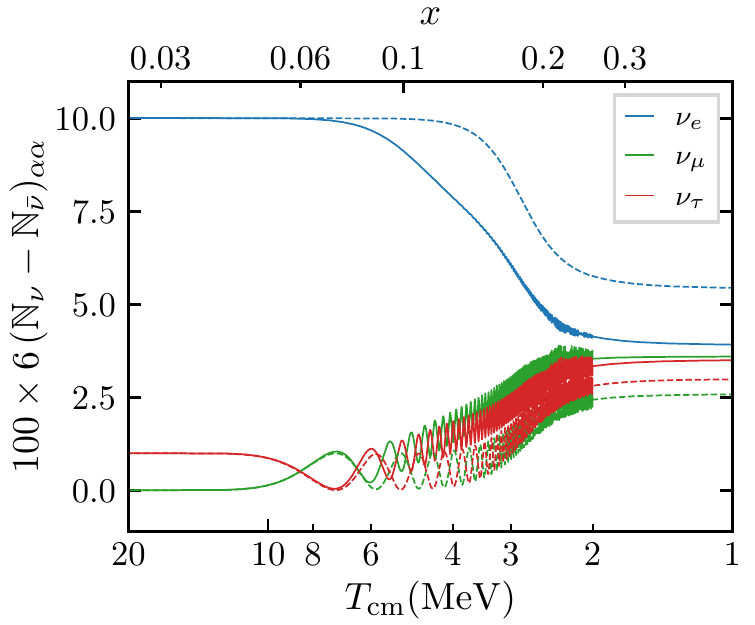}
        \includegraphics[]{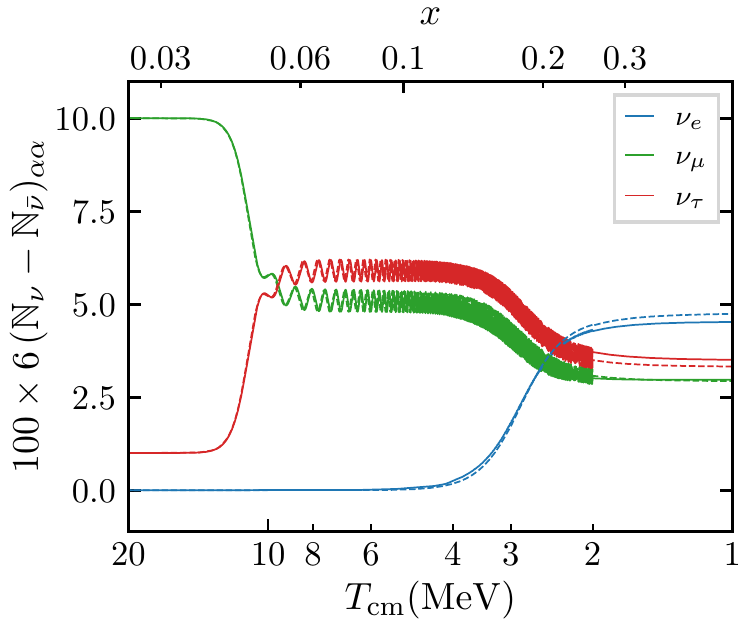}\\
        \includegraphics[]{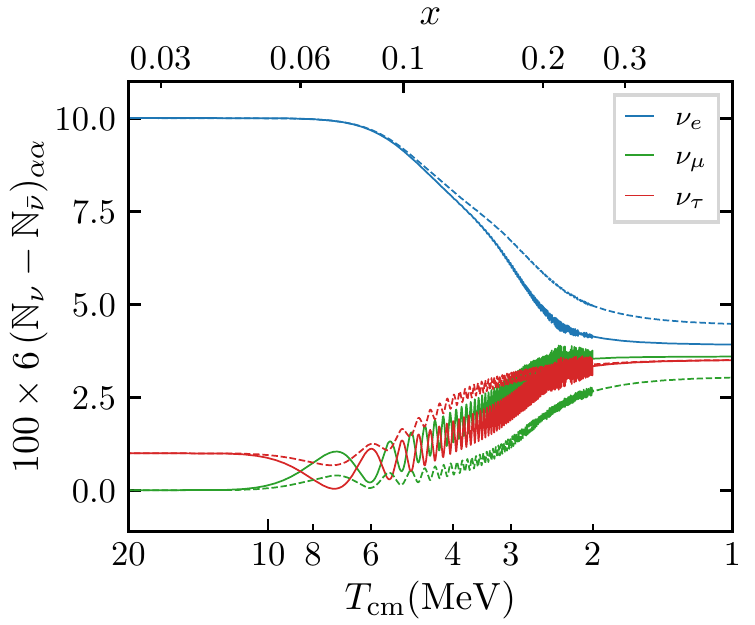}
        \includegraphics[]{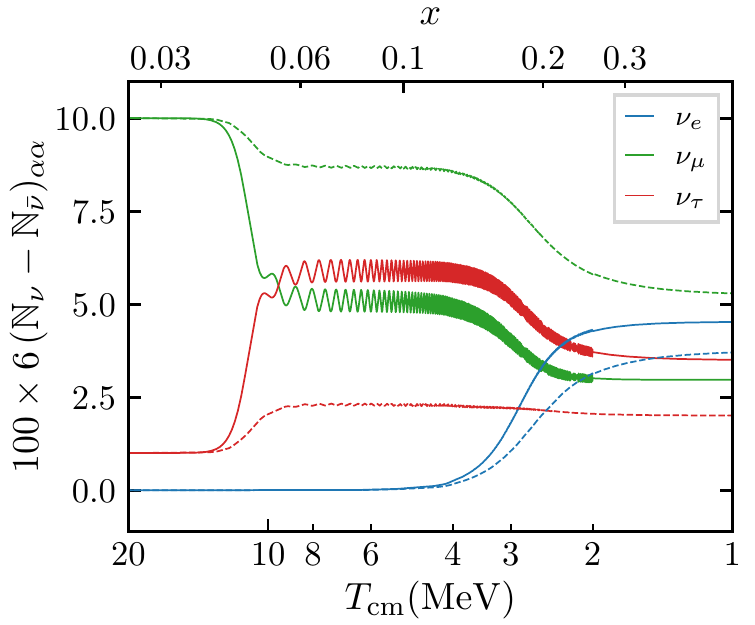}
	\caption{\label{figcompamgles} Comparison of the standard case (solid lines) with a modified setup (in dashed lines). In the upper plots, the modification is  $\theta_{13}=0$, and in the lower plots $\theta_{12}=\theta_{23}=\pi/8$, with everything else unchanged. Initial conditions are $\xi_\alpha=(0.1,0,0.01)$ on the left, and $\xi_\alpha=(0,0.1,0.01)$ on the right. At low temperatures, the green lines overlap on the top right subplot, and similarly for the red lines on the bottom left subplot.}
      \end{figure}

To that purpose, we compare in figure~\ref{figcompamgles} (upper plots) the standard case discussed above with modified setups. First, when $\theta_{13}=0$ (the rest being unchanged), we notice that equilibration is less efficient. Furthermore setting $\theta_{12} = \theta_{23} = \pi/8$, with $\theta_{13}$ at its standard value, the equilibration is also much less efficient as depicted in figure~\ref{figcompamgles} (lower plots). Therefore, the general result that asymmetries mostly tend to equilibrate crucially depends on the values of the mixing angles. Typically, for small values of the mixing angles $\theta_{12}$ and $\theta_{23}$, that is far away from $\pi/4$, equilibration is less efficient, and a non-vanishing value for $\theta_{13}$ also significantly helps the equilibration process as highlighted in~\cite{Dolgov_NuPhB2002,Mangano:2010ei,Mangano:2011ip}. Also, using $\theta_{23} = \pi/8$ instead of the larger standard value~\eqref{ValuesStandard} increases the geometric factor $\cos^2 \theta/\sin \theta$ of the adiabatic parameter~\eqref{Defgammatr2} by a factor $\simeq 3.6$. Therefore, the muon-driven transition is much more adiabatic and the resulting oscillations are suppressed (see the discussion in section~\ref{SecDescriptionMuonTransition}), as can be checked on the bottom plots of figure~\ref{figcompamgles}.

Evidently, even though we do not report it here, when all mixing angles vanish, equilibration entirely disappears. This highlights the importance of a consistent treatment of neutrino mixing when studying flavour equilibration in the early Universe. 

\subsection{Dependence on the Dirac phase}\label{SecDiracPhase}

We now examine the effect of the Dirac CP-violating phase $\delta$, that we discarded up until now in the PMNS matrix. Unless stated otherwise, all quantities in this section are considered in the case $\delta \neq 0$. $\vrho^{(\delta=0)}$ and $\bvrho^{(\delta=0)}$ refer to the solutions with vanishing Dirac phase, and we shall show how the general case with a non-zero phase can be deduced from it. It has been shown in \cite{Balantekin:2007es,Gava:2008rp,Gava:2010kz,Gava_corr} (see also appendix F of \cite{Froustey2020}) that the evolution with a non-vanishing
Dirac phase can be obtained from a transformation of the result obtained with a vanishing phase. More precisely, defining
$\check{S} \equiv R_{23} S R_{23}^\dagger$ (cf.~notations in section~\ref{subsec:Mixing_param}), we can define
\begin{equation}\label{CheckAll}
\check{\mathcal{H}}_{\rm lep} = \check{S}^\dagger \mathcal{H}_{\rm lep} \check{S} \, , \qquad \check{\mathcal{H}}_0 = \check{S}^\dagger \mathcal{H}_0 \check{S} \, , \qquad \check{\vrho} \equiv \check{S}^\dagger \vrho \check{S} \, , \qquad \check{\bvrho} \equiv \check{S}^\dagger \bvrho \check{S}\,,
\end{equation}
and similar transformations for the collision terms. Since $\check{S}$ is of the type \eqref{UtoU2} (see equation~\eqref{eq:Umat_SS} below), we infer from the property \eqref{MagicKFundamental} that 
\begin{equation}\label{MagiccheckK}
\check{\mathcal{K}}[\vrho,\bvrho] = \mathcal{K}[\check{\vrho},\check{\bvrho}] \quad \text{and} \quad \check{\bar{\mathcal{K}}}[\vrho,\bvrho] = \bar{\mathcal{K}}[\check{\vrho},\check{\bvrho}]\,. 
\end{equation}
Furthermore, given that 
\begin{equation}\label{MagicUU}
\check{S}^\dagger U = U^{(\delta=0)}S^\dagger\,,
\end{equation}
and $[\mathbb{M}^2,S]=0$, we deduce that $\check{\mathcal{H}}_0 = \mathcal{H}_0^{(\delta=0)}$. We then obtain that the evolution of $\check{\vrho}$ (resp. of
$\check{\bvrho}$) is the same as the evolution of $\vrho^{(\delta=0)}$ (resp. of $\bvrho^{(\delta=0)}$) when the replacements $\mathcal{H}_{\rm lep} \to \check{\mathcal{H}}_{\rm lep}$ and $\Hself \to \check{\Hself}$ have been performed, that is 
\begin{equation}
\frac{\partial \check{\vrho} }{\partial x} = - \ii [\Hvac^{(\delta=0)}+\check{\mathcal{H}}_{\rm lep} +
\check{\Hself},\check{\vrho}] + \mathcal{K}[\check{\vrho},\check{\bvrho}]\,,\qquad \frac{\partial \check{\bvrho} }{\partial x} = + \ii [\Hvac^{(\delta=0)}+\check{\mathcal{H}}_{\rm lep} -
\check{\Hself},\check{\bvrho}] + \bar{\mathcal{K}}[\check{\vrho},\check{\bvrho}]\,. 
\end{equation}
In the standard case, that is with vanishing initial chemical potentials, $\check{\vrho}$ and $\vrho$ have the same initial conditions. If we further neglect the mean-field effects of muons/antimuons, then $[\check{S},\Hlep]=0$, hence $\check{\mathcal{H}}_{\rm lep} = \Hlep$ and $\check{\vrho}=\vrho^{(\delta=0)}$ (likewise for antineutrinos) at all times, as shown in \cite{Gava:2010kz,Gava_corr}. From this property, we obtain $\vrho$ from $\vrho^{(\delta=0)}$ using the inverse transformation, that is we get $\vrho = \check{S}\vrho^{(\delta =0)} \check{S}^\dagger$, with a similar relation for antineutrinos. As detailed in \cite{Froustey2020}, it is equivalent to saying that both results are
exactly equal in their respective mass basis.

\begin{figure}[!ht]
	\centering
	\includegraphics[]{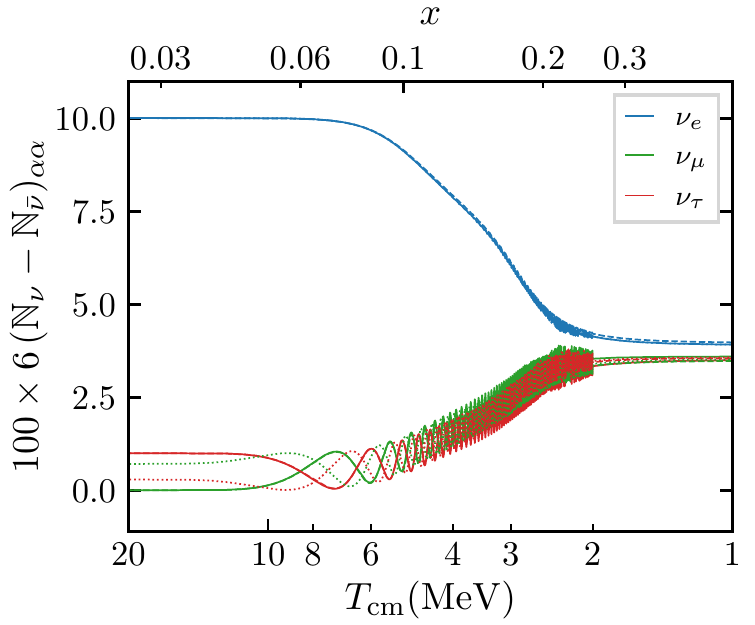}
	\includegraphics[]{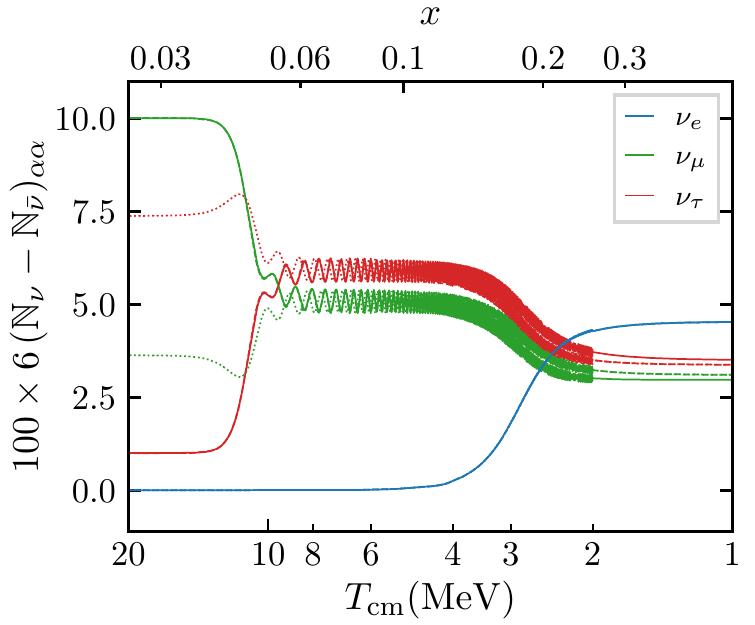}
	\caption{\label{figCP1} Effect of $\delta \neq 0$ on the evolution of asymmetries. The solid lines correspond to the standard case with $\delta=0$ (that is $\vrho^{(\delta=0)}$ and $\bvrho^{(\delta=0)}$). The dashed lines are the case with $\delta =245^\circ$ (central value in the most recent constraints~\cite{PDG}). The dotted lines correspond to the standard case results on which the transformation $\check{S}\vrho^{(\delta=0)} \check{S}^\dagger$ (and similarly for antineutrinos) has been applied. Initial conditions are $\xi_\alpha=(0.1,0,0.01)$ on the left and $\xi_\alpha=(0,0.1,0.01)$ on the right. Initially, the dashed lines are hidden behind the solid ones (see text). In the final stages the dotted lines are hidden behind the dashed lines hence showing that asymptotically $\vrho\simeq \check{S} \vrho^{(\delta=0)} \check{S}^\dagger$.
	}
\end{figure}

However in the presence of initial degeneracies, the initial conditions for $\vrho$ are not necessarily equal to those of $\check{\vrho}$. It is interesting to note that going from $\vrho$ to $\check{\vrho}$
amounts to a rotation in the vector description of the $\nu_\mu-\nu_\tau$
subspace. Indeed, both $S$  and $\check{S}$ are of the \eqref{UtoU2} type, and the associated ${\cal U}_S$ and ${\cal U}_{\check{S}}$ are expressed in terms of rotations as
\begin{equation}
\label{eq:Umat_SS}
\mathcal{U}_S = {\rm e}^{\ii \delta/2} \mathcal{R}_\z(\delta) \ , \qquad
\mathcal{U}_{\check{S}} = {\rm e}^{\ii \delta/2}
\mathcal{R}_\y (-2 \theta_{23}) \cdot
\mathcal{R}_\z(\delta) \cdot \mathcal{R}_\y^\dagger(-2 \theta_{23})\,.
\end{equation}
Forgetting the global ${\rm e}^{\ii \delta/2}$ factor which plays no role, we can use the property \eqref{LaMagieDesMaths} to interpret $\mathcal{U}_{\check{S}} $ as a rotation, when using the vector representation \eqref{MatrixToVector}. It corresponds to a rotation of angle $\delta$ around an axis whose direction is obtained from the rotation $\mathcal{R}_\y (-2 \theta_{23})$ of the $\z$-axis. However, using that $\theta_{23}^{\rm eff} \simeq \theta_{23}$ from \eqref{th23th23}, this axis is approximately the one subtended by the $\nu_\mu-\nu_\tau$ restriction of the vacuum Hamiltonian. This has interesting consequences.
 \begin{itemize}
 \item Before the electron-driven transitions, using~\eqref{CheckAll} with $U$ of the effective form~\eqref{UtoU}, and~\eqref{eq:Umat_SS} taking $\theta_{23}^{\rm eff} \simeq \theta_{23}$, we infer the approximate relations
  \begin{equation}
  \left(\check{\mathcal{H}}_0\right)_{\alpha\beta} \simeq \left(\mathcal{H}_0\right)_{\alpha\beta} \quad \implies \quad \left(\mathcal{H}_0\right)_{\alpha\beta}  \simeq \left(\mathcal{H}^{(\delta=0)}_0\right)_{\alpha\beta}\quad \text{for}\quad \alpha,\beta \in\{\mu,\tau \}\,.
  \end{equation}
  Therefore at early times the evolution of $\vrho$ and $\vrho^{(\delta=0)}$ are nearly completely similar (likewise for antineutrinos) as can be checked by comparing the solid and dashed lines of figure~\ref{figCP1} at high temperatures.
 \item Since the asymmetry vector precesses around that precise direction after the muon-driven transition, this amounts to the fact that after the muon-driven MSW transition and before the electron-driven ones, the oscillations of $\check{\vrho}$ (similarly for $\check{\bvrho}$) are simply phase shifted with respect to the ones of $\vrho^{(\delta=0)}$  (resp. $\bvrho^{(\delta=0)}$), by an angle $\delta$, as can be checked by comparing the dashed and dotted lines on figure~\ref{figCP1} (see also the left plot of figure~\ref{figComparison}).
\end{itemize}

Later, when the electron-driven transitions occur, the $\delta$-phase difference between $\check{\vrho}$ and $\vrho^{(\delta=0)}$ (likewise for antineutrinos) can only have an extremely marginal
effect because the oscillation frequency keeps increasing, and the amount of damping incurred in the magnitude of (the traceless part of) $\Anti$ is essentially only sensitive to the amplitude and axes of oscillations. Oscillations keep accelerating until synchronous oscillations disappear as we reach an average set by $\mathcal{H}_0$, which is captured by the \ATAOH scheme. Eventually the initial dephasing is lost, and it is impossible to distinguish between the final values of $\check{\vrho}$ and $\vrho^{(\delta=0)}$ (likewise for antineutrinos), therefore we can relate the final results to the case without Dirac phase by
\begin{equation}\label{FinalOperation}
\vrho \simeq \check{S}\vrho^{(\delta =0)} \check{S}^\dagger\,,\qquad \bvrho \simeq \check{S}\bvrho^{(\delta =0)} \check{S}^\dagger\,.
\end{equation}
There are two differences with the standard case without initial chemical potentials. On the one hand, \eqref{FinalOperation} is an approximate result and not an equality, based on the following approximations:
\begin{enumerate}
\item $\theta_{23}^{\rm eff} \simeq \theta_{23}$, which is guaranteed from equation~\eqref{th23th23} by $|\Delta m_{21}^2/\Delta m_{31}^2| \ll 1$ ;
\item the muon-driven MSW transition takes place well before the first electron-driven transition ($T_{\rm MSW}^{(\mu)} \gg T_{\rm MSW}^{(e),1}$), which is the case since $\me/m_\mu \ll 1$ ;
\item the amplitude and directions of oscillations are not meaningfully affected by the $\delta$-dephasing, and eventually this dephasing should not be observable as oscillations are averaged at large $x$.
\end{enumerate}
On the other hand, it is only valid at late times, whereas in the standard case it is valid at all times. It can be checked on figure~\ref{figCP1} that it is very accurate, as the dashed lines ($\vrho$) and dotted lines ($\check{S}\vrho^{(\delta =0)} \check{S}^\dagger$) are nearly indistinguishable at late times, and are only $\delta$-dephased at early times if oscillations develop.

\noindent Finally the property \eqref{FinalOperation} allows to understand the physical effects of the Dirac phase. 
\begin{itemize}
\item When converted into mass basis components, the property \eqref{MagicUU} and the relation~\eqref{FinalOperation} imply the (approximate) relation $\widetilde{\vrho} \simeq S \widetilde{\vrho}^{(\delta=0)} S^\dagger$. This means that in the \ATAOH approximation which holds at late times, we have $\widetilde \vrho = \widetilde \vrho^{(\delta=0)}$ and likewise for antineutrinos given that off-diagonal components in the matter basis vanish --- see appendix F of \cite{Froustey2020}. Hence the difference in the final state, when interpreted in flavour basis, is (approximately) only due to the different mass bases depending on the value of $\delta$. Structure formation, being sensitive to mass bases spectra, is therefore not affected by the Dirac phase. In addition, the trace of density matrices is conserved by \eqref{FinalOperation} such that $N_{\rm eff}$ is preserved, and cosmological expansion is not modified. We conclude that there is no sizeable gravitational signature of the Dirac phase.
\item Since $\check{S}$ is of the \eqref{UtoU2} type, the transformation \eqref{FinalOperation} affects the number densities of $\nu_\mu$ and $\nu_\tau$, that is $\vrho_{\mu\mu}$ and $\vrho_{\tau\tau}$, but the number density for $\nu_{e}$, that is $\vrho_{ee}$, is left invariant (likewise for antineutrinos). Only the coherence between the $e$ states and the $\mu$ and $\tau$ states, that is the $\vrho_{e\mu}$ and $\vrho_{e\tau}$ components, is affected. Therefore, there is also no perceptible effect on BBN because the neutron/proton freeze-out is sensitive only to the spectrum of electronic (anti)neutrinos.
\end{itemize}

\subsection{Equal but opposite asymmetries}

\begin{figure}[!ht]
\centering
\includegraphics[]{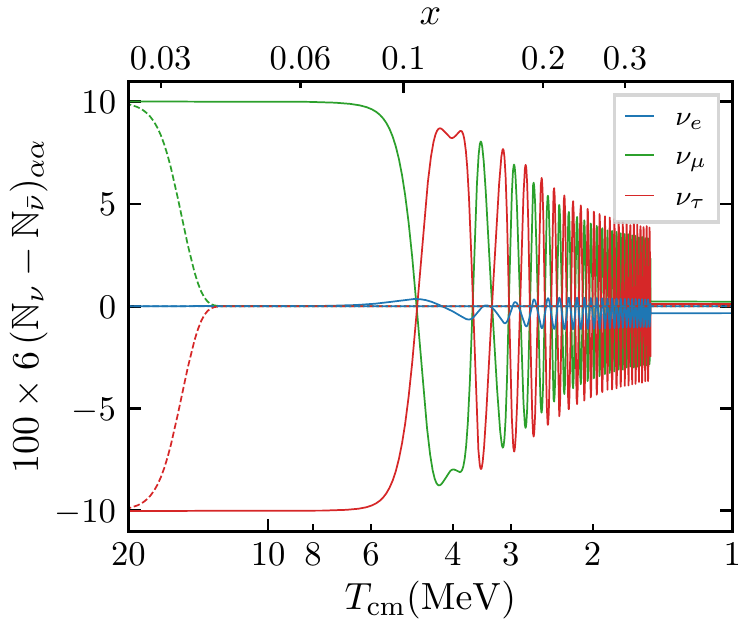}
\includegraphics[]{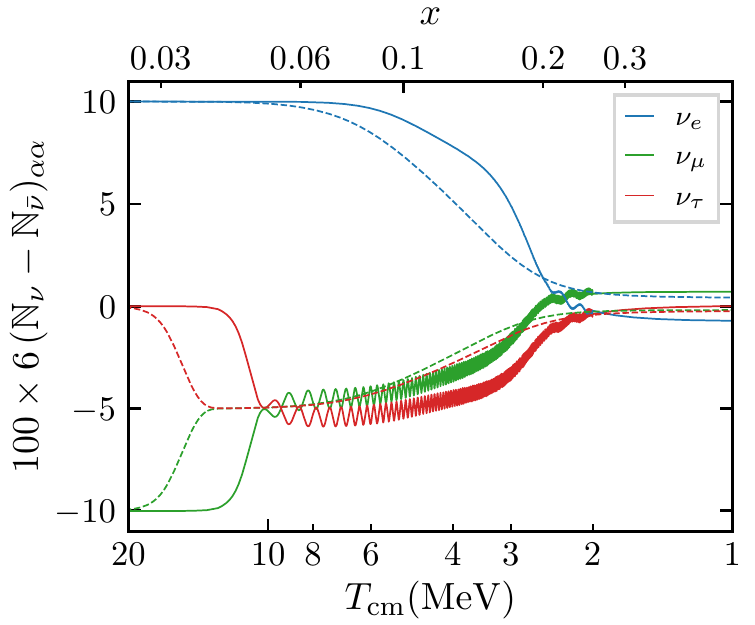}
	\caption{\label{figEBO} Special case of equal but opposite non-vanishing degeneracy parameters. The initial conditions on the left are $\xi_\alpha=(0,0.1,-0.1)$, and $\xi_\alpha=(0.1,-0.1,0)$ on the right. \emph{Numerical schemes:} \ATAOJH in solid lines, \ATAOH in dashed lines. On the left plot, we extended the \ATAOJH integration until $1.5 \, \mathrm{MeV}$ to see further the damping of the oscillations.}
\end{figure}

As noted in section~\ref{FreqSyncOsc}, the case of equal but opposite asymmetries is special because the leading order of synchronous oscillations vanishes --- meaning that the asymmetry should remain locked in its original configuration, --- but not the next-to-leading order. The results obtained in three-neutrino cases are depicted in figure~\ref{figEBO}. When $\xi_\mu+\xi_\tau=0$, we expect from the analysis of section~\ref{SecParticular} that oscillations in the $\nu_\mu-\nu_\tau$ space should start at $1.4 \,{\rm MeV}$, while we observe the first half-oscillation at $4\,{\rm MeV}$. These oscillations must therefore be triggered by the first electron-driven transition.

One can compare the left plot of figure~\ref{figEBO} with figure~9 of reference~\cite{Dolgov_NuPhB2002}: the damping of synchronous oscillations is considerably reduced in our calculation, which we attribute to the use of the exact collision term, instead of a damping approximation (see also the discussion in section~\ref{SecDiscussion}).

If the opposite initial degeneracy parameters are the electronic and muonic (or tauic) ones, there is no substantial difference with the “standard” case (figure~\ref{figStandard}). Indeed, the $\nu_\mu-\nu_\tau$ oscillations equilibrate partially (at least in the case of large $\xi_\tau$ taken in the figure) the asymmetries in the $\nu_\mu - \nu_\tau$ subspace, and the common asymmetry is then not the opposite of $\xi_e$.

\section{Discussion}\label{SecDiscussion}

\subsection{Transition from \ATAOJH to \ATAOH}\label{SecTransitionToATAOV}

We have shown that the \ATAOJH scheme works very well as long as $|\Hamil| \ll |\Hself|$. Let us discuss here in more details the end of the \ATAOJH regime, in a simplified case with only two flavours so as to use the vector formalism. When $\lvert \vec{\Hamil} \rvert$ and $\lvert \vec{\Hself} \rvert$ are of the same order, quasi-synchronous oscillations cease to exist. Therefore, in principle, only the QKE scheme can handle this regime. However, the individual $\vec{\vrho}(y)$ then evolve independently rather than collectively, so that extremely rapid precessions around $\vec{\Hamil}$ should take place for all momenta, and given the $y$-dependence of $\vec{\Hamil}$, they tend to have different frequencies. 
Given that $|\vec{\Hamil}| \propto  (\Delta m^2/(2 y \me H) \simeq (\Delta m^2/10^{-3}{\rm eV}^2)\times 8 \cdot 10^6 \times x^2/y$, the oscillating part in the spectrum is typically a trigonometric function whose phase $\phi \propto x^3/y$. This implies extremely fast oscillations in the spectrum (i.e. in the variable $y$) whenever $x \gg 1$, that is at cosmological times. It is expected that even a mild collision term can average them out. Even if this is not the case, we only aim at describing the average of this incredibly fast oscillating spectrum, since this is the only part that will survive any measurement or physical process. This means that after a transitory regime, the \ATAOH scheme must become a good approximation. Sadly, given the ${\cal O}(N^3)$ complexity for computing the collision term, it becomes numerically impossible to integrate this transitory regime. Furthermore the $y$-grid becomes necessarily too sparse to account for these spectral oscillations. Hence one must rely on a certain approximation to handle the transition from a period where the \ATAOJH scheme applies to a regime where \ATAOH is sufficient.

We chose to push the \ATAOJH scheme as far as possible, typically down to $2\,{\rm MeV}$ and then to switch immediately to a \ATAOH scheme. In doing so we necessarily miss some features of the transitory regime. If $\vec{\Anti}$ is already well aligned with $\vecHeff$, it is nonetheless expected to be a very good approximation. Hence it misses some physics essentially due to the oscillations in the $\nu_\mu-\nu_\tau$ space which have developed right after the muon-driven transition, and for which we have seen that the collision term is not very effective in dampening these oscillations towards $\vecHeff$. We can however estimate the nature of the error made. If we focus on the two-neutrino case describing the $\nu_\mu-\nu_\tau$ space after the muon-driven transition, the density matrices are in the \ATAOH scheme in the form given by \eqref{eq:vrho_ATAOJV}. If we neglect terms of order ${\cal O}(\xi^2)$, then $\abs{g(-\xi_1,y)- g(-\xi_2,y)}  \simeq \abs{g(\xi_1,y)- g(\xi_2,y)} $, that is we have essentially $\vec{{\vrho}} \propto \widehat{\Hamil+\Hself}$ and $\vec{{\bvrho}}  \propto \widehat{\Hamil-\Hself} $ with the same prefactor. When the ratio $|\Hamil|/|\Hself|$ grows this tends to displace both $\vec{{\vrho}}$ and $\vec{{\bvrho}}$ in the same direction, namely, the projection of $\vec{\Hamil}$ in a plane orthogonal to $\vec{\Hself}$ (in agreement with the next-to-leading order term of the expansion \eqref{ExpandSH} of $\widehat{\Hself}$). The net result is that around the end of validity of the \ATAOJH regime, neutrinos and antineutrinos of one flavour are converted to neutrinos and antineutrinos of the other flavour, but in similar proportions for neutrinos and antineutrinos, hence preserving the asymmetry. Of course, since the ratio $|\Hamil| / |\Hself|$ reaches unity earlier for small $y$, it is expected that this concerns more the small momenta $y$. Also, the more  $\vec{\Hamil}$ and $\vec{\Hself}$ are misaligned in the end of the \ATAOJH regime, the more this phenomenon takes place. This is nicely seen in figure~9 of \cite{Johns:2016enc}, a configuration solved numerically without collisions, where we observe that both the number of neutrinos and antineutrinos of a given flavour increase while the opposite takes place for the other flavour. By construction, our approach based on an instantaneous switching from \ATAOJH to \ATAOH cannot capture this phenomenon. 
Finding a method to handle this transitory regime when including the collision term is an upcoming numerical challenge for the computation of equilibration in the early universe.

\subsection{Comparison with the literature}

Our numerical results differ from the literature in several aspects, which we review here.

\paragraph{Collision term} It is clear that the muon-driven oscillations are less damped in our case, compared to the results for instance reported in \cite{Dolgov_NuPhB2002}. We explain this by the fact that the general collision term satisfies the “factorization” property \eqref{MagicKFundamental}, and thus \eqref{MagicKUJ} in the $\nu_\mu-\nu_\tau$ subspace.
This implies that oscillations developing in this subspace are only very mildly damped, as detailed in section~\ref{SecCollisions}. When relying on an approximate collision term, this property is lost. For computations where all entries of the collision term are based on a damping approximation, it is the repopulation term which fails to satisfy the property \eqref{MagicKFundamental}. On the other hand, for computations which use the full collision term for on-diagonal components, but a damping approximation for off-diagonal components (as is the case in \cite{Dolgov_NuPhB2002}), the property is lost precisely because not all components are computed in the same method and this introduces preferred directions in the collision term. In all cases, using an approximation for the collision term results in much more damping of the oscillations in the $\nu_\mu-\nu_\tau$ space compared to our results. We already mentioned the case where $\xi_\mu=-\xi_\tau$  of \cite{Dolgov_NuPhB2002} (figure 9), compared to our results in figure~\ref{figEBO}. One of the consequences is that it is not possible to consider that by $10\,{\rm MeV}$ we would have generically achieved $\xi_\mu=\xi_\tau$, as is assumed for instance in \cite{Pastor:2008ti,Mangano:2010ei,Castorina2012}.

\begin{figure}[!h]
	\centering
	\includegraphics[]{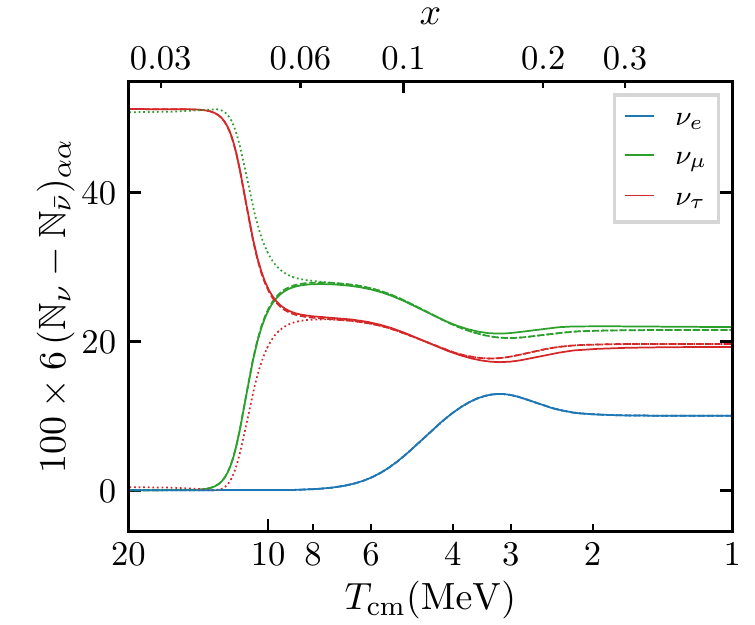}
	\includegraphics[]{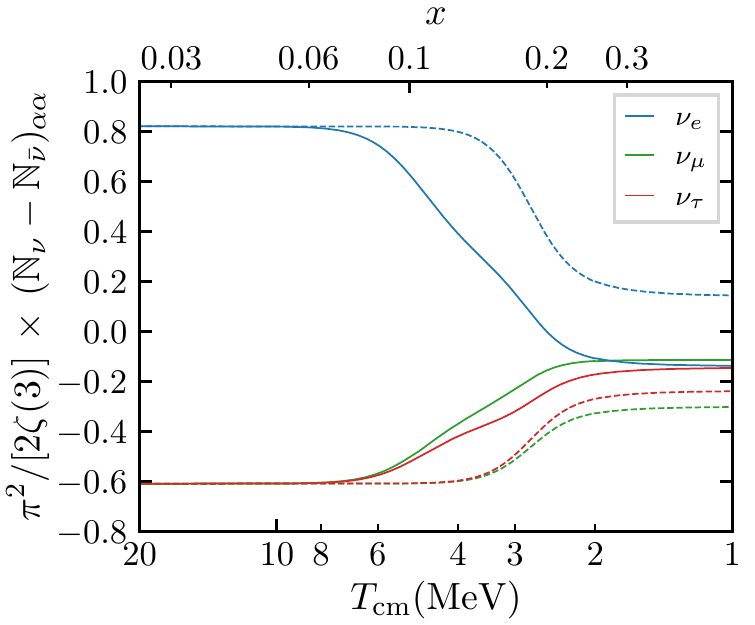}
	\caption{\label{figComparison} {\it Left plot:} $\xi_\alpha=(0,0,0.5)$, with $\delta=0$ in solid line and $\delta=\pi$ in dashed line. Dotted lines correspond to the case $\delta=0$ on which the transformation $\check{S}\vrho^{(\delta=0)} \check{S}^\dagger$ has been applied, and are hidden behind dashed lines at late times. It can be compared with figure~3  of \cite{Gava:2010kz}. {\it Right plot:} $\xi_\alpha = (1.0732,-0.833,-0.833)$, with $\theta_{13}=0.20$ in solid line and $\theta_{13}=0$ in dashed line. It corresponds to figure~1 of \cite{Mangano:2010ei}, noting that the initial conditions are $\eta_\mu=\eta_\nu = -0.61$ instead of the stated values $\eta_\mu=\eta_\nu = -0.41$.}
\end{figure}

\paragraph{Large mixing angle} Note that in references \cite{Dolgov_NuPhB2002,Gava:2010kz}, the large mixing angle value $\theta_{23}=\pi/4$ was used, which adds extra properties. Since $\cos(2 \theta_{23})=0$ we cannot use~\eqref{th23th23} to estimate $\theta_{23}^{\rm eff}$ after the muon-driven transition. Fortunately, in that case we get exactly 
\begin{equation}
\tan(2 \theta_{23}^{\rm eff}) =\frac{(\epsilon^{-1} + 1 )\cos^2 (\theta_{13}) +\sin^2 \theta_{12} \sin^2 \theta_{13} - \cos^2 \theta_{12}}{\sin(\theta_{13}) \sin(2 \theta_{12})} \,,
\end{equation}
hence we also find from $\epsilon = \Delta m_{21}^2  / \Delta m_{32}^2 \ll 1$ that $\theta_{23}^{\rm eff} \simeq \pi/4 $, and the vector representation of the vacuum Hamiltonian restricted to the $\nu_\mu-\nu_\tau$ space is approximately along the $\x$-axis. Therefore the precession direction corresponds to a state where the asymmetry is the same for $\nu_\mu$ and $\nu_\tau$. In that case, and given the extra damping incurred by the approximations in the collision term, it is expected that the equilibration of the degeneracy parameters $\xi_\mu$ and $\xi_\tau$ is very efficient right after the muon-driven transition. This is seen for instance in figures (7-10) of \cite{Dolgov_NuPhB2002}, or figure 3 of \cite{Gava:2010kz} whereas in its counterpart here (the left plot of \ref{figComparison}), $\xi_\mu \neq \xi_\tau$ after the muon-driven transition. Also this special choice of mixing angle explains why $\xi_\mu$ and $\xi_\tau$ remain equal on figure 1 of \cite{Mangano:2010ei} or \cite{Mangano:2011ip}, whereas in the right plot of figure~\ref{figComparison} they differ once the electron-driven MSW transitions are crossed, which results in a full equilibration being never achieved. Note that we find nonetheless the same influence of $\theta_{13}$, as discussed in section~\ref{sec:mixing_angles}.

\paragraph{CP phase} Furthermore our results about the effect of the Dirac phase differ with respect to \cite{Gava:2010kz,Gava_corr}. Although we confirm that the effect of the Dirac phase must be maximal when $\delta=\pi$ (strictly speaking there is no CP-violation in that specific case, as it is equivalent to $\delta=0$ and a change of sign in $\theta_{13}$), we find that it must necessarily be negligibly small given the structure of the equations (see section \ref{SecDiracPhase}) whereas it is found small but not negligible in \cite{Gava:2010kz}. To be specific, in figure 3 of \cite{Gava:2010kz} there is a small effect of the Dirac phase, whereas in the left plot of figure \ref{figComparison} no perceptible effect is found. Again these differences must find their origin in the differences for the treatment of the collision term, given that the property~\eqref{MagiccheckK} is not satisfied by an approximate repopulation term.

\section{Conclusion}

The complexity of the physics of neutrino evolution in the early Universe considerably increases when including initial degeneracies, a problem studied analytically and numerically in the last two decades. The \ATAOH scheme, developed in \cite{Froustey2020} and which relied on the adiabaticity of the evolution of the Hamiltonian governing the dynamics of $\vrho, \bvrho$ and the very fast scale of oscillations, was extended to the \ATAOJH scheme to account for non-vanishing chemical potentials.

A restriction to two-flavour systems showed the excellent accuracy of this method compared to a much longer QKE resolution, which we found to be at least ten times slower and even more when dealing with low temperatures. Even though our code can perform this “exact” QKE resolution, it is thus sufficient, notably if one wants to explore a wide range of parameters, to rely on the \ATAOJH scheme. Thanks to the \ATAOJH approximation, we recover the famous synchronous oscillations, but also predict and understand new results such as the existence of a phase of quasi-synchronous oscillations, that is an increased frequency regime ($\Omega(x) \propto x^2 \to x^6$) when the vacuum + mean-field Hamiltonian $\Hamil$ contribution becomes substantial compared to $\Hself$. The (non-)adiabaticity of the evolution of $\Anti$ during the lepton-driven transitions --- which depends on the degeneracies via the slowness factor~\eqref{eq:slowness} --- also allows to understand its qualitative behaviour, namely the (non-)efficiency of its alignment towards the vacuum Hamiltonian. In addition to their frequency, we can thus also estimate the beginning and the amplitude of the synchronous oscillations which develop afterwards.

We have shown that it is crucial to rely on the exact form of the collision term to fully take into account the physics of these oscillations --- approximate expressions previously overdamped degeneracy differences and led to a too rapid flavour equilibration. The ${\cal O}(N^3)$ complexity of the full collision term is the price to pay. Therefore, we argue that it is crucial to rely on a direct computation of the Jacobian to avoid worsening the problem. The method developed in \cite{Froustey2020} is extended to the situation with initial chemical potentials. Although it requires many more steps to implement it, as summarized in appendix~\ref{AppJacobJ}, it keeps the appealing ${\cal O}(N^3)$ complexity.

The \ATAOJH scheme fails when the vacuum potential is of the order of the self-interaction potential. In principle, the transition from the \ATAOJH to the \ATAOH regime should be solved numerically with the full QKE method, but in practice this is numerically daunting. Nevertheless, by switching directly at sufficiently low temperature from \ATAOJH to \ATAOH, we argued that errors incurred on the neutrino asymmetry should be minimized. 

In the standard case with three neutrinos, we have highlighted the influence of the various mixing parameters. Notably, the Dirac phase is found to have no perceptible effect as it essentially changes only the phase of synchronous oscillations, and its residual effect is accurately captured by the transformation~\eqref{FinalOperation}. In general, degeneracies tend to equilibrate partially, and this is due to the fact that the mixing angles $\theta_{12}$ and $\theta_{23}$ are not so different from maximal mixing ($\pi/4$), but this statement strongly depends on the non-vanishing of $\theta_{13}$. Given the complexity of the physics involved during the evolution of density matrices, the degree of equilibration depends non-trivially on the values of the initial degeneracies (e.g. equilibration is far from being achieved in the left plot of figure~\ref{figComparison}), and requires further systematic study. 

Moreover, our analysis is restricted to an homogeneous and isotropic cosmology, which brought a number of simplifications to the QKEs. A first step in the direction of relaxing the assumption of isotropy, while keeping the homogeneity, has recently been taken~\cite{Hansen:2020vgm} using a two-angle bin approximation. The generalisation to the three-flavour case with the full neutrino spectra and metric anisotropies, is also an major upcoming challenge.  

\acknowledgments

We thank the referee for his/her very thorough report on the first version of this paper, and Cristina Volpe for early discussions on this project and the physics of neutrino evolution in the early Universe.

\appendix

\section{Properties of Fermi-Dirac spectra}\label{AppFD}

We gather here some useful relations concerning Fermi-Dirac spectra. We have introduced in section~\ref{subsec:Anti} the distribution:
\begin{equation}
    g(\xi,y) = \frac{1}{e^{y-{\xi}}+1} \, .
\end{equation}
It corresponds to the spectrum of fermions of temperature $\Tcm$, which is an approximation since neutrinos are partially reheated by $e^\pm$ annihilations. However, throughout the analytical discussions of section~\ref{FreqSyncOsc} we do not make this distinction so as to simplify the reading (furthermore noting that for instance at $\Tcm = 5 \, \mathrm{MeV}$, $z_{\nu}-1 \simeq 10^{-4}$). 

\begin{subeqnarray}
\label{IntegralsFD}
\int [g(\xi,y) - g(-\xi,y)] \mathcal{D} y &=& \frac{\xi}{6} + \frac{\xi^3}{6\pi^2}\slabel{Intg1}\\
\int y \, [g(\xi,y) + g(-\xi,y)]  \mathcal{D} y &=& \frac{7 \pi^2}{120}+ \frac{\xi^2}{4} + \frac{\xi^4}{8\pi^2}\slabel{Intg2}\\
\int y^{-1} \, [g(\xi,y) + g(-\xi,y)]  \mathcal{D} y &=& \frac{1}{12} + \frac{\xi^2}{4\pi^2}\slabel{Intg3}\\
\int y^{-2} \, [g(\xi,y) - g(-\xi,y)] \mathcal{D} y &=& \frac{\xi}{2 \pi^2} \slabel{Intg4}
\end{subeqnarray}

\section{Evolution of the plasma temperature}\label{SectionPlasma}

For large temperatures, that is before we start the numerical resolution ($x<x_{\rm init}$), the evolution of the comoving plasma temperature is estimated assuming that neutrino spectra are thermal with the same temperature ($z_{\nu}=z$). Afterwards, the evolution of $z$ is computed using the full (anti)neutrino density matrices and the exact form for collisions between neutrinos and electrons/positrons. Including QED corrections~\cite{Heckler_PhRvD1994,Mangano2002,Bennett2020}, we use
\begin{equation}\label{D1Froustey2020}
\frac{\dd z}{\dd x} =  \frac{\displaystyle \frac{x}{z}J(x/z) - {\cal S}_\nu + G_1(x/z)}{ \displaystyle \frac{x^2}{z^2}J(x/z) + Y(x/z) + \frac{1}{4}\sum_\alpha Y_{\nu}(\xi_\alpha/z) +  \frac{2 \pi^2}{15} + G_2(x/z)} \, ,
\end{equation}
where we defined
\begin{equation}
\begin{aligned}
& S_\nu = 0\,, \quad Y_{\nu}(\zeta_\alpha) \equiv \frac{1}{\pi^2} \int_{0}^{\infty}{\dd \omega \, \omega^3 (\omega-\zeta_\alpha) \frac{\exp{(\omega-\zeta_\alpha)}}{[\exp{(\omega-\zeta_\alpha)}+1]^2}}\quad &\text{for}\quad x \leq x_{\rm init}\,, \\
&Y_\nu = 0 \,,\quad S_\nu \equiv \frac{1}{2 \pi^2 z^3} \int_{0}^{\infty}{\dd y \, y^3 \, \Tr \left[\mathcal{K}\right]}\ \quad &\text{for}\quad x > x_{\rm init}\,,
\end{aligned}
\end{equation}
with the notations of appendix D in \cite{Froustey2020} for the functions $J,\, Y, \, G_1, \, G_2$ and the collision term ${\cal K}$. The sum on $\alpha$ in the denominator of \eqref{D1Froustey2020} runs on $2 N_\nu$ elements, being all neutrinos and antineutrinos species. 

The starting condition $z_{\rm init}$ at $x_{\rm init}$ is found by solving the differential equation~\eqref{D1Froustey2020}, with the initial condition $z=1$ at $x=0$.\footnote{In principle $z$ increases at each species annihilation, and in particular we should consider $\mu^\pm$ annihilations since these leptons appear in the mean-field effects. This choice of initial conditions is however consistent with neglecting the interactions with $\mu^\pm$ in the collision term, which therefore do not reheat the plasma of neutrinos, photons and $e^\pm$.} When there are no neutrino degeneracies, it matches the condition found by all coupled species entropy conservation. However, it gives a slightly different $z_{\rm init}$ in the presence of initial degeneracies since entropy conservation is then violated. 

\section{Parameterizations of ${\rm SU}(2)$ and ${\rm SO}(3)$}\label{AppSU2SO3}

Any matrix of ${\rm SU}(2)$ can be expressed in terms of Euler angles as
\begin{equation}
\mathcal{R}_2(\alpha,\beta,\gamma) = \mathcal{R}_\z(\alpha)\cdot
\mathcal{R}_\y(\beta) \cdot  \mathcal{R}_\z(\gamma) = \begin{pmatrix}
{\rm e}^{-\ii(\alpha+\gamma)/2} \cos(\beta/2) & -{\rm e}^{-\ii(\alpha-\gamma)/2}\sin(\beta/2) \\
{\rm e}^{\ii(\alpha-\gamma)/2} \sin(\beta/2)& {\rm e}^{\ii(\alpha+\gamma)/2}\cos(\beta/2)
\end{pmatrix}
\end{equation}
with $\mathcal{R}_i(\theta) \equiv \exp(- \ii \theta \sigma_i/2)$. Similarly a ${\rm SO}(3)$ matrix is also expressed with Euler angles as
\begin{equation}
R_3(\alpha,\beta,\gamma) = R_\z(\alpha)\cdot R_\y(\beta) \cdot  R_\z(\gamma) 
\end{equation}
where $R_j(\theta) \equiv \exp(-\ii \theta \mathcal{J}^j)$ and
$(\mathcal{J}^i)_{jk} = -\ii \epsilon_{ijk}$. Both sets of matrices are related since they share the same Lie algebra, thanks to 
\begin{equation}\label{LaMagieDesMaths}
\mathcal{R}_2\cdot  \sigma_i \cdot \mathcal{R}_2^\dagger = \sigma_j\left(R_3\right)^j_{\,\,\,i}
\,,
\end{equation}
where it is implied that the Euler angles defining $\mathcal{R}_2$ and $R_3$ are the same. Therefore, the conjugation of the traceless part of a two-neutrino density matrix by an element ${\cal R}_2$ of ${\rm SU}(2)$ is equivalent to the associated rotation $R_3$ applied on its vector representation defined in equation~\eqref{MatrixToVector}.

\section{Adiabaticity parameter}\label{AppAdiabatic}

\begin{figure}[!ht]
	\centering
	\includegraphics[width=0.8 \textwidth]{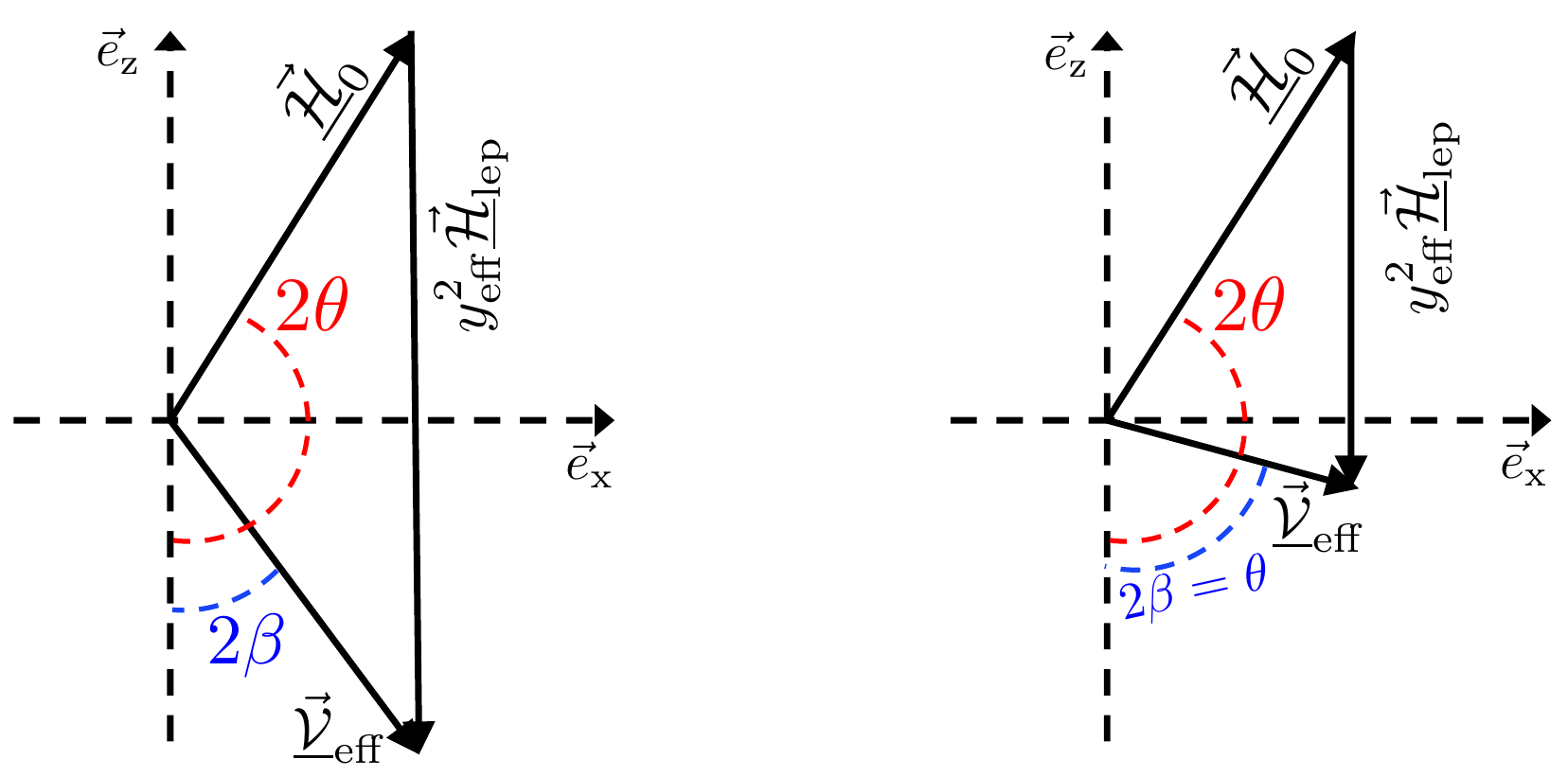}
	\caption{\label{FigAngles} Definition of the tipping angle $\beta$ on the left. The condition $\beta = \theta/2$ shown on the right corresponds to our definition of the MSW transition.}
\end{figure}

In the case with only two neutrinos, let us consider the evolution of the asymmetry vector $\vec{\Anti}$ without collisions and at leading order, which is dictated by equation~\eqref{eq:derivA_leading}. If the transition from a mean-field dominated to a vacuum dominated Hamiltonian, that is the MSW transition, is slow enough, then $\vec{\Anti}$ evolves adiabatically and follows $\vec{\underline{\mathcal{V}}}_{\rm eff} $. In order to assess the degree of (non-)adiabaticity, we thus need to quantify the speed at which the transition takes place. Let us first define a tipping angle $\beta$, illustrated in figure~\ref{FigAngles}, by
\begin{equation}
\cos(2\beta) = -\hat{\underline{\mathcal{V}}}_{\rm eff} \cdot \vec{e}_\z \quad \text{since }\quad \hat{\underline{\mathcal{V}}}_{\rm eff}(x \ll x_{\rm MSW})= -\vec{e}_\z\,.
\end{equation}

Initially the tipping angle vanishes, and long after the transition it reaches $\theta$. We define the location of the MSW transition $x_{\rm MSW}$ as the moment when the tipping angle takes half of its final value, $\beta_{\rm MSW} = \theta/2$, which corresponds to $|\vecHvacy| = y^2_{\rm eff}|\vecHlepy|$ (as can be checked by trigonometric manipulations). Let us then define an adiabaticity parameter as 
\begin{equation}\label{Defgammatr1}
\gamma \equiv \frac{|\vec{\Omega}|}{\partial_x (2 \beta)}\, ,
\end{equation}
with $\vec{\Omega} = F(\xi_1,\xi_2)\vec{\underline{\mathcal{V}}}_{\rm eff}$, whose value is estimated from\footnote{The next-to-leading order contribution to $\vec{\Omega}$ is not relevant here as we focus on cases where $x_{\rm MSW} \ll x_{\rm NLO}$.} \eqref{eq:derivA_leading}. A large $\gamma$ corresponds to a rate of change of the Hamiltonian direction ($2 \partial_x \beta$) much smaller than the instantaneous precession frequency ($|\vec{\Omega}|$), that is to a very adiabatic evolution. We find
\begin{equation}
\gamma^{-1} = -\frac{(\underline{\mathcal{H}}_0^\perp)^2 y_{\rm eff}^2}{\abs{F(\xi_1,\xi_2)}|\vec{\underline{\mathcal{V}}}_{\rm eff}|^3}  \partial_x\left(\frac{|\vecHlepy|}{\underline{\mathcal{H}}_0^\perp}\right)\,,\quad \text{where}\quad \underline{\mathcal{H}}_0^\perp \equiv |\vecHvacy| \sin(2 \theta)\,.
\end{equation}
We have used $\partial_x (2\beta) = -\sin^2(2 \beta)y^2_{\rm eff}\partial_x\left(|\vecHlepy|/\underline{\mathcal{H}}_0^\perp \right)$ and $\sin(2 \beta) = \underline{\mathcal{H}}_0^\perp/|\vec{\underline{\mathcal{V}}}_{\rm eff}|$.
In order to assess the adiabaticity of the transition, we must estimate how large $\gamma$ is at the transition. Since $|\vecHlepy|/ \underline{\mathcal{H}}_0^\perp \propto 1/x^6$, and using that at the transition we have $|\vec{\underline{\mathcal{V}}}_{\rm eff}| = 2 |\vecHvacy| \cos \theta$ (see the geometry on the right plot of figure~\ref{FigAngles}), the value of the adiabaticity parameter at the transition reduces to
\begin{equation}\label{Defgammatr2}
\gamma_{\rm MSW}  = \left.\abs{F(\xi_1,\xi_2)}\frac{2}{3} \frac{x |\vecHvacy| \cos^2 \theta}{\sin\theta}\right|_{x=x_{\rm MSW}}\,.
\end{equation}
Let us define the degree of non-adiabaticity through
\begin{equation}
P_{\rm n.a} \equiv \frac{1}{2}\left(1 \mp \widehat{\Anti} \cdot \widehat{\underline{\mathcal{V}}}_{\rm eff}\right)\,,
\end{equation}
with a $-$ sign (resp.~$+$ sign) if initially $\vec{\Anti}$ is aligned (resp.~anti-aligned) with $\vecHeff$ (i.e., $- \vec{e}_\z$). Its asymptotic value $P^\infty_{\rm n.a}$ when $x \gg x_{\rm MSW}$ estimates the misalignment of the final asymmetry vector due to lack of adiabaticity. Indeed if the transition is perfectly adiabatic, $\vec{\Anti}$ keeps tracking $\vec{\underline{\mathcal{V}}}_{\rm eff}$ and we always have $P_{\rm n.a}=0$. In general, the degree of non-adiabaticity needs not be much larger than unity to lead to a small $P^\infty_{\rm n.a}$, that is to a very non-adiabatic transition---see for instance the Landau-Zener approximation~\eqref{LZ}.

We note from equation~\eqref{Defgammatr2} that the adiabaticity parameter is of order $x \abs{F(\xi_1,\xi_2)}|\vecHvacy|$ evaluated at the transition, modulated by a geometric factor $(2/3) \cos^2 \theta/\sin \theta$. A transition is resonant when at some point the tipping angle goes through $2\beta = \pi/2$, that is through $\vec{\underline{\mathcal{V}}}_{\rm eff}$ having no component along $\vec{e}_\z$. Hence, a very non-resonant transition corresponds to $\theta \ll 1$, and in that case the geometric factor is enhanced by $1/\sin(\theta)$. It is less likely to encounter a small adiabaticity parameter because the tipping angle is small, and $|\vec{\Omega}|$ at the transition is larger than its final value (since $|\vec{\underline{\mathcal{V}}}_{\rm eff}|$ keeps decreasing). Conversely for a very resonant transition, $\pi/2-\theta \ll 1$, and the geometric factor is reduced by $\cos^2 \theta $ which is small, that is leads to a smaller adiabaticity parameter. This is partly because of the large tipping angle, but also because $|\vec{\Omega}|$ at the transition is much smaller than its final value (recall that at the transition $|\vec{\underline{\mathcal{V}}}_{\rm eff}| = 2 |\vecHvacy| \cos \theta$).

In the very resonant configuration ($\pi/2-\theta \ll 1$), the adiabaticity parameter for the dynamics of $\vec{\Anti}$ takes the simpler form
\begin{equation}\label{EqVeryResonant}
\gamma^{(\pi/2-\theta \ll 1)}_{\rm MSW} = \abs{F(\xi_1,\xi_2)}\underline{\mathcal{H}}_0^\perp \Delta x_{\rm MSW} \, , \ \ \text{with}\qquad \frac{1}{\Delta x_{\rm MSW}} \equiv y_{\rm eff}^2\left.\partial_x \left(\frac{|\vecHlepy|}{\underline{\mathcal{H}}_0^\perp}\right) \right|_{x=x_{\rm MSW}}\,.
\end{equation}

\paragraph{Landau-Zener formula} The Landau-Zener~\cite{Landau1932,Zener1932,Haxton:1986bc,Abazajian:2001nj,Johns:2016enc,Wittig} formula is an approximation of the degree of non-adiabaticity in this very resonant situation, using the approximation that the diagonal components of $\Hamil$ are linear in $x$ and that off-diagonal ones are constant, which reads
\begin{equation}\label{LZ}
P^\infty_{\rm n.a} \simeq \exp(-\pi \gamma_{\rm MSW}/2)\,.
\end{equation}

Note that the factors $F(\xi_1,\xi_2)$ and $y_{\rm eff}^2$ in \eqref{EqVeryResonant} are  specific to the fact that we consider the evolution of $\vec{\Anti}$. If we had considered the evolution of $\vec{\vrho}$ given by equation~\eqref{eq:QKE_2nu} without self-interactions nor collisions, that is $\partial_x \vec{\vrho} =  \vec{\Hamil}   \wedge \vec{\vrho} $, we would have obtained with a similar analysis (all quantities are written here for a given momentum $y$) the usual expression for the Landau-Zener adiabatic parameter
\begin{equation}\label{GammaLZ}
\gamma^{(\pi/2-\theta \ll 1)}_{\rm MSW} =\mathcal{H}_0^\perp \Delta x_{\rm MSW}\, , \quad \text{with} \qquad \frac{1}{\Delta x_{\rm MSW}} \equiv \left.\partial_x \left(\frac{|\vecHlep|}{\mathcal{H}_0^\perp}\right)\right|_{x=x_{\rm MSW}}
\end{equation}
where $\Hvac^\perp = \Hvac \sin(2 \theta)$. It is similar to equation~(9b) of \cite{Haxton:1986bc} and equation~(7.8) of \cite{Abazajian:2001nj}, 
the only difference being that $\Hvac^\perp$ is considered constant when dealing with solar neutrinos, whereas in the cosmological context it scales as $\propto x$, hence explaining why the expression \eqref{GammaLZ} for the transition width, $\Delta x_{\rm MSW}$, takes into account this evolution. However, note that this necessary modification for the expression of the adiabaticity parameter is absent from equation (28) of \cite{Johns:2016enc}, although this impacts only marginally the estimation of adiabaticity.

\section{Numerical implementation of ATAO schemes}\label{AppJacobJ}

In the next section we review in more details how the degrees of freedom in density matrices are serialized.  We then detail how the numerical method of \cite{Froustey2020} based on a direct computation of the Jacobian is extended to the \ATAOJH scheme. 

\subsection{Serialization in flavour space}

Since the spectra of density matrices are sampled on a grid of $N$ comoving momenta, the $\{y_n\}$, the variables which need to be solved for are the $\vrho_{\alpha\beta}(y_n)$, and $\bvrho_{\alpha\beta}(y_n)$. In the discretized numerical resolution, integrals are replaced by a quadrature method, that is by a weighted sum on the $y_n$, the weights depending on the chosen grid points. 

In order to alleviate the explanations, we will ignore the presence of antineutrinos and we shall consider that the variables are just the $\vrho_{\alpha\beta}(y_n)$, for which we use the short notation  $\vrho_{\alpha\beta,n}$.  In the \ATAOJH method, this is clearly wrong since an asymmetry matrix $\Anti_{\alpha\beta}$ requires that neutrinos and antineutrinos should have different distributions and one must always evolve both neutrinos and antineutrinos --- but that does not change the arguments presented here.

For each $n$, we serialize the matrix components.  This requires to define a basis ${P^a}$ for hermitian matrices with $N_\nu^2$ elements. These are divided into $N_\nu(N_\nu+1)/2$ basis matrices for the real components and $N_\nu(N_\nu-1)/2$ for the imaginary components. For instance when $N_\nu=2$ the matrices are 
\begin{equation}
P^1 = \begin{pmatrix} 
1&0\\
0&0\end{pmatrix} \, , \quad
P^2 = \begin{pmatrix} 
0&0\\
0&1\end{pmatrix} \, , \quad
P^3 = \begin{pmatrix} 
0&1\\
1&0\end{pmatrix} \, , \quad
P^4 = \begin{pmatrix} 
0& - \ii\\
\ii&0\end{pmatrix}\,.
\end{equation}
An inner product between two hermitian matrices is $(A,B) \equiv {\rm Tr}(A \cdot B^\dagger)$, hence the norms of the basis matrices are
\begin{equation}
\norm{P^a}^2 = \sum_{\alpha\beta} |{P^a}_{\alpha\beta}| ^2= {\rm Tr}(P^a \cdot P^{a\dagger})\,.
\end{equation}
Any density matrix is decomposed on serialized components as
\begin{equation}
\vrho_{\alpha \beta,n} \equiv \vrho_{a,n} {P^a}_{\alpha\beta}\,.
\end{equation}
The serialized components are also related to the components in the matter basis through
\begin{equation}\label{DefPaij}
\widetilde \vrho_{ij,n} = \vrho_{a,n} {P^{a}}_{ij}(n)\quad{\rm with}\quad {P^{a}}_{ij}(n) \equiv [U^\dagger(n) \cdot P^a \cdot U(n)]_{ij} \, ,
\end{equation}
where $U(n)$ stands for $U_{\Hcal}(y_n)$, $\Hcal$ being the appropriate Hamiltonian (it will depend on the numerical scheme chosen). Note that we use the same notation $P^a$ in the matter ($P^a_{ij}$) and flavour ($P^a_{\alpha \beta}$) bases, the difference being identified through the indices. Conversely the serialized components $\vrho_{a,n}$ are obtained from
\begin{equation}\label{Convertrhoarhoi}
\vrho_{a,n} =  \vrho_{\alpha\beta,n} {P^{\alpha\beta}}_{a}(n)=\frac{\widetilde{\vrho}_{ij,n} {P^{a}}_{ji}(n)}{\norm{P^a}^2 }\quad{\rm with}\quad {P^{\alpha\beta}}_{a}(n) \equiv \frac{ {P^{a\star}}_{\alpha\beta}(n)}{\norm{P^a}^2 } =\frac{ {P^{a}}_{\beta \alpha}(n) }{\norm{P^a}^2 }\,.
\end{equation}
In any ATAO scheme, we are only interested in the diagonal components of the matter basis, since by construction all off-diagonal components vanish, hence we define $\widetilde \vrho_{i,n} \equiv \widetilde \vrho_{ii,n}$ and obtain the following relations between the serialized (flavour) basis and the (diagonal) matter components
\begin{equation}
\begin{aligned}
\widetilde \vrho_{i,n} &= \vrho_{a,n} T^{a}_i(n)\quad &&\text{with}\quad T^{a}_i(n) \equiv {P^{a}}_{ii}(n) \, ,  \\
\vrho_{a,n}&= \widetilde \vrho_{i,n} T^i_{a}(n) \quad &&\text{with}\quad T^i_{a}(n) \equiv \frac{1}{\norm{P^a}^2} T^{a}_i(n)  \,.
\end{aligned}
\end{equation}
Since the $U(n)$ depend both on $x$ and on $y_n$, the $T^a_i(n)$ and $T^i_a(n)$ also depend on these variables, that is for each time step they must be computed for all points of the momentum grid.

\subsection{Direct computation of the Jacobian}

In this section we use a prime to denote a derivative with respect to $x$. Also we stress again that we do not mention the antineutrinos for the sake of clarity but all developments must be carried out taking them into account.

\paragraph{QKE scheme}

We must solve for the evolution of $z$ and of the flavour space serialized variables $\vrho_{a,n}$. Equation~\eqref{D1Froustey2020} dictates the evolution of $z$. The evolution of the $\vrho_{a,n}$ is governed by
\eqref{eq:QKE_compact}
\begin{equation}\label{eq:QKE_compactserialized}
\vrho_{a,n}' = M_{a,n}^c \vrho_{c,n}  + \mathcal{K}_{a,n} \, ,
\end{equation}
with 
\begin{equation}
M_{a,n}^c \equiv (\Hamil_{b,n} + \Hself_b) {C^{bc}}_a\quad\text{and}\quad {C^{bc}}_a \equiv -\ii[P^b,P^c]_{\alpha \beta} {P^{\alpha\beta}}_a\,.
\end{equation}
The first term in \eqref{eq:QKE_compactserialized} comes  from mean-field effects and the second term from collisions.
The associated Jacobian has the general structure
\begin{equation}\label{JacobNumQKE}
\begin{pmatrix}
0&0\\ \\
\displaystyle \left(\frac{\partial M_{a,n}^c }{\partial z} \vrho_{c,n} \right)& 
\displaystyle \left(\frac{\partial M_{a,n}^c }{\partial \vrho_{b,m}} \vrho_{c,n} +  M_{a,n}^b \delta^n_m \right)
 \end{pmatrix}+
 \begin{pmatrix}
\displaystyle \frac{\partial z'}{\partial z}& \displaystyle \frac{\partial z'}{\partial \vrho_{b,m}}\\ \\
\displaystyle \frac{\partial {\mathcal{K}}_{a,n}}{\partial z}& \displaystyle \frac{\partial {\mathcal{K}}_{a,n}}{\partial  \vrho_{b,m}}
 \end{pmatrix}
\end{equation}
where, as before, the first matrix is due to mean-field effects, and the second to collisions. The contributions from the mean-field effects require to calculate
\begin{equation}
 \frac{\partial M_{a,n}^c }{\partial z} =\frac{\partial \Hamil_{b,n}}{\partial z}{C^{bc}}_a\,, \qquad
 \frac{\partial M_{a,n}^c }{\partial \vrho_{b,m}} = \frac{\partial \Hself_d}{\partial \vrho_{b,m}} {C^{cd}}_a\,,
\end{equation}
and the quantity $\partial \Hself_d/\partial \vrho_{b,m}$ is read on the integral definition \eqref{DefJ}. The computation of the Jacobian associated with mean-field effects is at most ${\cal O}(N^2)$ when self-interactions are taken into account, and only ${\cal O}(N)$ when they are ignored or when there is no asymmetry. Let us now review the complexity of the remaining terms. 
\begin{itemize}
\item $\partial \mathcal{K}_{a,n}/ \partial \vrho_{b,m}$ is the time-consuming part. Since the complexity for computing the collision term is ${\cal O}(N^3)$, using a finite difference method would scale as ${\cal O}(N^4)$. A method to reduce the complexity to ${\cal O}(N_\nu^2 N^3)$ is detailed in \cite{Froustey2020}, hence considerably speeding the numerical resolution.
\item $\partial \mathcal{K}_{a,n}/\partial z$ is just the collision term where the contribution coming from the distributions of electrons/positrons is varied with respect to $z$. Hence it has the same ${\cal O}(N^3)$ complexity as the collision term.
\item $\partial z'/\partial z$ can be obtained from equation~\eqref{D1Froustey2020}. In practice we simply use a finite difference method.
\item $\partial z'/\partial \vrho_{b,m}$ is obtained from the chain rule as
\begin{equation}\label{dzpdrho}
\frac{\partial z'}{\partial \vrho_{b,m}} = \frac{\partial z'}{\partial \mathcal{K}_{a,n}} \frac{\partial \mathcal{K}_{a,n}}{\partial \vrho_{b,m}}
\end{equation}
and we only need the variation $\partial z'/\partial \mathcal{K}_{a,n}$ which is easily read on equation~\eqref{D1Froustey2020}. Indeed, since only the trace of the collision term sources $z'$, the only serialized components $a$ leading to a non-vanishing $\partial z'/\partial \mathcal{K}_{a,n}$ are those for which ${\rm Tr}(P^a) \neq 0$.
\end{itemize}

\paragraph{\ATAOH scheme}

In the \ATAOH scheme we integrate $z$ with \eqref{D1Froustey2020}, and the diagonal components $\widetilde \vrho_i$ with \eqref{BasicATAOH}, that is
\begin{equation}\label{QKEcompact}
\widetilde{\vrho}'_{i,n}= \widetilde{\mathcal{K}}_{i,n}\,.
\end{equation}
Note that this is a very compact notation which hides the fact that what is known in general are the $\mathcal{K}_{\alpha \beta}(y_n)$ which depend on the $\vrho_{\alpha\beta}(y_n)$. Hence at each step one must transform the matter basis components $\widetilde \vrho_{i,n}$ to the flavour basis, compute the collision terms, and convert back into the matter basis, and keep only the diagonal terms. The relation between the diagonal matter basis components and the (flavour basis) serialized components reads
\begin{equation}
\label{eq:relation_K_Ktilde}
\widetilde{\mathcal{K}}_{i,n} = \mathcal{K}_{a,n} T^a_i(n).
\end{equation}

The general form of the Jacobian is then
\begin{equation}\label{JacobNumATAOH}
 \begin{pmatrix}
\displaystyle \frac{\partial z'}{\partial z}& \displaystyle \frac{\partial z'}{\partial \widetilde\vrho_{j,m}}\\ \\
\displaystyle \frac{\partial \widetilde{\mathcal{K}}_{i,n}}{\partial z}& \displaystyle \frac{\partial \widetilde{\mathcal{K}}_{i,n}}{\partial \widetilde \vrho_{j,m}}
 \end{pmatrix}
\end{equation}
since in this scheme there are no mean-field effects to solve for, as they are hidden in the evolution of the matter basis. Again a finite difference method to compute $\partial \widetilde{\mathcal{K}}_{i,n}/\partial \widetilde \vrho_{j,m}$ would be of complexity ${\cal O}(N^4)$, but using the method detailed in \cite{Froustey2020} it is reduced to a complexity ${\cal O}(N_\nu N^3)$. This is even slightly faster (reduced by a factor $N_\nu$) than for computing  $\partial \mathcal{K}_{a,n}/ \partial \vrho_{b,m}$ because there are only $N_\nu$ diagonal matter components instead of $N_\nu^2$ flavour components. Both Jacobian blocks are related thanks to
\begin{equation}\label{RelateJacob}
\frac{\partial \widetilde{\mathcal{K}}_{i,n}}{\partial \widetilde\vrho_{j,m}} = T^{a}_i(n) \frac{\partial {\mathcal{K}}_{a,n}}{\partial \vrho_{b,m}} T_{b}^j(m) \, . 
\end{equation}
Let us review the other blocks in \eqref{JacobNumATAOH}. First we use the chain rule
\begin{equation}\label{dzpdrhoi}
\frac{\partial z'}{\partial \widetilde \vrho_{j,m}} = \frac{\partial z'}{\partial \widetilde{\mathcal{K}}_{i,n}} \frac{\partial \widetilde{\mathcal{K}}_{i,n}}{\partial \widetilde\vrho_{j,m}}\,,
\end{equation}
where $\partial z' / \partial \widetilde{\mathcal{K}}_{i,n} $ is easily read from equation~\eqref{D1Froustey2020} since only the trace of the collision term sources $z'$. And finally $\partial \widetilde{\mathcal{K}}_{i,n} / \partial z $ is similar to the computation of a collision term, but with the contribution from the $e^\pm$ distribution varied upon $z$.

\paragraph{\ATAOJH scheme}

The general \ATAOJH equation \eqref{BasicATAOJH}, when written explicitly in matter basis components and using the previous notation, is also of the form \eqref{QKEcompact}. However in the \ATAOJH we also solve at the same time the evolution of $\Anti_{\alpha\beta}$ given by \eqref{dJdx}. Note that it depends on the full collision term $\mathcal{K}_{\alpha \beta}(y_n)$ (or the $\mathcal{K}_{a,n}$ in serialized basis) and not just on the diagonal components in the matter basis $\widetilde{\mathcal{K}}_{i,n}$, contrary to the evolution of the $\widetilde \vrho_i$ in \eqref{QKEcompact}.

Since we supplement $z$ and  the $\widetilde\vrho_{i,n}$ with the $N_\nu^2$ variables $\Anti_a$, this extends the size of the Jacobian. We now show how the new blocks in the Jacobian can be computed, and that this preserves the ${\cal O}(N^3)$ complexity.  The general form of the Jacobian is 
\begin{equation}\label{JacobNum}
 \begin{pmatrix}
\displaystyle \frac{\partial z'}{\partial z}& \displaystyle \frac{\partial z'}{\partial \widetilde\vrho_{j,m}}& \displaystyle \frac{\partial z'}{ \partial \Anti_b}\\ \\
\displaystyle \frac{\partial \widetilde{\mathcal{K}}_{i,n}}{\partial z}& \displaystyle \frac{\partial \widetilde{\mathcal{K}}_{i,n}}{\partial \widetilde \vrho_{j,m}}& \displaystyle\frac{\partial \widetilde{\mathcal{K}}_{i,n}}{\partial \Anti_b}\\ \\
 \displaystyle \frac{\partial \Anti'_a}{\partial z}& \displaystyle \frac{\partial \Anti'_a}{\partial \widetilde\vrho_{j,m}}& \displaystyle \frac{\partial \Anti'_a}{\partial \Anti_b}
 \end{pmatrix} \, ,
\end{equation}
and only the blocks in the right column or the bottom line are specific to the \ATAOJH scheme. 
As detailed hereafter, in order to compute these new blocks we shall need the computation of $\partial \mathcal{K}_{a,n}/\partial \vrho_{b,m}$, which is what is needed when computing the Jacobian in the QKE method. The block $\partial \widetilde{\mathcal{K}}_{i,n}/\partial \widetilde \vrho_{j,m}$ is then deduced through \eqref{RelateJacob}.

We also need to know how the $U(n)$ vary when the components of $\Anti$ are varied. Let us define the set of anti-hermitian matrices
\begin{equation}\label{DefWan}
W^{a,n} \equiv \frac{\partial U(n)}{\partial \Anti_a} \cdot U^\dagger(n) = - U(n) \cdot \frac{\partial U^\dagger(n)}{\partial \Anti_a}\,.
\end{equation}
which allow to know how the flavour components vary when $\Anti$ varies, for fixed matter basis components. They are obtained thanks to 
\begin{equation}\label{CrypticW}
\left(W^{a,n}\right)_{ij}= \frac{\sqrt{2} G_F}{(xH)}\left(\frac{m_e}{x}\right)^3   \frac{\left[U^\dagger(n) \cdot P^a \cdot U(n)\right]_{ij}}{(\Hamil_{j,n} + \Hself_{j,n} -\Hamil_{i,n} - \Hself_{i,n})}\quad \text{for}\quad i\neq j\,, 
\end{equation}
where the $\left(W^{a,n}\right)_{ij}$ are the components of $W^{a,n}$ in the matter basis, that is they are defined as $\left[U^\dagger(n)\cdot W^{a,n} \cdot U(n)\right]_{ij}$\,. The $(\Hamil+\Hself)_{j,n}$ are the diagonal components of $\Hamil+\Hself$ in the matter basis, which are by definition its eigenvalues. The $W^{a,n}$ are then found by transforming~\eqref{CrypticW} to the flavour basis with $U(n)$. Using their definition~\eqref{DefWan}, one then finds 
\begin{equation}\label{drhoadAb}
\frac{\partial \vrho_{a,n}}{\partial \Anti_b} =\vrho_{c,n} [W^{b,n}, P^c]_{\alpha \beta} {P^{\alpha \beta}}_a(n)\,.
\end{equation}

We now have all the tools to compute the blocks in the Jacobian \eqref{JacobNum} that are specific to the presence of $\Anti$. 
\begin{itemize}
\item $\partial \widetilde{\mathcal{K}}_{i,n}/\partial \Anti_b$ 

This is deduced from~\eqref{eq:relation_K_Ktilde}. Using the Leibniz rule, we deduce
\begin{equation}\label{dKidAb}
\frac{\partial \widetilde{\mathcal{K}}_{i,n}}{\partial \Anti_b} = \frac{\partial \mathcal{K}_{a,n}}{\partial \Anti_b}T^a_i(n) + \mathcal{K}_{a,n} \frac{\partial T^a_i(n)}{\partial \Anti_b}\quad \text{with} \quad \frac{\partial T^a_i(n)}{\partial \Anti_b} = - \left(U^\dagger(n) \cdot [ W^{b,n}, P^a ]\cdot U(n)\right)_{ii} \, .
\end{equation}

\item $\partial z'/\partial A_b$

We only need to apply the chain rule using equation \eqref{dKidAb} since
\begin{equation}
\frac{\partial z'}{\partial \Anti_b} = \frac{\partial z'}{\partial \widetilde{\mathcal{K}}_{i,n}} \frac{\partial \widetilde{\mathcal{K}}_{i,n}}{\partial \Anti_b}\,,
\end{equation}
with $\partial z'/\partial \widetilde{\mathcal{K}}_{i,n}$ already needed for equation \eqref{dzpdrhoi}.

\item $\partial \Anti'_a/\partial z$

This is similar to the treatment of equation~\eqref{dJdx} with the replacement $\Hamil \to \partial \Hamil/\partial z$ and $\mathcal{K} \to \partial \mathcal{K}/\partial z$.

\item $\partial \Anti'_a/\partial \widetilde \vrho_{j,n}$
 
Let us define the following derivative :
\begin{equation}\label{TotalStrangeDer}
\frac{\partial \Anti'_a}{\partial \vrho_{b,n}} \equiv \left(\left. \frac{\partial \Anti'_a}{\partial \vrho_{b,n}}\right|_{\rm mf} + \frac{\partial \Anti'_a}{\partial \mathcal{K}_{c,m}} \frac{\partial \mathcal{K}_{c,m}}{\partial \vrho_{b,n}}\right)\,,
\end{equation}
where it is understood that the first term corresponds to the mean-field term, that is the first term on the rhs of equation~\eqref{dJdx}, and the second term is the indirect contribution via the dependence of the collision term.  Since the integrals appearing in equation~\eqref{dJdx} are computed with a quadrature method, that is a sum on the grid of comoving momenta with appropriate weights, these derivatives select only one term in these sums (the one corresponding to the comoving momentum $y_n$). We then immediately get from equation~\eqref{Convertrhoarhoi} and the chain rule
 \begin{equation} 
\frac{\partial \Anti'_a}{\partial \widetilde\vrho_{j,n}} = \frac{\partial \Anti'_a}{\partial \vrho_{b,n}}T^j_b(n)\,.
\end{equation}

\item $\partial \Anti'_a/\partial \Anti_b$

Finally, using again the derivative \eqref{TotalStrangeDer}, we find a simple expression for the last block
\begin{equation}\label{dApdA}
\frac{\partial \Anti'_a}{\partial \Anti_b} = \frac{\partial \Anti'_a}{\partial \vrho_{c,n}} \frac{\partial \vrho_{c,n}}{\partial \Anti_b} \,.
\end{equation}
 
\end{itemize}
In practice, pairs of indices like ${i,n}$ or ${a,n}$ are also serialized (e.g. with $I = n N_\nu + i$ and $A = n N_\nu^2 +a$), such that all products with implicit summations in this section appear as matrix multiplications when implemented in the code.

To summarize we need to compute the Jacobian as in the QKE method, which gives $\partial \mathcal{K}_{a,n}/\partial \vrho_{b,m}$. It corresponds to the variation of all flavour components in the collision term with respect to variations in all flavour components of the density matrices. We also need to compute the $T^a_i(n)$ from equation~\eqref{DefPaij}, and the $W^{a,n}$ from equation~\eqref{CrypticW}. We then deduce from equation~\eqref{drhoadAb} the $\partial \vrho_{a,n}/\partial \Anti_b$. Finally, knowing $\partial z'/\partial \widetilde{\mathcal{K}}_i(n)$, $\left.\partial \Anti'_a/\partial \vrho_{b,n}\right|_{\rm mf}$ and $\partial \Anti'_a/\partial \mathcal{K}_{c,m}$ from the equations governing the evolution of $z$ and $\Anti_{\alpha\beta}$, we can compute the five new blocks of the Jacobian as described from equations~\eqref{dKidAb} to~\eqref{dApdA}. The step which is the most time-consuming is the first one, that is the computation of $\partial \mathcal{K}_{a,n}/\partial \vrho_{b,m}$, whose complexity is ${\cal O}(N^3)$. Since this is already the longest step in the direct computation of the Jacobian in the QKE scheme, we deduce that for large $N$ the direct computation of the Jacobian in the \ATAOJH method takes roughly the same time as the direct computation of the Jacobian in the QKE method.

\bibliographystyle{JHEP}
\bibliography{BiblioATAOJ}

\providecommand{\href}[2]{#2}\begingroup\raggedright\begin{thebibliography}{10}

\bibitem{Dodelson_Turner_PhRvD1992}
S.~{Dodelson} and M.~S. {Turner}, \emph{{Nonequilibrium neutrino statistical
  mechanics in the expanding Universe}},
  \href{https://doi.org/10.1103/PhysRevD.46.3372}{\emph{Phys. Rev. D}
  {\bfseries 46} (1992) 3372}.

\bibitem{Dolgov1992}
A.~Dolgov and M.~Fukugita, \emph{{Nonequilibrium effect of the neutrino
  distribution on primordial helium synthesis}},
  \href{https://doi.org/10.1103/PhysRevD.46.5378}{\emph{Phys. Rev. D}
  {\bfseries 46} (1992) 5378}.

\bibitem{Dolgov_NuPhB1997}
A.~D. Dolgov, S.~H. Hansen and D.~V. Semikoz, \emph{{Nonequilibrium corrections
  to the spectra of massless neutrinos in the early universe}},
  \href{https://doi.org/10.1016/S0550-3213(97)00479-3}{\emph{Nucl. Phys. B}
  {\bfseries 503} (1997) 426}
  [\href{https://arxiv.org/abs/hep-ph/9703315}{{\ttfamily hep-ph/9703315}}].

\bibitem{Esposito_NuPhB2000}
S.~{Esposito}, G.~{Miele}, S.~{Pastor}, M.~{Peloso} and O.~{Pisanti},
  \emph{{Non equilibrium spectra of degenerate relic neutrinos}},
  \href{https://doi.org/10.1016/S0550-3213(00)00554-X}{\emph{Nucl. Phys. B}
  {\bfseries 590} (2000) 539}
  [\href{https://arxiv.org/abs/astro-ph/0005573}{{\ttfamily
  astro-ph/0005573}}].

\bibitem{Mangano2002}
G.~{Mangano}, G.~{Miele}, S.~{Pastor} and M.~{Peloso}, \emph{{A precision
  calculation of the effective number of cosmological neutrinos}},
  \href{https://doi.org/10.1016/S0370-2693(02)01622-2}{\emph{Phys. Lett. B}
  {\bfseries 534} (2002) 8}
  [\href{https://arxiv.org/abs/astro-ph/0111408}{{\ttfamily
  astro-ph/0111408}}].

\bibitem{Mangano2005}
G.~{Mangano}, G.~{Miele}, S.~{Pastor}, T.~{Pinto}, O.~{Pisanti} and P.~D.
  {Serpico}, \emph{{Relic neutrino decoupling including flavour oscillations}},
  \href{https://doi.org/10.1016/j.nuclphysb.2005.09.041}{\emph{Nucl. Phys. B}
  {\bfseries 729} (2005) 221}
  [\href{https://arxiv.org/abs/hep-ph/0506164}{{\ttfamily hep-ph/0506164}}].

\bibitem{Grohs2015}
E.~Grohs, G.~M. Fuller, C.~T. Kishimoto, M.~W. Paris and A.~Vlasenko,
  \emph{Neutrino energy transport in weak decoupling and big bang
  nucleosynthesis},
  \href{https://doi.org/10.1103/PhysRevD.93.083522}{\emph{Phys. Rev. D}
  {\bfseries 93} (2016) 083522}
  [\href{https://arxiv.org/abs/1512.02205}{{\ttfamily 1512.02205}}].

\bibitem{Relic2016_revisited}
P.~F. de~Salas and S.~Pastor, \emph{{Relic neutrino decoupling with flavour
  oscillations revisited}},
  \href{https://doi.org/10.1088/1475-7516/2016/07/051}{\emph{JCAP} {\bfseries
  07} (2016) 051} [\href{https://arxiv.org/abs/1606.06986}{{\ttfamily
  1606.06986}}].

\bibitem{Gariazzo_2019}
S.~Gariazzo, P.~de~Salas and S.~Pastor, \emph{{Thermalisation of sterile
  neutrinos in the early Universe in the 3+1 scheme with full mixing matrix}},
  \href{https://doi.org/10.1088/1475-7516/2019/07/014}{\emph{JCAP} {\bfseries
  07} (2019) 014} [\href{https://arxiv.org/abs/1905.11290}{{\ttfamily
  1905.11290}}].

\bibitem{Akita2020}
K.~Akita and M.~Yamaguchi, \emph{{A precision calculation of relic neutrino
  decoupling}},
  \href{https://doi.org/10.1088/1475-7516/2020/08/012}{\emph{JCAP} {\bfseries
  08} (2020) 012} [\href{https://arxiv.org/abs/2005.07047}{{\ttfamily
  2005.07047}}].

\bibitem{Froustey2020}
J.~Froustey, C.~Pitrou and M.~C. Volpe, \emph{{Neutrino decoupling including
  flavour oscillations and primordial nucleosynthesis}},
  \href{https://doi.org/10.1088/1475-7516/2020/12/015}{\emph{JCAP} {\bfseries
  12} (2020) 015} [\href{https://arxiv.org/abs/2008.01074}{{\ttfamily
  2008.01074}}].

\bibitem{Bennett:2020zkv}
J.~J. Bennett, G.~Buldgen, P.~F. De~Salas, M.~Drewes, S.~Gariazzo, S.~Pastor
  et~al., \emph{{Towards a precision calculation of $N_{\rm eff}$ in the
  Standard Model II: Neutrino decoupling in the presence of flavour
  oscillations and finite-temperature QED}},
  \href{https://doi.org/10.1088/1475-7516/2021/04/073}{\emph{JCAP} {\bfseries
  04} (2021) 073} [\href{https://arxiv.org/abs/2012.02726}{{\ttfamily
  2012.02726}}].

\bibitem{FrousteyTAUP2021}
J.~Froustey, \emph{{Precision calculation of neutrino evolution in the early
  Universe}}, \href{https://doi.org/10.1088/1742-6596/2156/1/012013}{\emph{J.
  Phys. Conf. Ser.} {\bfseries 2156} (2021) 012013}
  [\href{https://arxiv.org/abs/2110.11296}{{\ttfamily 2110.11296}}].

\bibitem{Froustey2019}
J.~Froustey and C.~Pitrou, \emph{Incomplete neutrino decoupling effect on big
  bang nucleosynthesis},
  \href{https://doi.org/10.1103/PhysRevD.101.043524}{\emph{Phys. Rev. D}
  {\bfseries 101} (2020) 043524}
  [\href{https://arxiv.org/abs/1912.09378}{{\ttfamily 1912.09378}}].

\bibitem{Simha:2008mt}
V.~Simha and G.~Steigman, \emph{{Constraining The Universal Lepton Asymmetry}},
  \href{https://doi.org/10.1088/1475-7516/2008/08/011}{\emph{JCAP} {\bfseries
  0808} (2008) 011} [\href{https://arxiv.org/abs/0806.0179}{{\ttfamily
  0806.0179}}].

\bibitem{Fields:2019pfx}
B.~D. Fields, K.~A. Olive, T.-H. Yeh and C.~Young, \emph{{Big-Bang
  Nucleosynthesis after Planck}},
  \href{https://doi.org/10.1088/1475-7516/2020/03/010}{\emph{JCAP} {\bfseries
  03} (2020) 010} [\href{https://arxiv.org/abs/1912.01132}{{\ttfamily
  1912.01132}}].

\bibitem{Oldengott:2017tzj}
I.~M. Oldengott and D.~J. Schwarz, \emph{{Improved constraints on lepton
  asymmetry from the cosmic microwave background}},
  \href{https://doi.org/10.1209/0295-5075/119/29001}{\emph{Europhys. Lett.}
  {\bfseries 119} (2017) 29001}
  [\href{https://arxiv.org/abs/1706.01705}{{\ttfamily 1706.01705}}].

\bibitem{Planck18}
{\scshape Planck} collaboration, \emph{{Planck 2018 results. VI. Cosmological
  parameters}},
  \href{https://doi.org/10.1051/0004-6361/201833910}{\emph{Astron. Astrophys.}
  {\bfseries 641} (2020) A6}
  [\href{https://arxiv.org/abs/1807.06209}{{\ttfamily 1807.06209}}].

\bibitem{Pitrou_2018PhysRept}
C.~Pitrou, A.~Coc, J.-P. Uzan and E.~Vangioni, \emph{Precision big bang
  nucleosynthesis with improved helium-4 predictions},
  \href{https://doi.org/https://doi.org/10.1016/j.physrep.2018.04.005}{\emph{Phys.
  Rept.} {\bfseries 754} (2018) 1}
  [\href{https://arxiv.org/abs/1801.08023}{{\ttfamily 1801.08023}}].

\bibitem{Neutrino_Cosmology}
J.~Lesgourgues, G.~Mangano, G.~Miele and S.~Pastor, \emph{{Neutrino
  Cosmology}}. Cambridge University Press, 2013,
  \href{https://doi.org/10.1017/CBO9781139012874}{10.1017/CBO9781139012874}.

\bibitem{Davidson_Leptogenesis}
S.~Davidson, E.~Nardi and Y.~Nir, \emph{{Leptogenesis}},
  \href{https://doi.org/10.1016/j.physrep.2008.06.002}{\emph{Phys. Rept.}
  {\bfseries 466} (2008) 105}
  [\href{https://arxiv.org/abs/0802.2962}{{\ttfamily 0802.2962}}].

\bibitem{McDonald:1999in}
J.~McDonald, \emph{{Naturally large cosmological neutrino asymmetries in the
  MSSM}}, \href{https://doi.org/10.1103/PhysRevLett.84.4798}{\emph{Phys. Rev.
  Lett.} {\bfseries 84} (2000) 4798}
  [\href{https://arxiv.org/abs/hep-ph/9908300}{{\ttfamily hep-ph/9908300}}].

\bibitem{March-Russell:1999hpw}
J.~March-Russell, H.~Murayama and A.~Riotto, \emph{{The Small observed baryon
  asymmetry from a large lepton asymmetry}},
  \href{https://doi.org/10.1088/1126-6708/1999/11/015}{\emph{JHEP} {\bfseries
  11} (1999) 015} [\href{https://arxiv.org/abs/hep-ph/9908396}{{\ttfamily
  hep-ph/9908396}}].

\bibitem{Gu:2010dg}
P.-H. Gu, \emph{{Large Lepton Asymmetry for Small Baryon Asymmetry and Warm
  Dark Matter}}, \href{https://doi.org/10.1103/PhysRevD.82.093009}{\emph{Phys.
  Rev. D} {\bfseries 82} (2010) 093009}
  [\href{https://arxiv.org/abs/1005.1632}{{\ttfamily 1005.1632}}].

\bibitem{Pastor:2001iu}
S.~Pastor, G.~G. Raffelt and D.~V. Semikoz, \emph{{Physics of synchronized
  neutrino oscillations caused by self-interactions}},
  \href{https://doi.org/10.1103/PhysRevD.65.053011}{\emph{Phys. Rev. D}
  {\bfseries 65} (2002) 053011}
  [\href{https://arxiv.org/abs/hep-ph/0109035}{{\ttfamily hep-ph/0109035}}].

\bibitem{Dolgov_NuPhB2002}
A.~D. {Dolgov}, S.~H. {Hansen}, S.~{Pastor}, S.~T. {Petcov}, G.~G. {Raffelt}
  and D.~V. {Semikoz}, \emph{{Cosmological bounds on neutrino degeneracy
  improved by flavor oscillations}},
  \href{https://doi.org/10.1016/S0550-3213(02)00274-2}{\emph{Nucl. Phys. B}
  {\bfseries 632} (2002) 363}
  [\href{https://arxiv.org/abs/hep-ph/0201287}{{\ttfamily hep-ph/0201287}}].

\bibitem{Abazajian2002}
K.~N. Abazajian, J.~F. Beacom and N.~F. Bell, \emph{{Stringent Constraints on
  Cosmological Neutrino Antineutrino Asymmetries from Synchronized Flavor
  Transformation}},
  \href{https://doi.org/10.1103/PhysRevD.66.013008}{\emph{Phys. Rev. D}
  {\bfseries 66} (2002) 013008}
  [\href{https://arxiv.org/abs/astro-ph/0203442}{{\ttfamily
  astro-ph/0203442}}].

\bibitem{Wong2002}
Y.~Y.~Y. Wong, \emph{{Analytical treatment of neutrino asymmetry equilibration
  from flavor oscillations in the early universe}},
  \href{https://doi.org/10.1103/PhysRevD.66.025015}{\emph{Phys. Rev. D}
  {\bfseries 66} (2002) 025015}
  [\href{https://arxiv.org/abs/hep-ph/0203180}{{\ttfamily hep-ph/0203180}}].

\bibitem{Bell:1998ds}
N.~F. Bell, R.~R. Volkas and Y.~Y.~Y. Wong, \emph{{Relic neutrino asymmetry
  evolution from first principles}},
  \href{https://doi.org/10.1103/PhysRevD.59.113001}{\emph{Phys. Rev. D}
  {\bfseries 59} (1999) 113001}
  [\href{https://arxiv.org/abs/hep-ph/9809363}{{\ttfamily hep-ph/9809363}}].

\bibitem{Pastor:2008ti}
S.~Pastor, T.~Pinto and G.~G. Raffelt, \emph{{Relic density of neutrinos with
  primordial asymmetries}},
  \href{https://doi.org/10.1103/PhysRevLett.102.241302}{\emph{Phys. Rev. Lett.}
  {\bfseries 102} (2009) 241302}
  [\href{https://arxiv.org/abs/0808.3137}{{\ttfamily 0808.3137}}].

\bibitem{Gava:2010kz}
J.~Gava and C.~Volpe, \emph{{CP violation effects on the neutrino degeneracy
  parameters in the Early Universe}},
  \href{https://doi.org/10.1016/j.nuclphysb.2010.04.024}{\emph{Nucl. Phys. B}
  {\bfseries 837} (2010) 50} [\href{https://arxiv.org/abs/1002.0981}{{\ttfamily
  1002.0981}}].

\bibitem{Gava_corr}
M.~C. Volpe, \emph{{Corrigendum to “CP violation effects on the neutrino
  degeneracy parameters in the Early Universe” [Nucl. Phys. B 837 (2010)
  50–60)]}},
  \href{https://doi.org/https://doi.org/10.1016/j.nuclphysb.2020.115035}{\emph{Nucl.
  Phys. B} {\bfseries 957} (2020) 115035}.

\bibitem{Mangano:2010ei}
G.~Mangano, G.~Miele, S.~Pastor, O.~Pisanti and S.~Sarikas, \emph{{Constraining
  the cosmic radiation density due to lepton number with Big Bang
  Nucleosynthesis}},
  \href{https://doi.org/10.1088/1475-7516/2011/03/035}{\emph{JCAP} {\bfseries
  03} (2011) 035} [\href{https://arxiv.org/abs/1011.0916}{{\ttfamily
  1011.0916}}].

\bibitem{Johns:2016enc}
L.~Johns, M.~Mina, V.~Cirigliano, M.~W. Paris and G.~M. Fuller, \emph{{Neutrino
  flavor transformation in the lepton-asymmetric universe}},
  \href{https://doi.org/10.1103/PhysRevD.94.083505}{\emph{Phys. Rev. D}
  {\bfseries 94} (2016) 083505}
  [\href{https://arxiv.org/abs/1608.01336}{{\ttfamily 1608.01336}}].

\bibitem{Barenboim:2016shh}
G.~Barenboim, W.~H. Kinney and W.-I. Park, \emph{{Resurrection of large lepton
  number asymmetries from neutrino flavor oscillations}},
  \href{https://doi.org/10.1103/PhysRevD.95.043506}{\emph{Phys. Rev. D}
  {\bfseries 95} (2017) 043506}
  [\href{https://arxiv.org/abs/1609.01584}{{\ttfamily 1609.01584}}].

\bibitem{SiglRaffelt}
G.~{Sigl} and G.~{Raffelt}, \emph{{General kinetic description of relativistic
  mixed neutrinos}},
  \href{https://doi.org/10.1016/0550-3213(93)90175-O}{\emph{Nucl. Phys. B}
  {\bfseries 406} (1993) 423}.

\bibitem{McKellarThomson}
B.~H.~J. McKellar and M.~J. Thomson, \emph{{Oscillating neutrinos in the early
  universe}}, \href{https://doi.org/10.1103/PhysRevD.49.2710}{\emph{Phys. Rev.
  D} {\bfseries 49} (1994) 2710}.

\bibitem{BlaschkeCirigliano}
D.~N. Blaschke and V.~Cirigliano, \emph{Neutrino quantum kinetic equations: The
  collision term},
  \href{https://doi.org/10.1103/PhysRevD.94.033009}{\emph{Phys. Rev. D}
  {\bfseries 94} (2016) 033009}
  [\href{https://arxiv.org/abs/1605.09383}{{\ttfamily 1605.09383}}].

\bibitem{GiuntiKim}
C.~Giunti and C.~W. Kim, \emph{{Fundamentals of Neutrino Physics and
  Astrophysics}}. Oxford University Press, 2007.

\bibitem{PDG}
{\scshape Particle Data Group} collaboration, \emph{{Review of Particle
  Physics}}, \href{https://doi.org/10.1093/ptep/ptaa104}{\emph{Prog. Theor.
  Exp. Phys.} {\bfseries 2020} (2020) 083C01}.

\bibitem{MSW_W}
L.~Wolfenstein, \emph{{Neutrino Oscillations in Matter}},
  \href{https://doi.org/10.1103/PhysRevD.17.2369}{\emph{Phys. Rev. D}
  {\bfseries 17} (1978) 2369}.

\bibitem{MSW_MS}
S.~Mikheyev and A.~Smirnov, \emph{{Resonance Amplification of Oscillations in
  Matter and Spectroscopy of Solar Neutrinos}}, {\emph{Sov. J. Nucl. Phys.}
  {\bfseries 42} (1985) 913}.

\bibitem{Mirizzi2012}
A.~Mirizzi, N.~Saviano, G.~Miele and P.~D. Serpico, \emph{Light sterile
  neutrino production in the early universe with dynamical neutrino
  asymmetries}, \href{https://doi.org/10.1103/PhysRevD.86.053009}{\emph{Phys.
  Rev. D} {\bfseries 86} (2012) 053009}
  [\href{https://arxiv.org/abs/1206.1046}{{\ttfamily 1206.1046}}].

\bibitem{Malkus:2014iqa}
A.~Malkus, A.~Friedland and G.~C. McLaughlin, \emph{{Matter-Neutrino Resonance
  Above Merging Compact Objects}},
  \href{https://arxiv.org/abs/1403.5797}{{\ttfamily 1403.5797}}.

\bibitem{Samuel:1993uw}
S.~Samuel, \emph{{Neutrino oscillations in dense neutrino gases}},
  \href{https://doi.org/10.1103/PhysRevD.48.1462}{\emph{Phys. Rev. D}
  {\bfseries 48} (1993) 1462}.

\bibitem{Kostelecky:1993yt}
V.~A. Kostelecky, J.~T. Pantaleone and S.~Samuel, \emph{{Neutrino oscillation
  in the early universe}},
  \href{https://doi.org/10.1016/0370-2693(93)90156-C}{\emph{Phys. Lett. B}
  {\bfseries 315} (1993) 46}.

\bibitem{Kostelecky:1993dm}
V.~A. Kostelecky and S.~Samuel, \emph{{Neutrino oscillations in the early
  universe with an inverted neutrino mass hierarchy}},
  \href{https://doi.org/10.1016/0370-2693(93)91795-O}{\emph{Phys. Lett. B}
  {\bfseries 318} (1993) 127}.

\bibitem{Kostelecky:1993ys}
V.~A. Kostelecky and S.~Samuel, \emph{{Nonlinear neutrino oscillations in the
  expanding universe}},
  \href{https://doi.org/10.1103/PhysRevD.49.1740}{\emph{Phys. Rev. D}
  {\bfseries 49} (1994) 1740}.

\bibitem{Grohs:2016cuu}
E.~Grohs, G.~M. Fuller, C.~T. Kishimoto and M.~W. Paris, \emph{{Lepton
  asymmetry, neutrino spectral distortions, and big bang nucleosynthesis}},
  \href{https://doi.org/10.1103/PhysRevD.95.063503}{\emph{Phys. Rev. D}
  {\bfseries 95} (2017) 063503}
  [\href{https://arxiv.org/abs/1612.01986}{{\ttfamily 1612.01986}}].

\bibitem{Mangano:2011ip}
G.~Mangano, G.~Miele, S.~Pastor, O.~Pisanti and S.~Sarikas, \emph{{Updated BBN
  bounds on the cosmological lepton asymmetry for non-zero $\theta_{13}$}},
  \href{https://doi.org/10.1016/j.physletb.2012.01.015}{\emph{Phys. Lett. B}
  {\bfseries 708} (2012) 1} [\href{https://arxiv.org/abs/1110.4335}{{\ttfamily
  1110.4335}}].

\bibitem{Balantekin:2007es}
A.~B. Balantekin, J.~Gava and C.~Volpe, \emph{{Possible CP-violation effects in
  core-collapse supernovae}},
  \href{https://doi.org/10.1016/j.physletb.2008.03.038}{\emph{Phys. Lett. B}
  {\bfseries 662} (2008) 396}
  [\href{https://arxiv.org/abs/0710.3112}{{\ttfamily 0710.3112}}].

\bibitem{Gava:2008rp}
J.~Gava and C.~Volpe, \emph{{Collective neutrino oscillations in matter and CP
  violation}}, \href{https://doi.org/10.1103/PhysRevD.78.083007}{\emph{Phys.
  Rev. D} {\bfseries 78} (2008) 083007}
  [\href{https://arxiv.org/abs/0807.3418}{{\ttfamily 0807.3418}}].

\bibitem{Castorina2012}
E.~Castorina, U.~Franca, M.~Lattanzi, J.~Lesgourgues, G.~Mangano, A.~Melchiorri
  et~al., \emph{{Cosmological lepton asymmetry with a nonzero mixing angle
  $\theta_{13}$}},
  \href{https://doi.org/10.1103/PhysRevD.86.023517}{\emph{Phys. Rev. D}
  {\bfseries 86} (2012) 023517}
  [\href{https://arxiv.org/abs/1204.2510}{{\ttfamily 1204.2510}}].

\bibitem{Hansen:2020vgm}
R.~S.~L. Hansen, S.~Shalgar and I.~Tamborra, \emph{{Neutrino flavor mixing
  breaks isotropy in the early universe}},
  \href{https://doi.org/10.1088/1475-7516/2021/07/017}{\emph{JCAP} {\bfseries
  07} (2021) 017} [\href{https://arxiv.org/abs/2012.03948}{{\ttfamily
  2012.03948}}].

\bibitem{Heckler_PhRvD1994}
A.~F. {Heckler}, \emph{{Astrophysical applications of quantum corrections to
  the equation of state of a plasma}},
  \href{https://doi.org/10.1103/PhysRevD.49.611}{\emph{Phys. Rev. D} {\bfseries
  49} (1994) 611}.

\bibitem{Bennett2020}
J.~J. Bennett, G.~Buldgen, M.~Drewes and Y.~Y. Wong, \emph{{Towards a precision
  calculation of the effective number of neutrinos $N_{\rm eff}$ in the
  Standard Model I: The QED equation of state}},
  \href{https://doi.org/10.1088/1475-7516/2020/03/003}{\emph{JCAP} {\bfseries
  03} (2020) 003} [\href{https://arxiv.org/abs/1911.04504}{{\ttfamily
  1911.04504}}].

\bibitem{Landau1932}
L.~D. {Landau}, \emph{{On the theory of energy transmission in collisions.
  II}}, {\emph{Phys. Zs. Sowjet} {\bfseries 2} (1932) 46}.

\bibitem{Zener1932}
C.~{Zener}, \emph{{Non-Adiabatic Crossing of Energy Levels}},
  \href{https://doi.org/10.1098/rspa.1932.0165}{\emph{Proceedings of the Royal
  Society of London Series A} {\bfseries 137} (1932) 696}.

\bibitem{Haxton:1986bc}
W.~C. Haxton, \emph{{Analytic Treatments of Matter Enhanced Solar Neutrino
  Oscillations}}, \href{https://doi.org/10.1103/PhysRevD.35.2352}{\emph{Phys.
  Rev. D} {\bfseries 35} (1987) 2352}.

\bibitem{Abazajian:2001nj}
K.~Abazajian, G.~M. Fuller and M.~Patel, \emph{{Sterile neutrino hot, warm, and
  cold dark matter}},
  \href{https://doi.org/10.1103/PhysRevD.64.023501}{\emph{Phys. Rev. D}
  {\bfseries 64} (2001) 023501}
  [\href{https://arxiv.org/abs/astro-ph/0101524}{{\ttfamily
  astro-ph/0101524}}].

\bibitem{Wittig}
C.~Wittig, \emph{{The Landau-Zener formula }}, {\emph{J. Phys. Chem. B}
  {\bfseries 109 (17)} (2005) 8428}.

\end{thebibliography}\endgroup

\end{document}